\documentclass{mn2e}

\usepackage[dvips]{graphicx}

\begin{document}

\title[Baryons and dark matter haloes]{The joint evolution of baryons and dark matter haloes}

\author[Pedrosa, Tissera, Scannapieco]{Susana Pedrosa$^{1,2}$, Patricia B. Tissera$^{1,2}$, Cecilia Scannapieco$^{3}$\\
$^1$  Consejo Nacional de Investigaciones Cient\'{\i}ficas y T\'ecnicas, CONICET, Argentina.\\
$^2$ Instituto de Astronom\'{\i}a y F\'{\i}sica del Espacio, Casilla de Correos 67, Suc. 28, 1428, Buenos Aires, Argentina.\\
$^3$ Max-Planck Institute for Astrophysics, Karl-Schwarzchild Str. 1, D85748, Garching, Germany\\
}

\maketitle

\begin{abstract}

We have studied the dark matter (DM) distribution in a  $\approx 10^{12} h^{-1}  M_{\sun}$ mass halo extracted from a simulation consistent with the concordance cosmology, where the physics regulating the transformation of gas into stars was allowed to change producing galaxies with different morphologies.  The presence of baryons produces the  concentration of the DM halo with respect to its corresponding dissipationless run, but we found that this response does not only depend on the amount of baryons gathered in the central region but also on the way they have been assembled. DM and baryons affect each other in a complex way so the formation history of a galaxy plays an important role on its final total mass distribution. Supernova (SN) feedback regulates the star formation and triggers galactic outflows not only in the central  galaxy but also in its satellites. Our results suggest that, as the effects of SN feedback get stronger, satellites get less massive and can even be more easily disrupted by dynamical friction, transferring less angular momentum. We found indications that this angular momentum could be acquired not only by the outer part of the DM halo but also by the inner ones and  by the stellar component in the central galaxy. The latter effect produces stellar migration which contributes to change the inner potential well, probably  working against further DM contraction.  As a consequence of the action of these processes, when the  halo hosts a galaxy with an important disc structure formed by smooth gas accretion, it is more concentrated than when it hosts a spheroidal  system which experienced more massive mergers and interactions. We also found that in the later case, the halo has less radial velocity anisotropy than when the halo hosts a disc galaxy. In most of our runs with baryons, we do not  detect the inversion of the 
velocity dispersion characteristic of the dissipationless haloes.
 We have found that rotation velocities for the systems that were able to develop a disc structure  are in good agreement with the observations and none of them have been formed satisfying the adiabatic contraction hypothesis.

\end{abstract}


\section{Introduction}
Over the last  decades numerical simulations have became a powerful tool to study the validity of cosmological models.
$\Lambda$-CDM scenarios have been found to be able to reproduce successfully the global properties of the observed structure  at large scales.
However, at galactic scale, the so-called concordance ($\Lambda$-CDM) paradigm has been challenged by several  observational results. In this respect, the cuspy inner profile obtained in many of the numerical simulations has been claimed  not to be consistent with the rotation curves observed for low surface brightness (Flores \& Primack 1994; Moore 1994; Dutton, van den Bosch \& Courteau 2008; Salucci, Yegorova \& Drory 2008) and dwarf galaxies (e.g. Gnedin \& Zhao 2002). There have been many attempts to explain the possible erasement of the inner cusp via different mechanisms: interactions through dynamical friction with the substructure (e.g. El-Zant et al. 2001, Tonini, Lapi \& Salucci 2006), SN feedback  (Mashchenko et al. 2006) etc., but the problem is not yet solved.
· 

Another important  pending issue is the overabundance of small dark matter (DM) subhaloes (e.g. Moore et al. 1999; Stadel et al. 2009). The total number of subhaloes found in the simulations is much larger than the number of known satellite galaxies surrounding the Milky Way.
The inner shape of the DM  density profiles and the abundance of subhaloes are of particular interest because they can be  used to perform several observational diagnostics such as gravitational lensing (Zackrisson \& Riahm 2009). Also, the prospect of detecting DM particles annihilation makes it essential a deep understanding of the small scale distribution of the DM in the central regions of our Galaxy (Diemand et al. 2008, Springel et al. 2008). 

Although the ``universality'' of the spherically-averaged density profiles (Navarro, Frenk and White 1996, hereafter NFW) of $\Lambda$-CDM  haloes has gained a broad consensus over the past couple of decades, new evidences based on high resolution simulations have been found against it (Navarro et al. 2004, Merrit et al. 2006, Gao et al. 2008). Merrit et al. (2006) tested different fitting functions and found that the profiles were better described by a de Vaucouleurs' law instead of the NFW two parameter formula. They also found a systematic variation in the profile shape with halo mass. Navarro  et al. (2008, hereafter N08) analysed different galaxy-sized haloes simulated with unprecedent high resolution and found small but significant deviations from the so-called NFW universal profile. 

The contraction of the DM haloes due to the infall and condensation of baryons in the central regions is a well accepted process
 (e.g. Barnes \& White 1984). Many attempts to predict their effects have been made through  models based on the adiabatic contraction (AC) hypothesis  such as the one developed by Blumenthal et al. (1986, hereafter B86). However, it has been shown that this kind of models overestimates the level of contraction. More recently,  Gnedin et al. (2004) and Sellwood \& McGaugh (2005) developed AC models based on the work of Young (1980) which considered the possibility of radial motions. However, all  AC-based models  missed the hierarchical characteristic of the assembly of the galaxy as suggested by the current large scale observations (e.g. S\'anchez et al. 2006), which might have non negligible consequences on the final distribution of
the DM (e.g. Debatistta et al. 2008).

In a fully cosmological context, simulations  have already shown how the DM haloes concentrate  when  baryons are included, providing hints of a possible dependence on the assembly history  (e.g. Tissera \& Dom\'{\i}nguez-Tenreiro et al. 1998; Gnedin et al. 2004; O\~norbe et al. 2007). Recently Romano-D\'{\i}az et al. (2008) analysed the evolution of the central DM profile   in cosmologically grown galactic haloes, claiming that  when baryons are present the cusp is gradually levelled off. They suggest that this effect could be associated with the action of the subhaloes that heat up the cusp region of the DM halo through dynamical friction, and force it to expand (Ma \& Boylan-Kolchin 2004; Debatistta et al. 2008). 

In order to help shed light on these issues, we study a set of intermediate resolution cosmological simulations where different baryonic structures have been able to  form from identical initial conditions via the modification of the physics of baryons. 
The set of simulations studied in our work are those analysed by Scannapieco et al. (2008, hereafter S08) where  the star formation activity and the SN feedback were modified in order to study the role played by each of these processes. As a consequence, a variety of galaxies were obtained, each one exhibiting different morphological and dynamical properties. These experiments allow us to analyse how the dark matter evolves when baryons are assembled in a different fashion but governed by the same underlying merger tree. 
The first results of this analysis were reported by  Pedrosa, Tissera \& Scannapieco (2009) where it is clearly shown that the final structure of the dark matter halo  depends on the way baryons are put together and not solely on the amount of baryons gathered in the centre. 
We also found hints for a  re-distribution of angular momentum related to the accretion of satellites.  
In this paper, we extend this work and analyse in detail the DM haloes considering the evolution of baryons.

 This paper is organized as follows. In Section 2, we describe the numerical experiments and summarize the main features of the simulated galaxies. In Section 3, we analyse the DM density profiles. In Section 4, we study the interaction with satellites. In Section 5, rotation curves and the AC hypothesis are analysed. In Section 6 we summarize our main results.

\section{The Numerical Experiments}

We analysed a set of six realizations of a  $\approx 10^{12} h^{-1}$  M$_{\odot}$ mass  halo, run with an
extended version of the code GADGET-2  according to Scannapieco et al. (2005, 2006). This extended GADGET-2 code was designed to improve the representation of the ISM and SN feedback by including a new multiphase model for the gas component, metal-dependent
 cooling, chemical enrichment and energy feedback by SN events.

The initial  condition corresponds to an $\approx 10^{12} h^{-1}\ $M$_{\odot}$ halo extracted from a cosmological simulation and re-simulated with higher resolution. This halo was required to have no major mergers since $z=1$. The simulations have been run from $z=38$ to $z=0$ and are consistent with 
 a $\Lambda$CDM universe with 
$\Omega_{\Lambda}=0.7$, 
$\Omega_{m}=0.3$, $\Omega_{b}=0.04$, 
$\sigma_{8}=0.9$
 and $H_{0}= 100 h \ {\rm km} \ {\rm s}^{-1}\ {\rm Mpc}^{-1}$, with $h=0.7$. The dark matter particle mass 
is $1.6\times 10^{7} h^{-1}\ $M$_{\odot}$ while initially the gas mass particle is  $2.4\times 10^{6} h^{-1}\  $M$_{\odot}$.
 The maximum gravitational softening used is $\epsilon_{g}=0.8 h^{-1}$ kpc. 

The analysed simulations  have the same initial condition but have been run using  different input
parameters for the Star Formation (SF) and SN feedback models as it can be seen in Table ~\ref{tab1}. The version of {\small GADGET-2} used to run these
simulations
includes the multiphase model for the interstellar medium, the SF algorithm and SN feedback
presented by Scannapieco et al. (2005, 2006). This set of simulations was performed by 
S08 who varied the star formation efficiency ($c$), the fraction of SN energy ($\epsilon_{\rm c}$) injected into the cold phase (and correspondingly the fraction of SN energy that is pumped into the hot phase) and the total energy released during a SN explosion ($E_{\rm SN}$). 
As a result, DM haloes host baryonic structures with different morphologies since the transformation of gas into stars has been 
regulated differently in each run.

In order to be able to assess the effects of galaxy formation  on the dark matter haloes, for this work, we  performed a pure gravitational run (DM-only) of the same initial condition used by S08, with a dark matter  particle of $1.84\times 10^{7} h^{-1}\ $M$_{\odot}$.

\subsection{SF and SN feedback parameters and resulting morphology}

The hydrodynamical simulations used in this work have been analysed in detail by S08, particularly the different properties of the
simulated galaxy at $z=0$. In this section, we only summarize their main characteristics to facilitate the interpretation 
of our results.

The first point to note is that when  SN feedback is not included, the gas collapses and concentrates at the centre of the potential well very efficiently. In this case, the star formation  follows the gas collapse and there is no mechanism to  regulate the SF activity. As a result, a stellar spheroidal component is formed very early,  consuming most of the gas reservoir and preventing the formation of a disc structure at later times, as it  is the case in the NF halo.
 On the contrary, when the SN feedback is included, the SF activity gets self-regulated as a consequence of the heating and pressurizing of the interstellar gas. In this case, disc-like components can be formed, populated mainly by young stars (S08). 
Depending on the combination of SF and SN parameters, the baryons settle down determining structures with different morphologies
and disc components of different sizes. In some cases, the systems are dominated by a spheroidal component.

S08 adopted the standard value of $E_{\rm SN}=10^{51}$ erg per SN event and, then, varied it from 0.3 to 3  $\times10^{51}$ erg as shown in Table ~\ref{tab1}. As larger values for the SN energy are assumed, the SF is more strongly quenched and more violent
winds are able to develop,  resulting in lower final stellar and gas masses. In E-3, where  the unrealistic value of $3\times10^{51}$ erg per SN is assumed,  most of the gas is blown out, producing  the most DM dominated system  with the less concentrated DM profile in our set of hydrodynamical simulations.
Alternatively, S08 varied the fraction  $\epsilon_{\rm c}$ of energy pumped into the cold phase producing a stronger inhibition of the 
star formation activity due to increased strength of the induced galactic winds (F-0.9). This particular combination of parameters
results in a very extended disc structure.
A decrease in the star formation efficiency (C-0.01) produces a slower rate of transformation of gas into stars with weaker starbursts so that the energy injected into the ISM is not able to generate strong galactic winds. The regulation of the star formation
activity is not enough to prevent an important early consumption of gas into stars.

In summary, the final simulated galaxies have the following characteristics.
At $z=0$, the galaxy formed in  NF run is dominated by an extended spheroid, with most of its stars
formed at $ z > 2$. The E-0.7 and F-0.9 runs have been able to produce galaxies with important disc components
as a result of the regulation of SF by SN feedback. These systems have a half mass radius ($r_{d}$) of 5.72 kpc $h^{-1}$ and 9.74 kpc $h^{-1}$, respectively. And, they also show the largest disc to spheroid mass ratios ($D/S$): 0.82 for E-0.7 and 0.98 for F-0.9. The E-0.3 run was able to develop a small ($r_{\rm d}=4.75$) and thick disc ($D/S = 0.6$). In the case of C-0.01 run the disc component is almost negligible. In the E-3 run, the large amount of energy assumed per SN  triggers  violent outflows which expel a significant amount of the gas content of the main galaxy. 
The stellar masses of the final simulated galaxies vary from 15.9 to 1.3 $\times 10^{10} h^{-1} $M$_{\odot}$, as summarized in Table \ref{tab1}.

\begin{table*}
  \begin{center}
  \caption{Main characteristics of  DM haloes and their main central galaxies. For each simulation we provide the values of SF and SN parameters: $\epsilon_{\rm c}$, $c$ and $E_{SN} $ (see Section 2 for details).
 We show the virial radius $r_{\rm 200}$, the virial mass $M_{\rm 200}$, total stellar mass $M_{\rm s}$ of the central galaxy,
the  total-mass-to-stellar
mass ratio $M_{\rm t}/M_{\rm s}$, the $n$ and $r_{-2}$ Einasto parameters, the
 inner logarithm slope 
$\gamma_{\rm inner}$, the logarithmic slope of the baryonic circular velocity $LS$ and the ratio 
$V_{\rm max}/V_{\rm 200}$. $M_{\rm t}$ and $M_{\rm s}$ are evaluated within twice 
the  optical radius defined as that  enclosing $83\%$ of the baryons in the central region. Bootstraps errors for $n$ 
and $r_{-2}$ are shown within parenthesis.}
  \label{tab1}
 {\scriptsize
  \begin{tabular}{|l|c|c|c|c|c|c|c|c|c|c|c|c|}\hline
{\bf Run}  & {\bf $\epsilon_{\rm c}$}  & {\bf $E_{\rm SN}$}  & {\bf $c$}  & {\bf $r_{\rm 200}$} & {\bf $M_{\rm 200}$} & {\bf $M_{\rm s}$} &  {\bf $M_{\rm t}/M_{\rm s}$} &  {\bf $n$} & {\bf $r_{-2}$} & {\bf $\gamma_{\rm inner}$} & {$ LS $} & {$V_{\rm max}/V_{\rm 200} $}\\
   & & & &kpc $h^{-1}$& $10^{10} h^{-1}\ $M$_{\odot}$& $10^{10} h^{-1}\ $M$_{\odot}$ & & &kpc $h^{-1}$ &&\\ \hline
NF & - & - & 0.1 & 217.5 & 224.4 & 15.9 &  4.1 &  5.973 (1)& 16.89 (1) & 1.24  & -0.22 & 1.46 \\ 
F-0.9 & 0.9 & 1 & 0.1 & 212.0 & 212.3 & 6.7 &  7.7 &  6.476 (1)& 17.83 (1) & 1.24  & -0.08 & 1.25 \\ 
C-0.01 & 0.5 & 1 & 0.01 & 207.7 & 202.2 & 10.0 &  3.7 &  6.499 (1)& 16.56 (1) & 1.24 & -0.20 & 1.43 \\ 
E-0.3 & 0.5 & 0.3 & 0.1 & 214.7 & 220.8 & 13.2 &  3.8 &  6.764 (1)& 14.76 (1) & 1.24 & -0.17 & 1.47 \\ 
E-0.7 & 0.5 & 0.7 & 0.1 & 211.1 & 209.0& 7.5 &  5.1 &  6.887 (4)& 15.35 (1)& 1.28 & -0.08 & 1.33 \\ 
E-3 & 0.5 & 3 & 0.1 & 205.8 & 194.0 & 1.3 & 37.2 & 5.585 (1)& 21.90 (1)& 1.15 & -0.19 & 1.17 \\ 
DM-only & - & - & - & 217.0 & 227.4 & - & - & 5.239 (3) & 24.06 (1) & 1.08 & - & - \\ 

  \end{tabular}
  }
 \end{center}
\vspace{1mm}
\end{table*}

\section{Dark matter density profiles}

All analysed DM haloes have been identified at their virial radius ($r_{200}$) defined as  the one that enclosed an sphere of mean density $200 \times \rho_{crit}$ where $\rho_{crit}$ is the closure density of the Universe.
We have checked that, at $z=0$, all virialised haloes are  relaxed as indicated by a relax parameter of $\approx  0.002$ (Neto et al. 2007).
On average,  DM  haloes have more than a total of 120000 particles within their virial radius.

In order to construct the DM profiles, we first eliminate the substructures within $r_{200}$, which affects the profiles mainly in the outer regions near $r_{200}$. 
The determination of the centre of mass  of the haloes is of great importance for the analysis of the dark matter distribution since a displacement could produce an artificial erasement of the inner cusp. We used the shrinking sphere method proposed by  Power et al. (2003) in order to find it. We calculate spherically-averaged density profiles between three times $\epsilon_{g}$ and the virial radius. We use three $\epsilon_{g}$ as the inner radius to increase the numerical robustness of our fittings. 

We calculate spherically-averaged DM profiles using logarithmic binning of the DM distribution cleaned of substructures. 
Then, we fit the NFW, Jaffe (Jaffe, W. 1983)  and Einasto (Einasto, J. 1965) expressions to the DM profiles finding that the Einasto's model provides the best fit in all cases. The Einasto's formula can be written as:

\begin{equation}
ln\left( \frac{\rho\left( r\right) }{\rho_{-2}} \right) = \left(\ -2n\right) \left[ \left( \frac{r}{r_{-2}}\right) ^{1/n}-1\right]  
\label{eq1}
\end{equation}

where  $n$, $r_{-2}$  and $\rho_{-2}$ indicate the sharpness of the profiles and the radius and density where their logarithmic slope takes the isothermal value. The number of free parameters is reduced to two by imposing the extra constrain of 
obtaining the total mass at the virial radius. 
 The fitting values obtained for the different haloes are shown in Table \ref{tab1}. 
We estimated bootstrap errors for $n$ and $r_{-2}$ by fitting the Einasto's formula to 100 
randomly-generated realizations
of the DM  profiles and by estimating the standard dispersion over the generated set of  parameters. 

\begin{figure}
\resizebox{8cm}{!}{\includegraphics{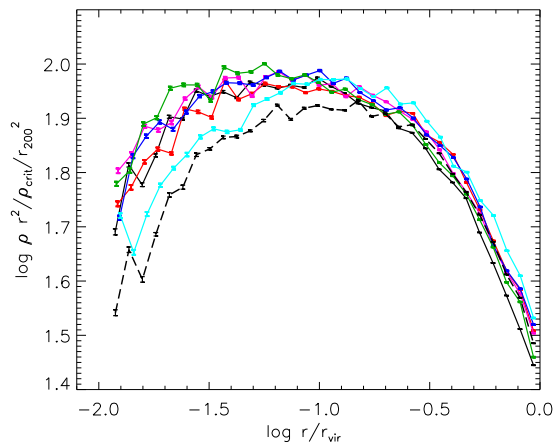}}
\hspace*{-0.2cm}
\caption{Spherically-averaged DM profiles for the NF (black line), E-0.7 (magenta line), F-0.9 (red line), E-0.3 (green line), C-0.01 (blue line), E-3 (cyan line)  and  DM-only (black thick dashed line)  experiments. The  inner most bin corresponds to three times the  gravitational softenings.
The errorbars have been estimated by the boostrap resampling technique. } 
\label{fig1}
\end{figure}

\begin{figure*}
\hspace*{-0.2cm}\resizebox{7cm}{!}{\includegraphics{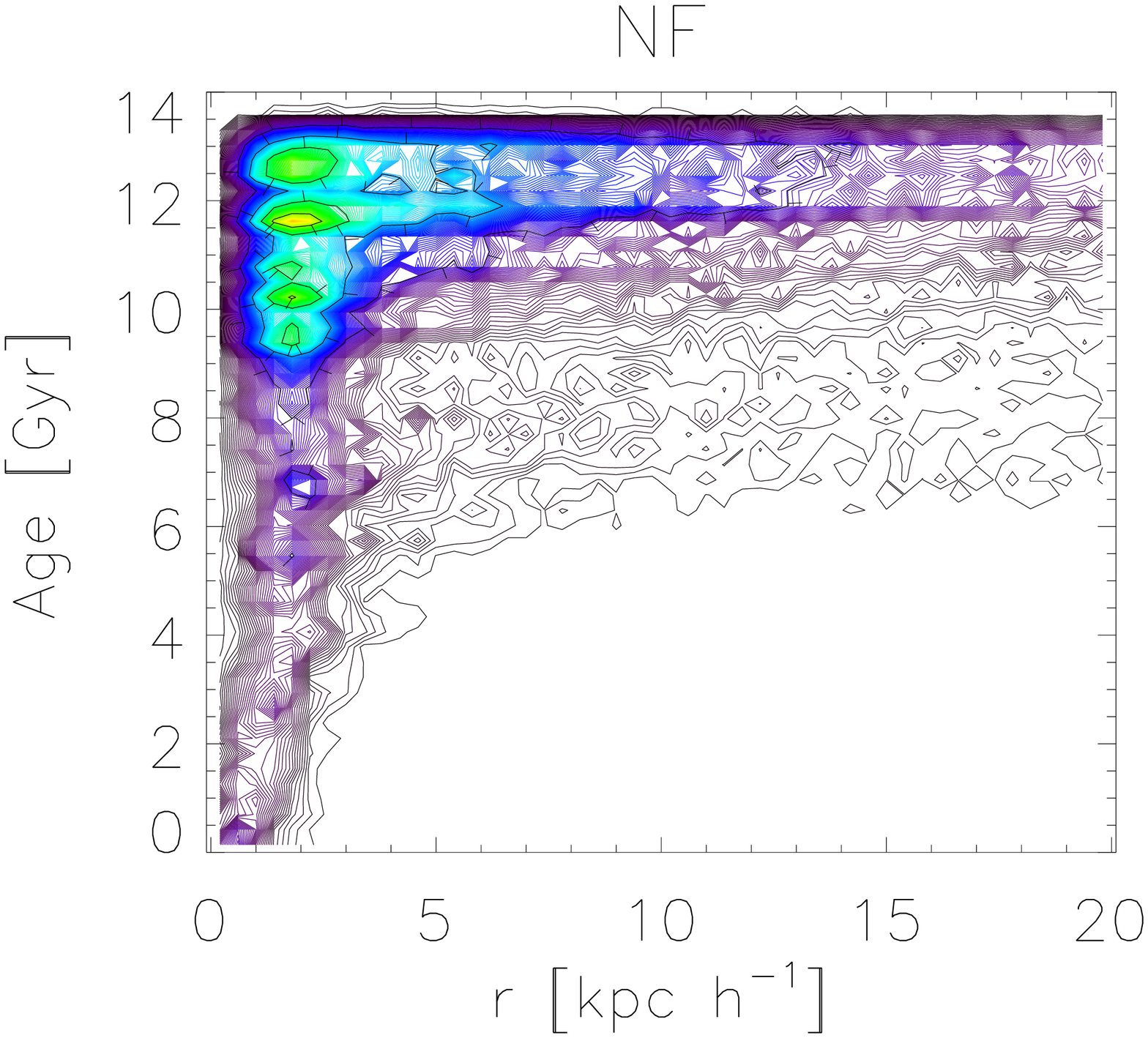}}
\hspace*{-0.2cm}\resizebox{7cm}{!}{\includegraphics{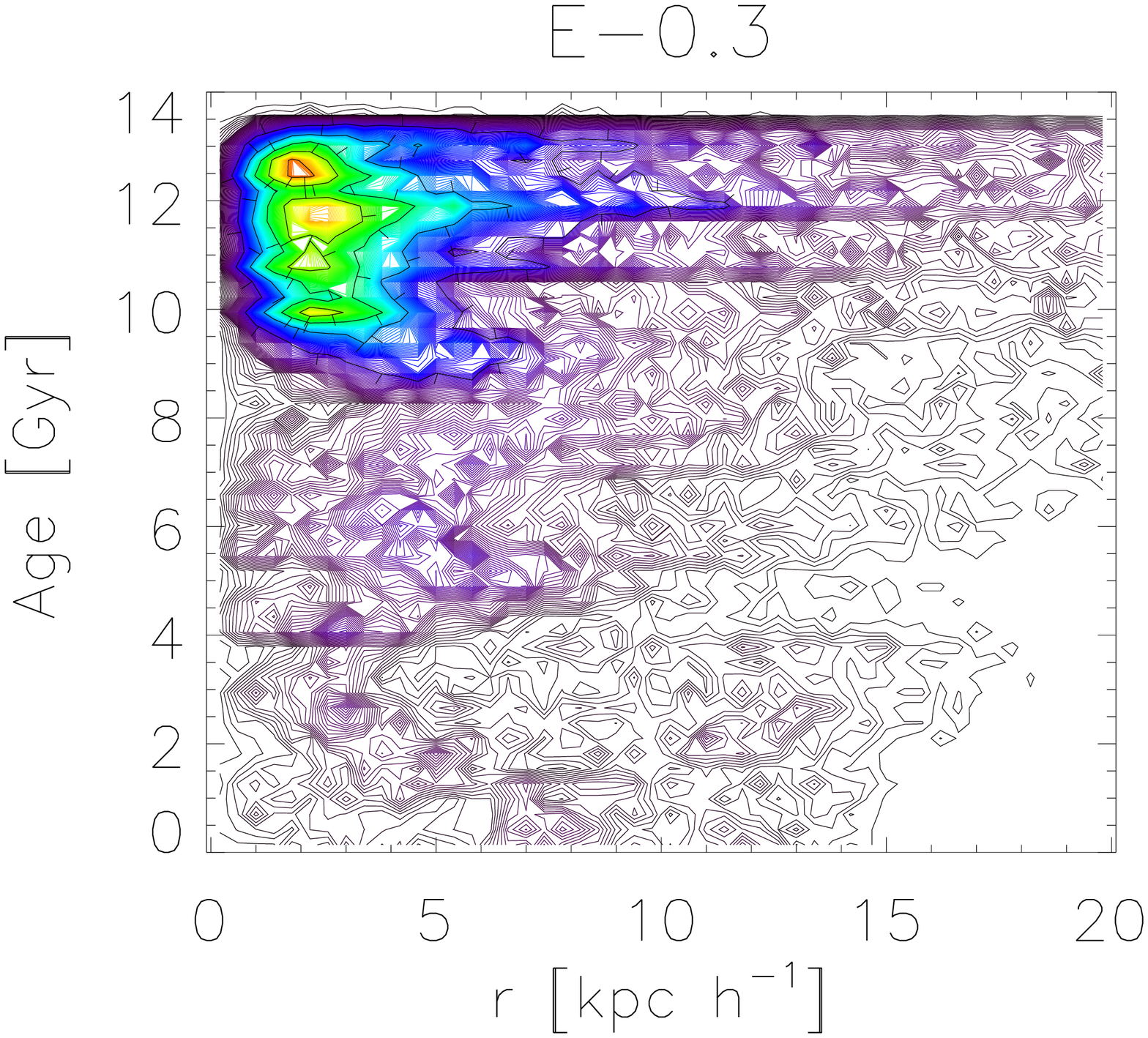}}
\hspace*{-0.2cm}\resizebox{7cm}{!}{\includegraphics{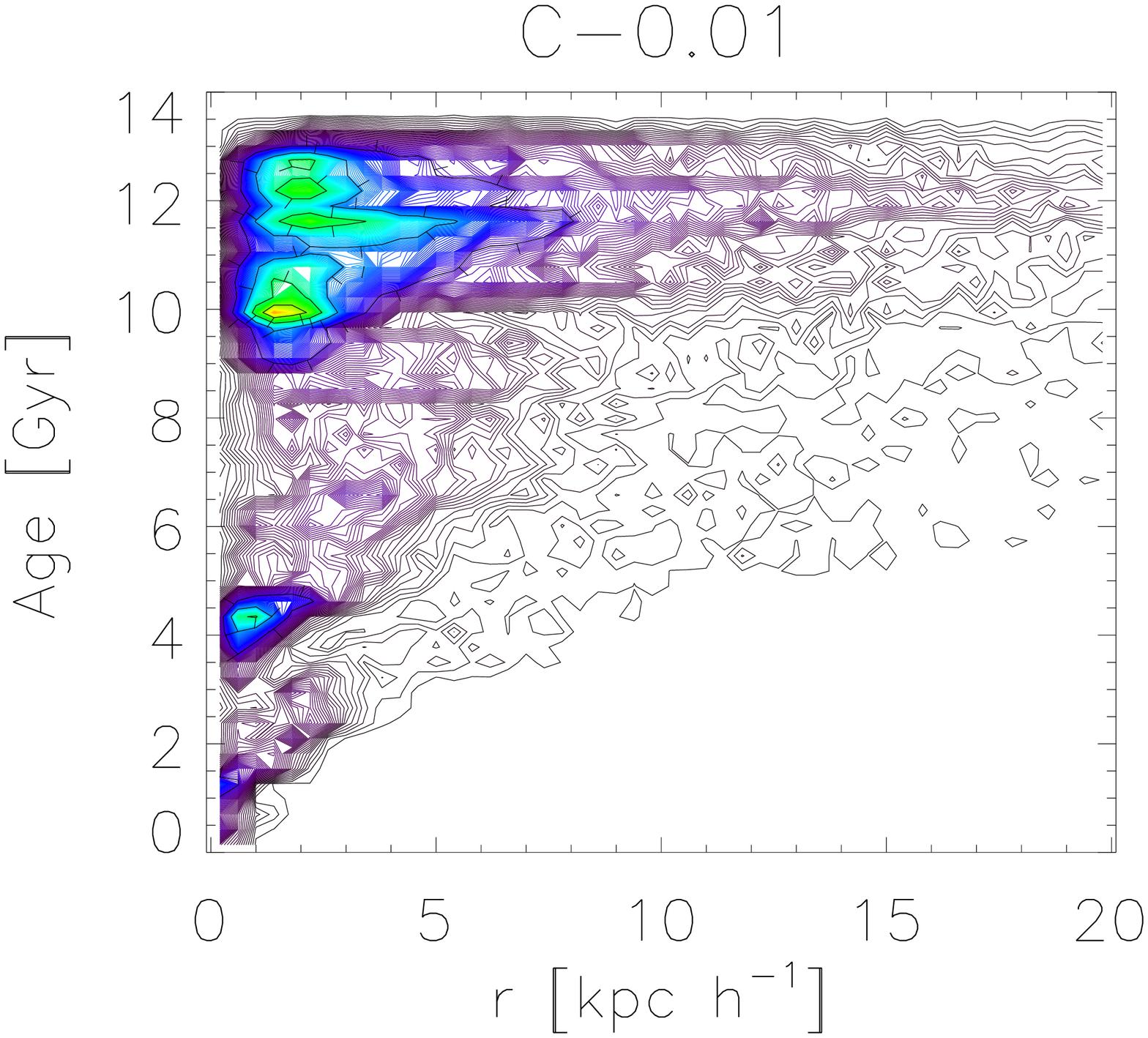}}
\hspace*{-0.2cm}\resizebox{7cm}{!}{\includegraphics{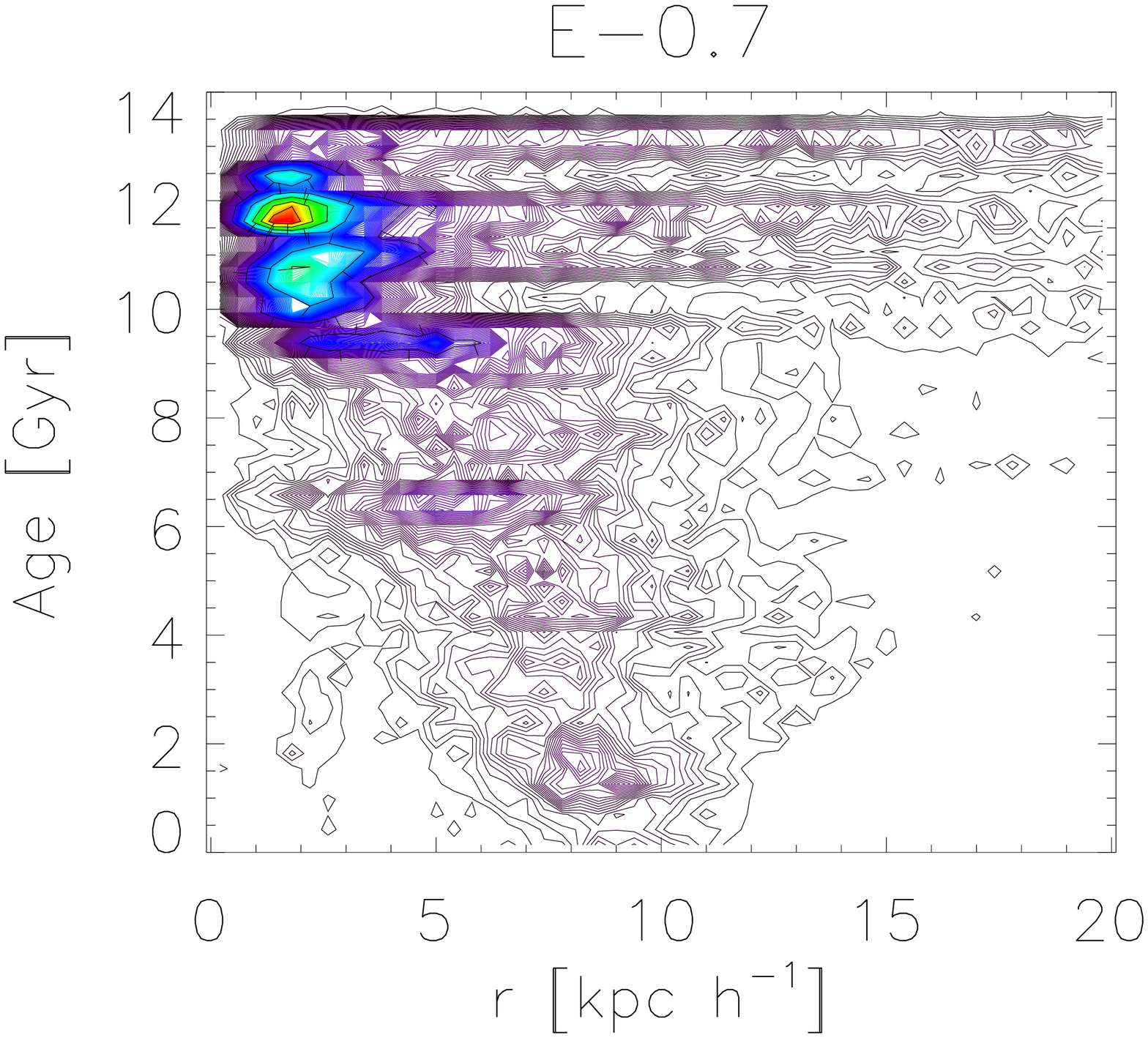}}
\hspace*{-0.2cm}\resizebox{7cm}{!}{\includegraphics{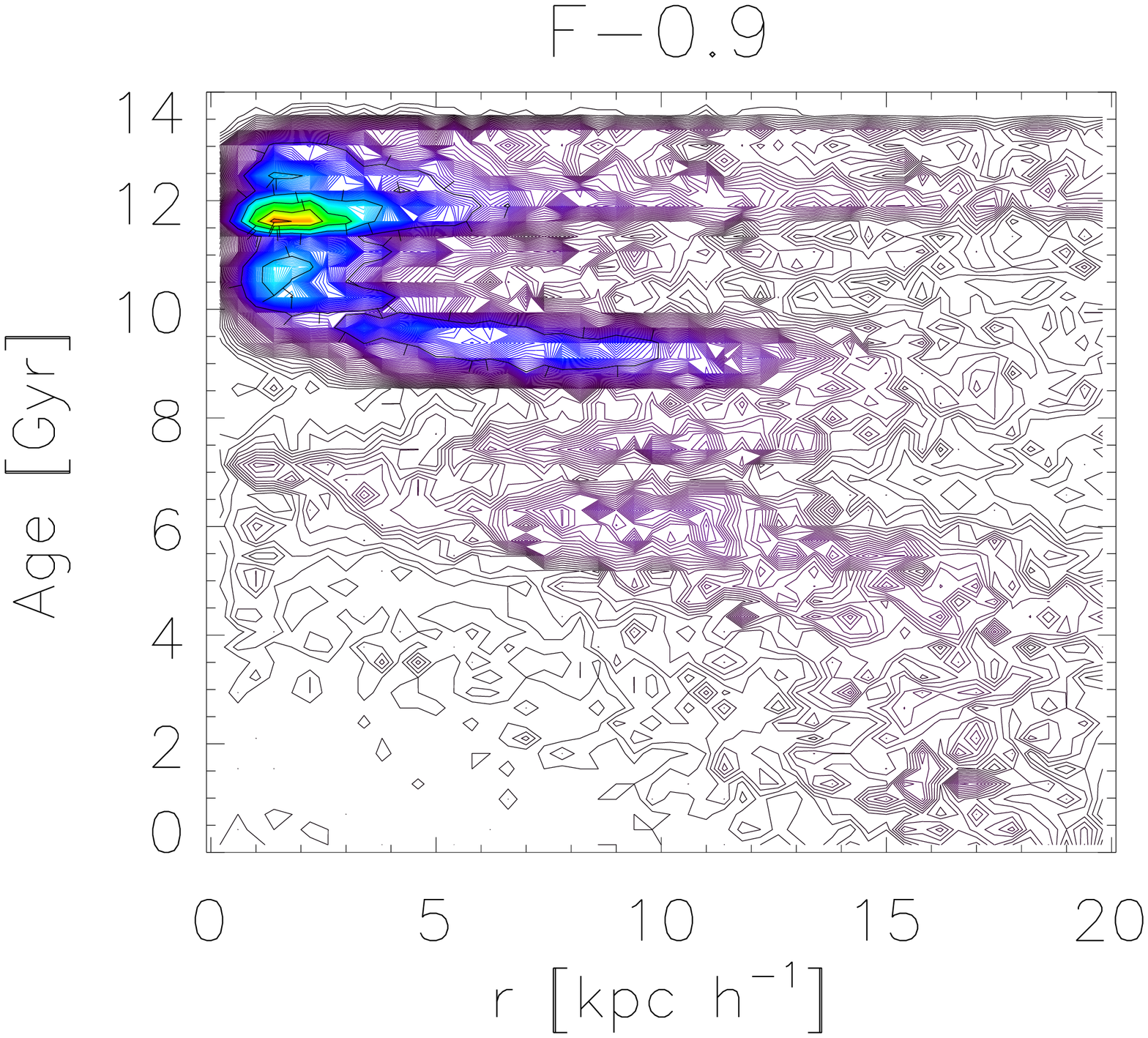}}
\hspace*{-0.2cm}\resizebox{7cm}{!}{\includegraphics{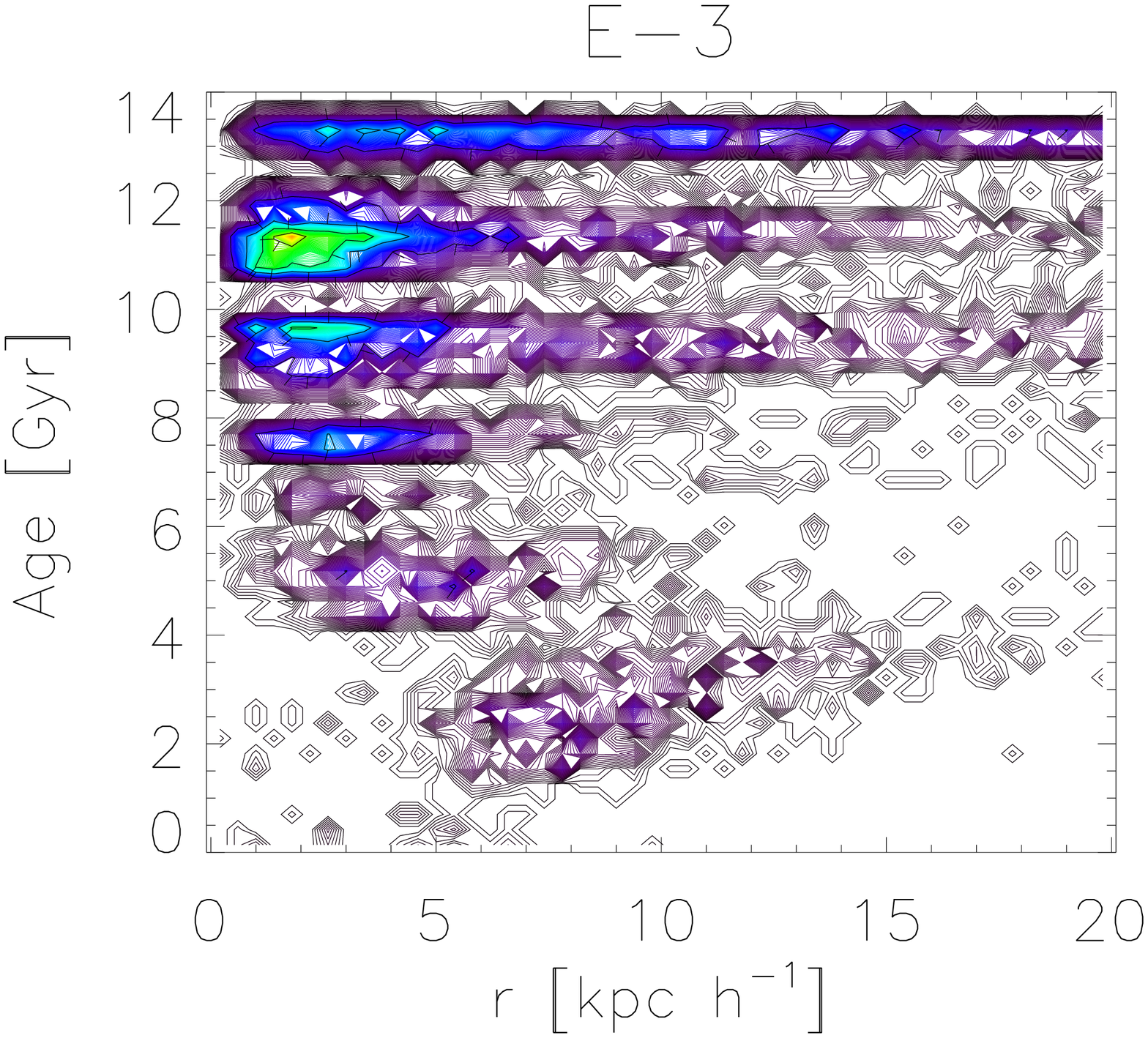}}
\hspace*{-0.2cm}\\
\caption{Age-radial distance maps of the stars in the NF (upper left panel), E-0.3 (upper right panel), C-0.01 (middle left panel), E-0.7 (middle right panel), F-0.9 (lower left panel) and E-3 (lower right panel) experiments. This figure shows the inside-out formation of the disc in E-0.7 and F-0.9, and the outside-in formation of the spheroid in NF and C-0.01.} 
\label{histories}
\end{figure*}

When baryons are present, the shape of the DM profiles in the central regions changes significantly in comparison to the DM-only case  as it can be appreciated from   Fig.~\ref{fig1}. In the case of the DM-only run, its profile is sharper (i.e. smaller $n$ value)
than those of haloes with baryons, indicating the increase in the DM concentration in the latter cases. Interestingly, all  haloes, except for the E-3 (the most DM dominated) and the DM-only cases, present a nearly isothermal behaviour in the region dominated by baryons, in agreement with the results found by Tissera et al. (2009) who analysed several dark matter haloes with higher numerical resolution as part of the Aquarius project. 

However, as it can be seen from Table \ref{tab1}, each DM halo has different fitting parameters. In order to understand the origin of these differences, we  correlate their properties with those of the galaxy they  host. In fact, the comparison between the  E-0.7 and the NF profiles shows that the DM distribution in E-0.7  is more concentrated than in NF, although it hosts a galaxy a factor of two less massive
than the later.  This finding suggests that the total amount of baryons collected within the central region of a halo is  not the only relevant factor  affecting the response of the DM to the presence of baryons.

In Fig. ~\ref{histories}, we display the density-contour maps of  the age of the stars associated to each simulated galaxy as a function of tridimensional
distance to the centre of mass. These maps provide a picture of both the star formation history and the final stellar mass distribution in each galaxy.
Those systems with an important disc structure (F-0.9 and E-0.7) have most of the stars  younger than 8 Gyr located
outside the central region, while systems dominated by an old stellar spheroidal component (NF, E-0.3, C-0.01) have most of their stars in the inner region.

The NF and  E-0.7 runs have produced very different galaxies as shown in  Fig.\ref{histories}.
  The galaxy in the  NF run is dominated by
old stars, determining a spatially extended spheroid with  $78\%$ of the final stellar mass older than 10 Gyr.
The system in  E-0.7 has a compact old spheroid and an important disc component populated by younger stars. 
This disc is able to survive the interaction with satellites at lower redshifts. 
Their different star formation histories and morphologies are the result of the action of the  SN feedback in E-0.7 which was successful at regulating the transformation of gas into stars, preventing the formation of an early, extended spheroid and assuring the existence of gas to form a disc later on. The SN feedback has also affected the formation of stars in the satellite systems that merged with the main object as we will discussed in detail in Section 4. 

A similar  trend is  found when comparing the E-0.3 and C-0.01 runs. In these two cases the feedback was not that efficient at regulating the SF activity. As a result the final stellar masses are only slightly lower than the NF case.
 But interestingly, the E-0.3  has a more concentrated DM profile than  C-0.01. We note that while C-0.01 formed an old extended spheroid,  E-0.3 has an old  spheroid but it was able to develop  a smaller and thicker disc. 

For the F-0.9 run, we found that it has also a more concentrated profile than the NF case. And, again, this can be linked to a similar pattern in  its formation history: F-0.9 and E-0.7 both have compact old spheroids and  extended, inside-out-formed discs.
In these last two runs, the total amount of stars within the central region is approximately a factor of two lower than
in the case of the NF run. 

In E-3, where an extreme  value for the energy per SN was assumed, we obtained the less concentrated  DM profile among the cases with baryons. This profile is weakly more concentrated than the DM-only one. 
As expected, the galaxy in E-3 is also the most DM dominated one in the central region, as shown by the total
to stellar mass ratio $M_{\rm t}/M_{\rm s}=37$ (Table~\ref{tab1}). 
In this simulation, most of the gas has been blown away and  only a
 small fraction of stars has been formed in a bursty fashion as shown in Fig. ~\ref{histories}.

\subsection{Velocity Dispersion}

\begin{figure}
\hspace*{-0.2cm}\resizebox{7cm}{!}{\includegraphics{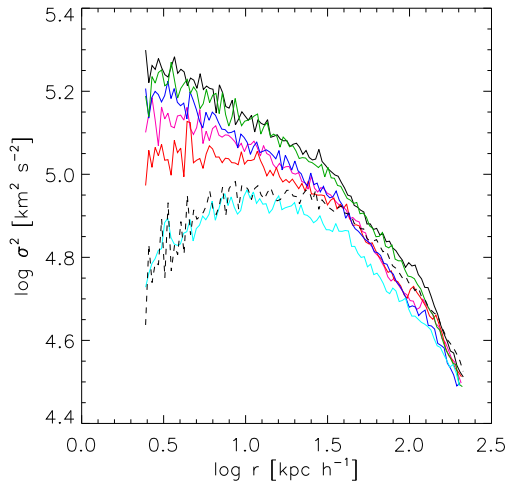}}
\hspace*{-0.2cm}\\
\caption{Velocity dispersion  as a function of radius from three times the gravitational softening: NF (black line), E-0.7 (magenta line), F-0.9 (red line), E-0.3 (green line), C-0.01 (blue line), E-3 (cyan line)  and  DM-only (black thick dashed line).} 
\label{Sigma}
\end{figure}

In Fig.~\ref{Sigma}, we plotted the velocity dispersion $\sigma$ as a function of the radius from three times the gravitational softening. We found that when baryons are included the $\sigma^2$ profiles increase  in the central regions compared to the DM-only cases so that
the 'temperature inversion' typical of the NFW profiles is lost (e.g. Tissera et al. 1998; Romano-D\'{\i}az et al. 2008; see also Tissera et al. 2009 for high resolution simulations).
  N08 found a similarity between the $\sigma ^2$ profiles and $\rho r^2$ and they proposed that this may be due to a scaling relation between densities and velocity dispersions in haloes. We found that this similarity only holds for the E-3 run, where most of the baryons have been blown away due to the strong galactic winds produced as a consequence of the extreme high energy assumed per SN.

From Fig.~\ref{Sigma} and Table ~\ref{tab1}, we can see that there is a correlation between the inner slope of the $\sigma^2$ profile and the stellar mass in the simulated galaxies, so that the higher the mass, the steeper the inner profile. From the analysis of the profiles of the progenitor system as a function of redshift, we found that, in the DM-only case, the temperature inversion is present from at least $z \approx 2$ to $z=0$, but in the runs with baryons, the inversion profile is never at place (Romano-D\'{\i}az et al. 2008).

\subsubsection{Velocity Anisotropy}

\begin{figure}
\hspace*{-0.2cm}\resizebox{7cm}{!}{\includegraphics{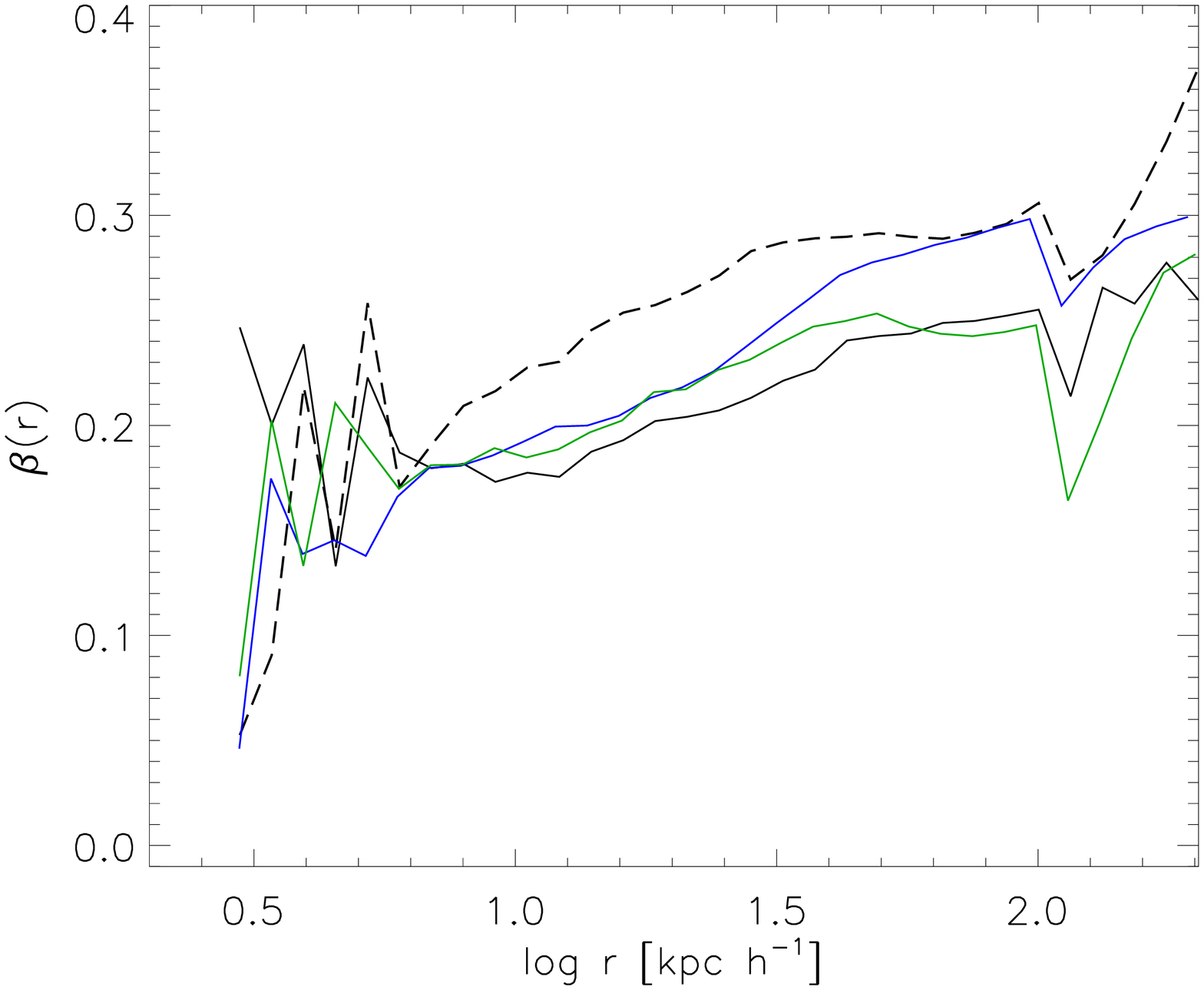}}
\hspace*{-0.2cm}\resizebox{7cm}{!}{\includegraphics{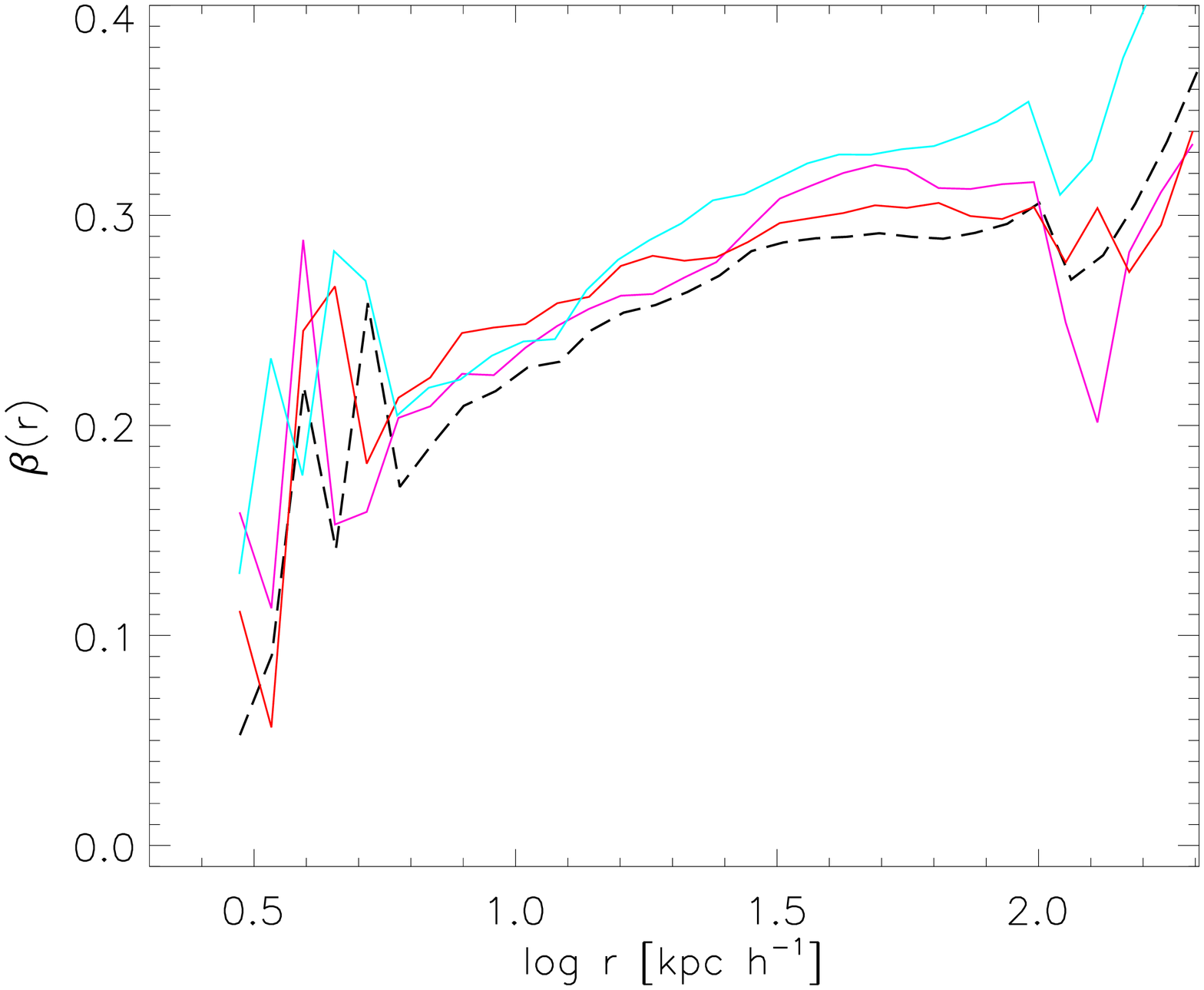}}
\hspace*{-0.2cm}\\
\caption{ Velocity anisotropy parameter $\beta$ as a function of radius  for the NF (black line), E-0.3 (green line), C-0.01 (blue line) runs (upper panel), and, for E-0.7 (magenta line), F-0.9 (red line) and E-3 (cyan line) runs (lower panel), at $z=0$. In both panels, the dashed line corresponds to the DM-only run.} 
\label{Beta}
\end{figure}

To provide a measure of the velocity structure of the haloes, we calculate the anisotropy parameter $\beta = 1 - \frac{\sigma_{t}^{2}}{2 \sigma_{r}^{2}}$, where $\sigma_{t}$ and $\sigma_{r}$ are the dispersions of the tangential and radial motions. For an isotropic distribution, $\beta$ should be zero while for a system dominated by radial motions, it should have positive values. Fig.~\ref{Beta} shows the anisotropy parameter as a function of radius for the haloes at $z=0$. When baryons are included the central anisotropy tends to increase slightly in most cases, although because of the high level of noise, this result should be confirmed by higher numerical resolution. Interestingly, from $r \approx 10$ kpc $h^{-1}$, we found those haloes  hosting spheroid-dominated galaxies to have weaker level of radial anisotropy compared with the DM-only case (upper panel of Fig.~\ref{Beta}). Conversely, those haloes hosting disc-dominated systems have higher velocity anisotropies (lower panel of Fig.~\ref{Beta}).

\section{Interaction with satellites}

\begin{figure}
\resizebox{8cm}{!}{\includegraphics{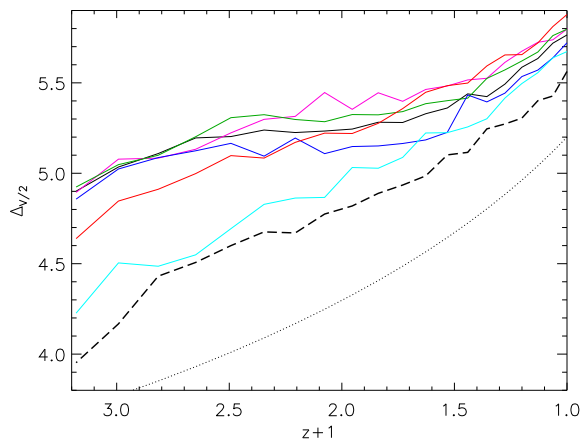}}
\hspace*{-0.2cm}
\caption{Central halo mass concentration $\Delta_{v/2}$ as function of redshift for NF (black line), E-0.7 (magenta line), F-0.9 (red line), E-0.3 (green line), C-0.01 (blue line), E-3 (cyan line)  and  DM-only (black dashed line) haloes. 
The dotted line is the expected growth for a 
constant density due to the expansion of the Universe.} 
\label{delta_v2}
\end{figure}

The analysis of the DM profiles suggests a connection between the DM evolution and the history of formation of the baryonic structures. We can follow the formation of a halo and its galaxy with time but it is not possible to reliably estimate the
DM profiles when the system get smaller because of the high numerical noise present in our simulations.
Then, in order to assess the evolution of the different runs,  we calculated the concentration parameter proposed by Alam, Bullock \& Weinberg (2002), $\Delta_{v/2}$, as a function of redshift. This parameter measures the mean DM density normalized to the cosmic closure density within the radius at which the circular rotational speed due to the DM alone rises to half its maximum value. This parameter has the advantage of being independent of a specific density profile and, as it is an integrated quantity, it can be estimated more robustly at any time. In  Fig.~\ref{delta_v2}, we show $\Delta_{v/2}$  as a function of redshift for the progenitor haloes for our set of simulations.  The dotted line shows the expected relation for a constant density perturbation due to the expansion of the Universe alone (i.e. hereafter critical relation). All haloes increase their concentration as they grow with time.
The DM-only run has the lowest  concentration, as expected, at all times. We note that the relation flattens between $z \approx 1$ and $z \approx 1.8$, coinciding with the close approaching of satellites as we will discuss in more detail later on.
From Fig.~\ref{delta_v2},  we can see that haloes  hosting baryons follow different paths between them. The NF run shows a DM halo
which is always more concentrated than the DM-only one but it has a stronger flattening of the relation during the same period of time. The other haloes do not show such a strong change in the slope, except for C-0.01 run. Note that a flat slope
is indicating an expansion of the mass distribution in the central regions.
Another interesting case is that of the F-0.9 halo which host the most extended and important disc. This system does not show a 
change in the slope and follows the critical  relation  even closer than the DM-only case.
A similar behaviour can be observed for  E-3 which shows
a slightly higher concentration driven by the presence of the baryons that have been able to settle in. 
In these runs, the entrances of satellites presumably cause weaker effects on the dark matter distribution compared to the  DM-only run, since they are less massive due to the strong action of SN feedback and are easily disrupted as they fall in.

Recall that the merger trees of these haloes are the same with the only difference being the  fraction of 
baryons and the gas reservoir in each substructure, which depend on the SF and the SN feedback parameters adopted in each run. So, in order to understand the origin of the different evolution of  halo concentration, we 
analyse the satellites within the virial radius of the progenitor objects as a function of redshift

It has been shown in previous works (e.g. Barnes \& Hernquist 1992) that when two systems collide, orbital angular momentum can be transfered from the baryonic clumps to the dark matter haloes. Hence, it might be possible that if the properties
of the satellite distribution were affected by the choice of SF and SN parameters, 
 they might also transfer different amount of angular momentum to the DM. 

\begin{figure*}
\hspace*{-0.2cm}\resizebox{5cm}{!}{\includegraphics{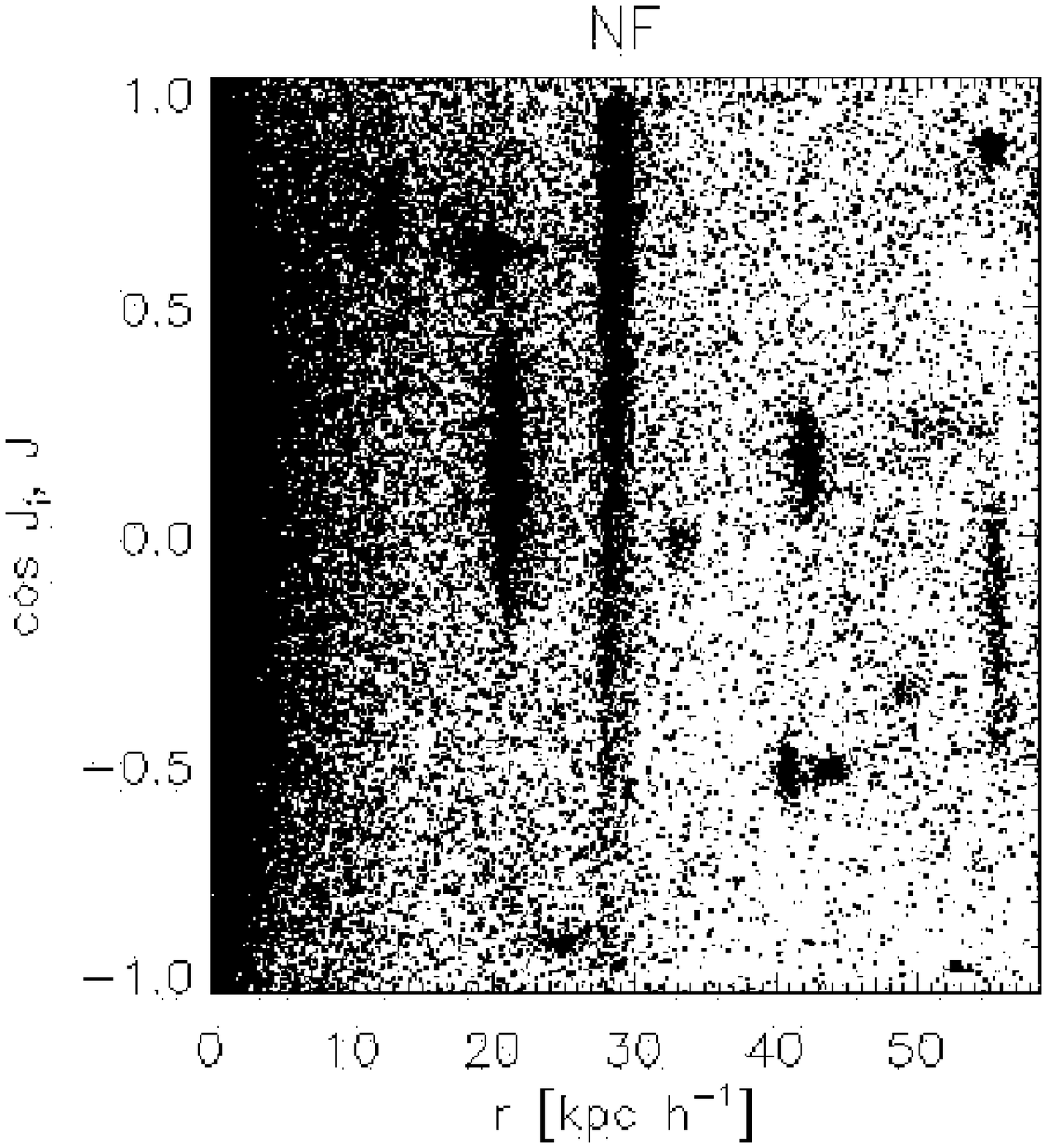}}
\hspace*{-0.2cm}\resizebox{5cm}{!}{\includegraphics{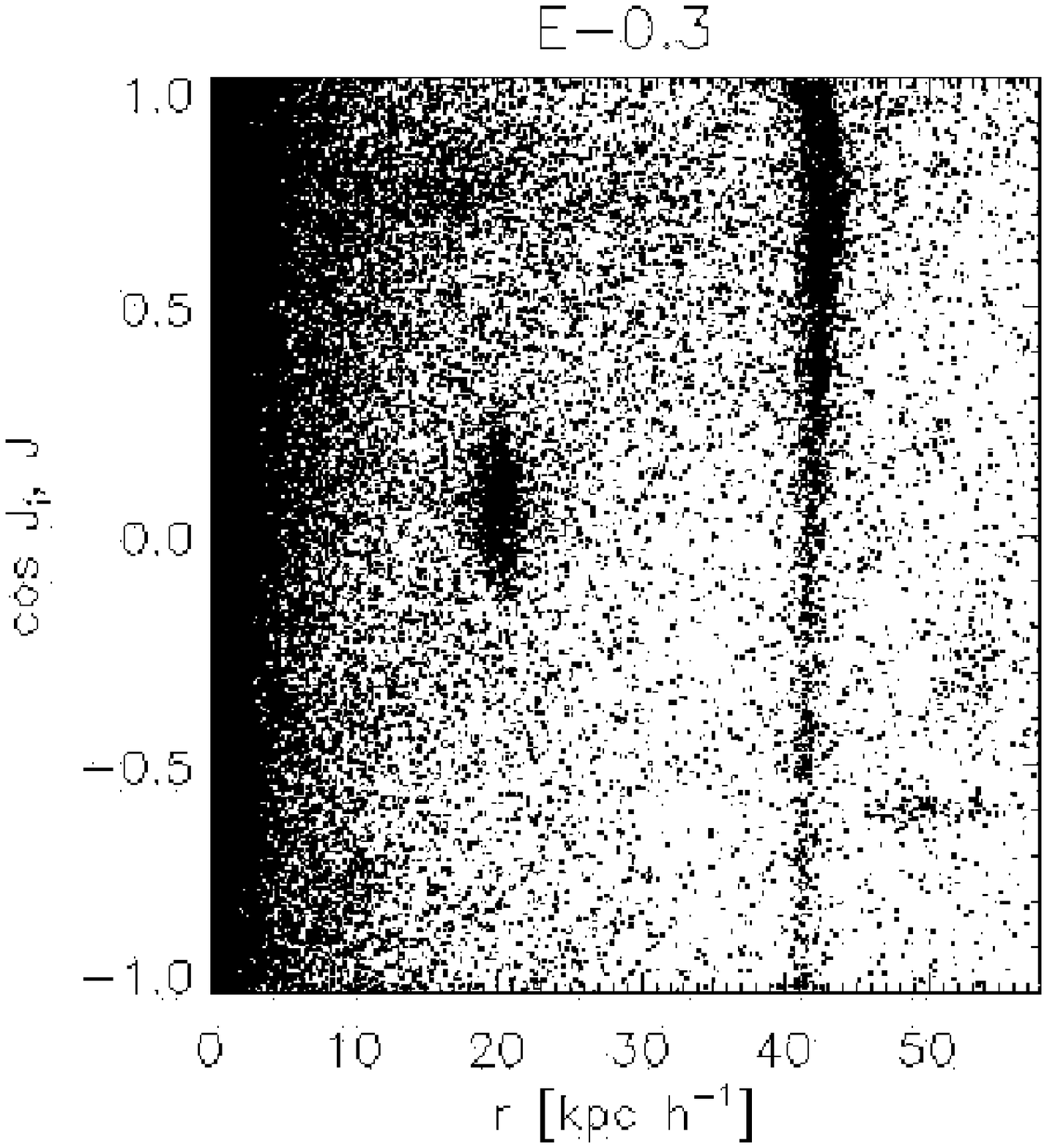}}
\hspace*{-0.2cm}\resizebox{5cm}{!}{\includegraphics{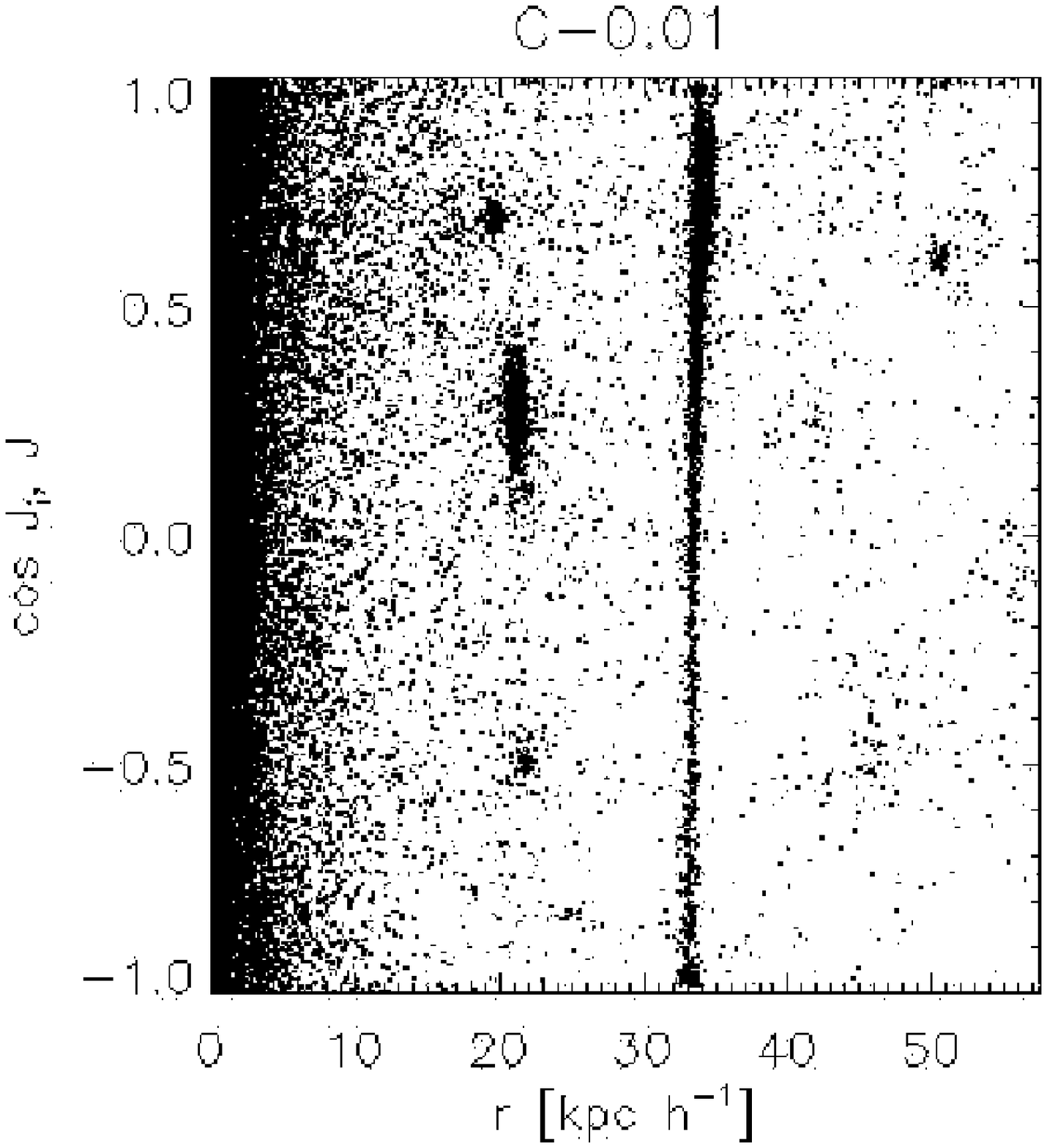}}
\hspace*{-0.2cm}\resizebox{5cm}{!}{\includegraphics{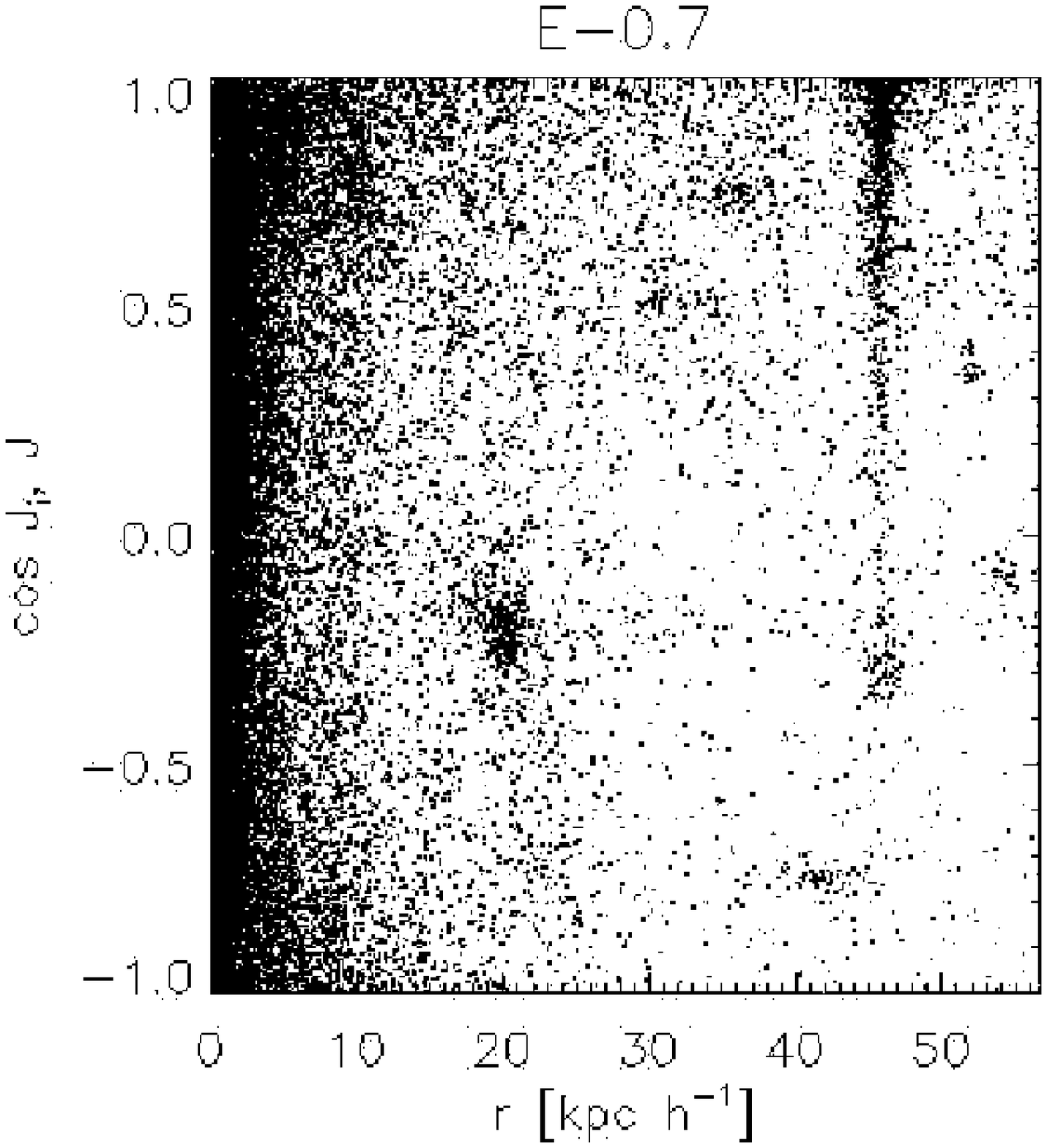}}
\hspace*{-0.2cm}\resizebox{5cm}{!}{\includegraphics{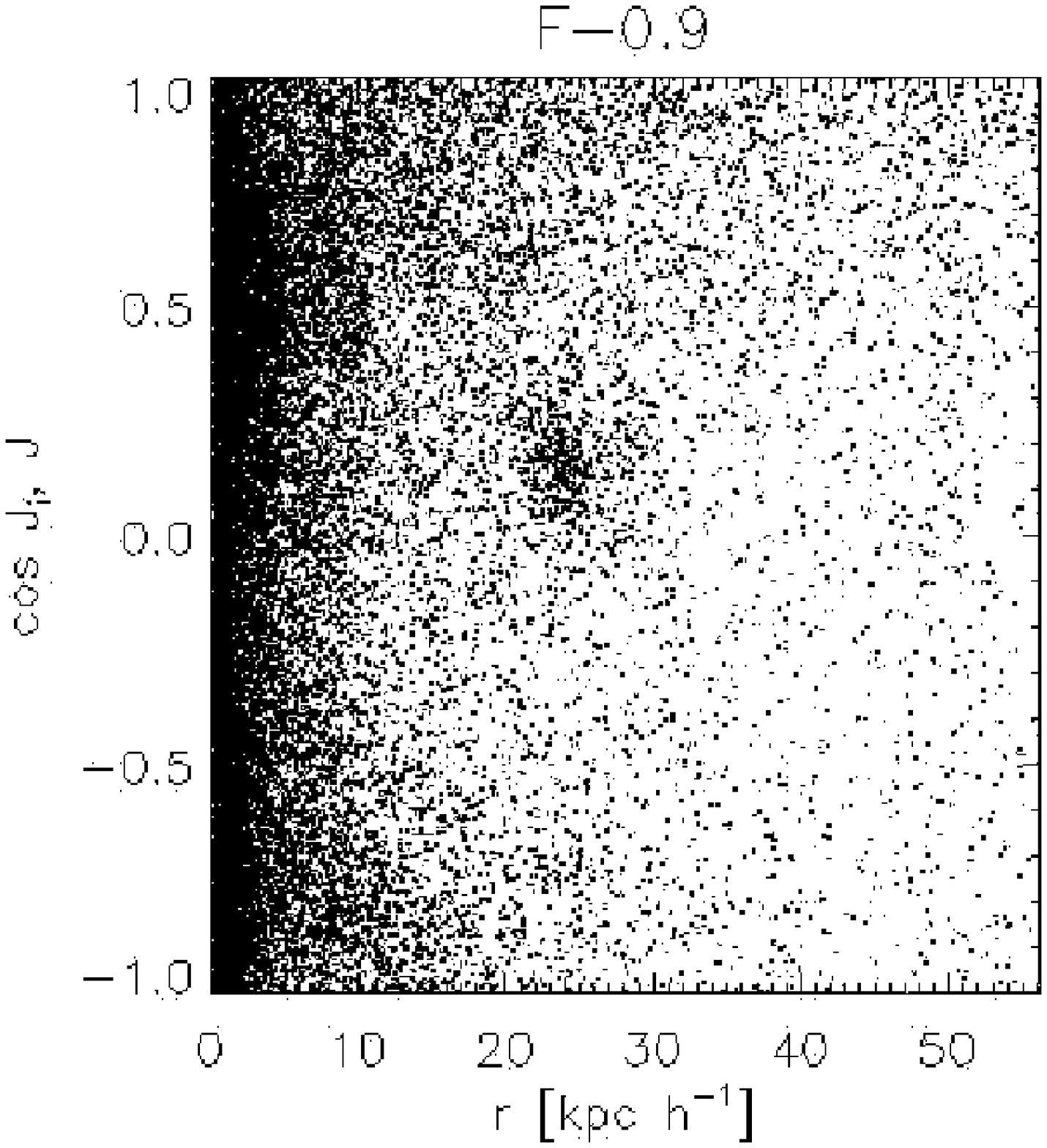}}
\hspace*{-0.2cm}\resizebox{5cm}{!}{\includegraphics{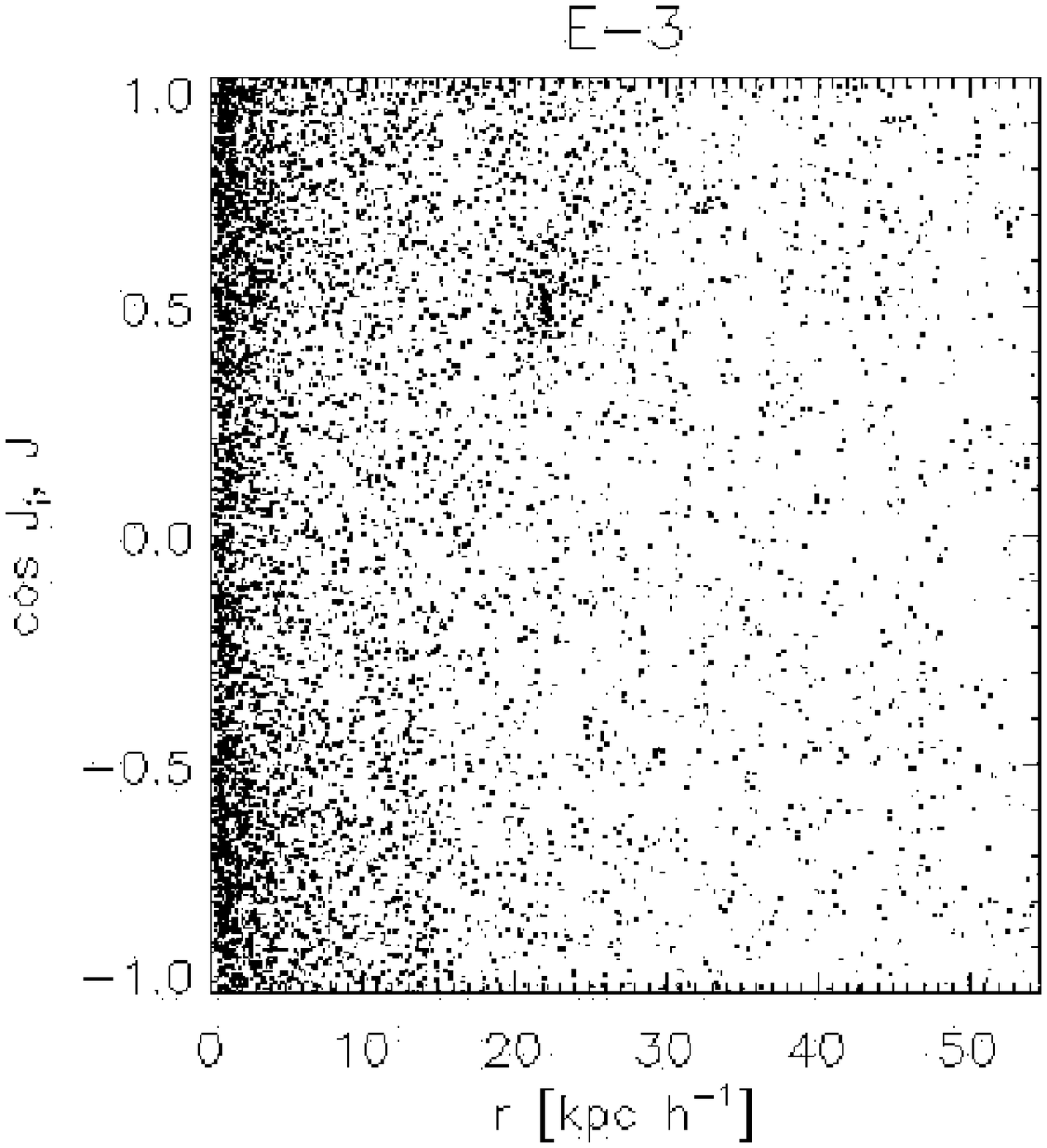}}
\hspace*{-0.2cm}\\
\caption{Cosine of the angle between the total  angular momentum of the simulated galaxies and the angular momentum of each stellar particles within the virial radius at $z \approx 1.6$ for   NF , E03 and C-0.01  (upper panels) and  E-0.7 , F-0.9 and  E-3 (lower panel).} 
\label{coseno}
\end{figure*}

To illustrate the differences  in the satellite distribution in each run, in Fig.~\ref{coseno} we show the distribution  of the cosine of the angle between the total angular momentum ($J$)  of the  main stellar system and the angular momentum of each stellar particle ($J_i$) at $z \approx 1.6$, when the $\Delta_{v/2}$
shows a change in the slope (Fig.~\ref{delta_v2}). As we can see,  at this redshift there is no disc structure within any of the systems (i.e. there are no particles ordered at cosine $\approx 1$). And secondly, the distribution of stellar clumps  surrounding the main systems is very different.  It can be seen that, in the NF case, the satellites are clearly more massive and have been able to survive further in the halo since stars are more gravitationally bounded.
Those systems that later on are able to develop a disc component (E-0.7 and F-0.9) show smaller stellar satellites.

\begin{figure*}
\hspace*{-0.2cm}\resizebox{5cm}{!}{\includegraphics{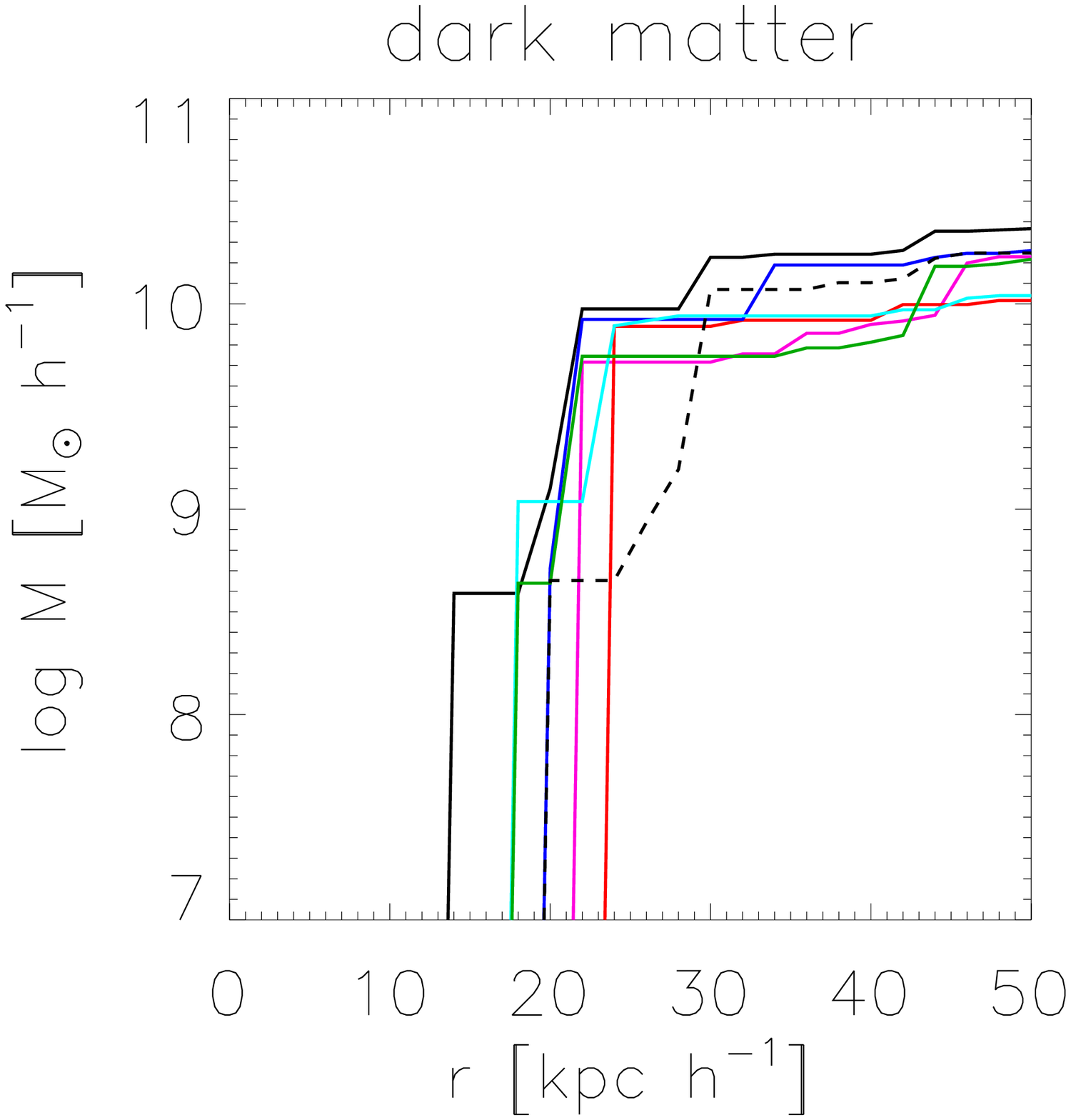}}
\hspace*{-0.2cm}\resizebox{5cm}{!}{\includegraphics{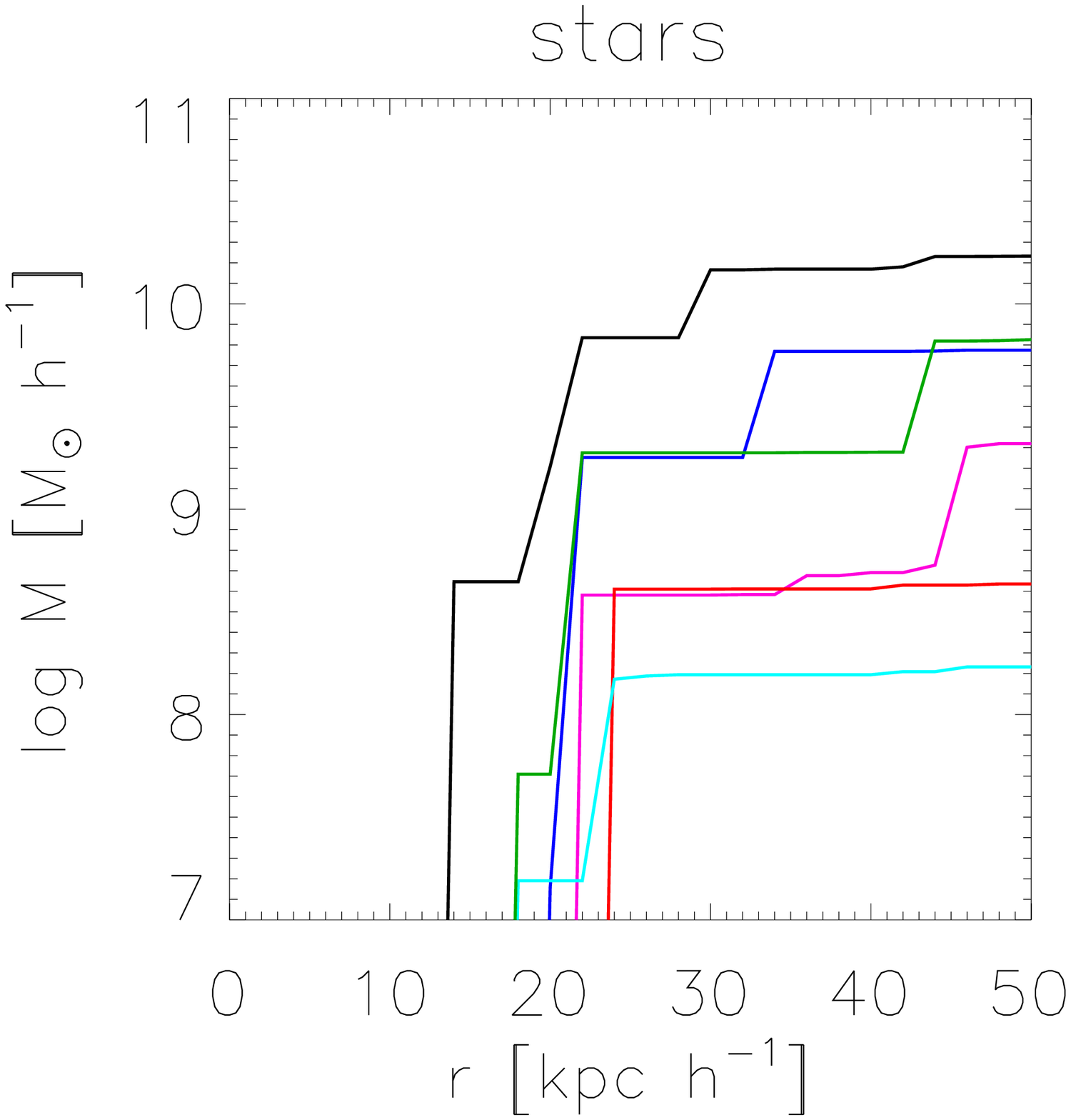}}
\hspace*{-0.2cm}\resizebox{5cm}{!}{\includegraphics{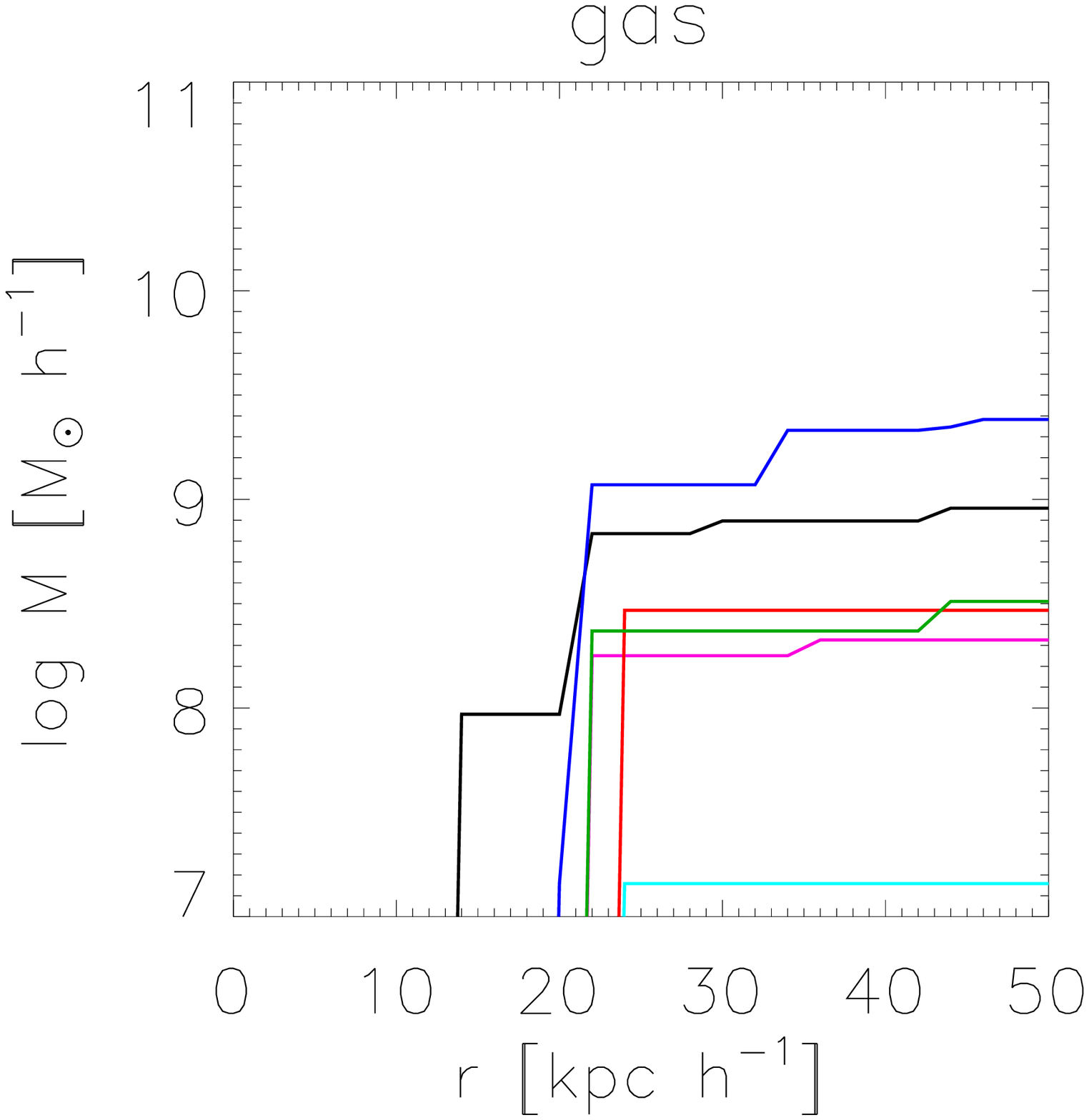}}
\hspace*{-0.2cm}\\
\caption{Integrated  DM, stellar and gas mass of  the satellites within the virial radius at $z \approx 1.6$ (see Fig.8 for color code).} 
\label{massint}
\end{figure*}

This trend can be quantified from Fig.~\ref{massint} where we show the cumulative mass of the satellites for each component (dark matter, gas and stars) as a function of the distance to the centre of mass at $z \approx 1.6$. For the stellar component, the satellites of the NF run are the most massive one, while those in  E-0.7 and F-0.9 are less massive due to the SN feedback action.
The most diffuse and smallest satellites are found in the E-3 halo, as expected.
The gas component behaves similarly to the stellar one, so that  those systems with weaker or no SN feedback have also the larger fraction of gas   per satellite (i.e. systems run with strong feedback blow away  important fractions of gas).

\begin{figure*}
\hspace*{-0.2cm}\resizebox{7cm}{!}{\includegraphics{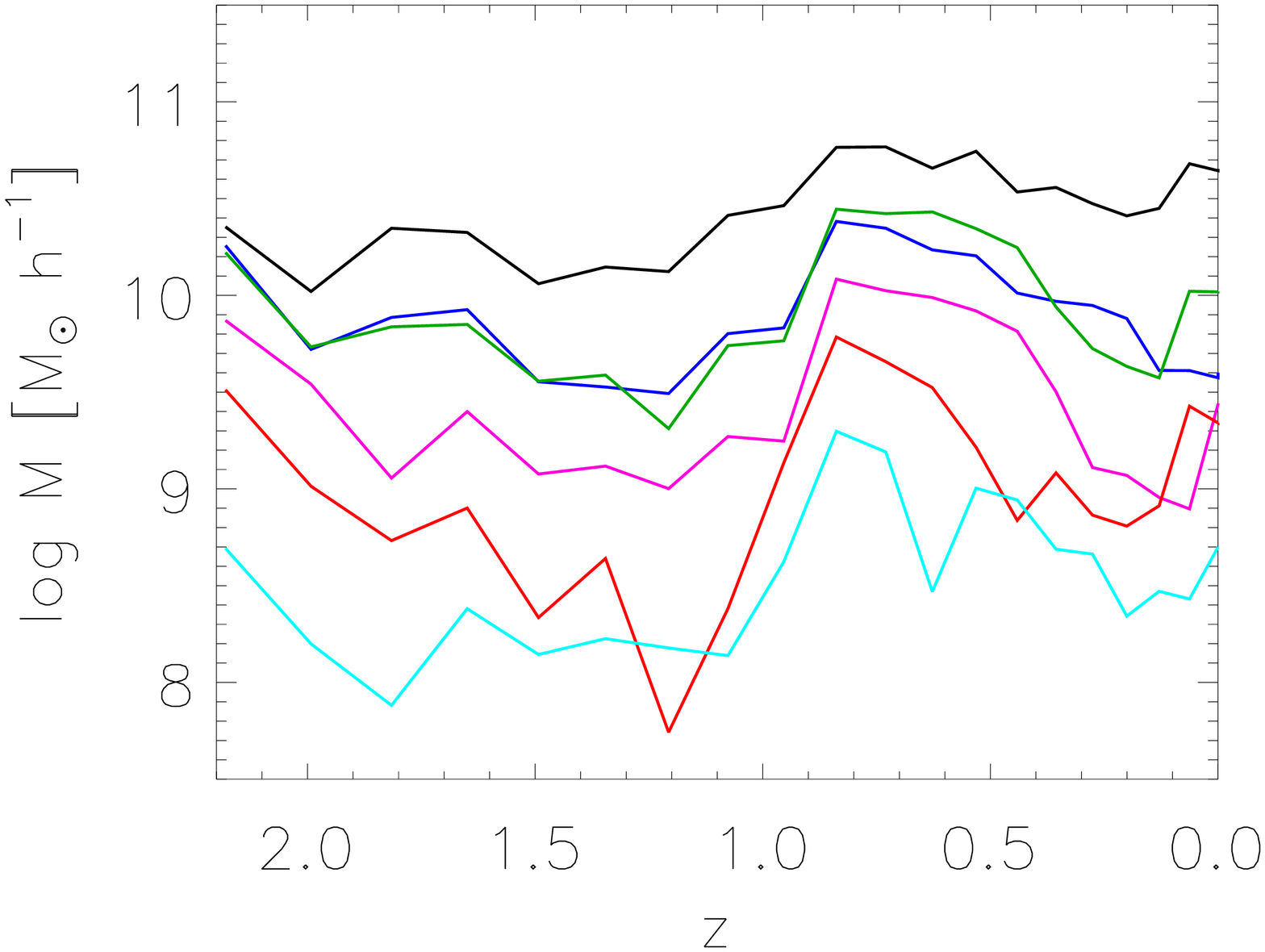}}
\hspace*{-0.2cm}\resizebox{7cm}{!}{\includegraphics{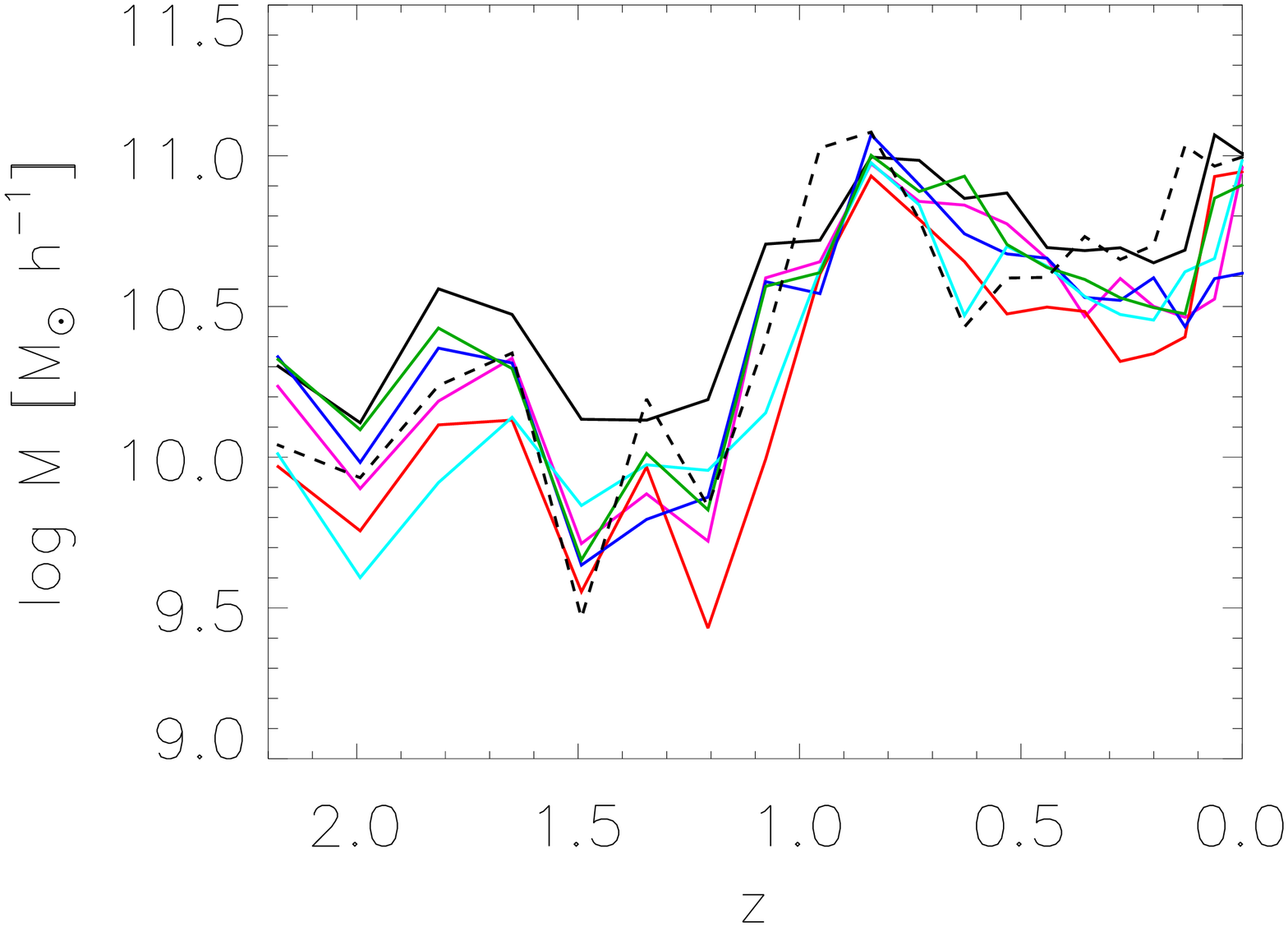}}
\hspace*{-0.2cm}\\
\caption{Baryonic (left panel) and DM (right panel) total mass in satellites  within the virial radius as a function of the redshift for the  NF (black line), E-0.7 (magenta line), F-0.9 (red line), E-0.3 (green line), C-0.01 (blue line), E-3 (cyan line)  and  DM-only (black thick dashed line) runs.} 
\label{sat_mas}
\end{figure*}

Fig.~\ref{sat_mas} shows the total baryonic (left) and dark matter (right) mass of the satellites within the virial radius as a function of the redshift. It can be seen that both components, baryonic and DM, present noticeable mergers or disintegration episodes from $z \approx 2$ indicated by a decrease in the total mass in the identified substructure within the virial radius at a given redshift. An increase of the total mass in satellites implies that new substructure has entered the virial radius. From this figure we can also see that the DM associated to the satellites varies slightly from halo to halo as expected since they all share the same merger tree. However the baryonic mass shows more important differences. In the case of the NF halo, the total baryonic mass in satellites within the virial radius changes weakly with time. However, the rest of the haloes not only have a lower baryonic content, but also experience larger changes as a function of redshift.
After $z  \approx 2$, the main period of satellites accretion is around $z \approx 0.8$. Before and after that time, there are mainly  merger events with the main galaxy  or satellite tidal disruptions which feed the background halo.  We can not differentiate between these processes with the help of this plot since it provides the total mass associated to the subhaloes which can be individualized at a given redshift. However, we carried out a thorough analysis of these subhaloes following their progenitors in time in each run to be sure that we
were actually quantifying these effects, without being contaminated by fly-bye intruders.

In order to improve our understanding of  the angular momentum content of the central galaxy and the DM,  we analysed
 the  specific angular momentum of each mass component of the systems defined at $z \approx 1.6$  as a function of time. 
We chose the systems at this redshift as a reference one because, from this time,
 we detect the larger differences in the evolution of the central mass concentration of our haloes
(Fig.~\ref{delta_v2}) and also the most  important interaction events with satellites (Fig.~\ref{sat_mas}).
We selected   the stars and gas  components within 1.5 times the optical radius ($\approx$ 12  $h^{-1}$ kpc)  and of the DM within the virial radius, without including subhaloes, at $z \approx 1.6$. We estimated the cumulative  mass in  bins containing a growing fraction of the total selected mass. We analysed the specific angular momentum content of each mass component  with redshift.

\begin{figure*}

\hspace*{-0.2cm}\resizebox{6cm}{!}{\includegraphics{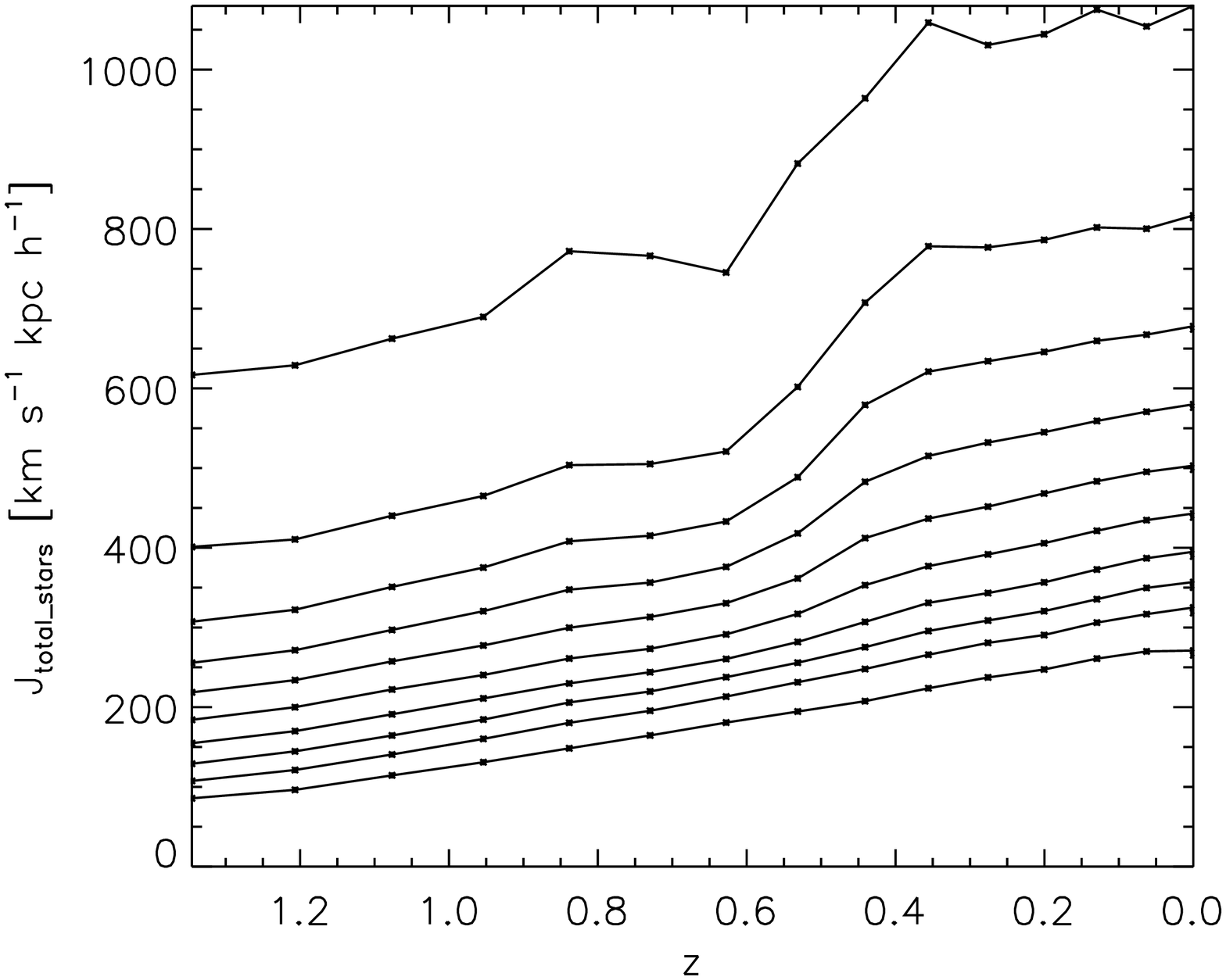}}
\hspace*{-0.2cm}\resizebox{6cm}{!}{\includegraphics{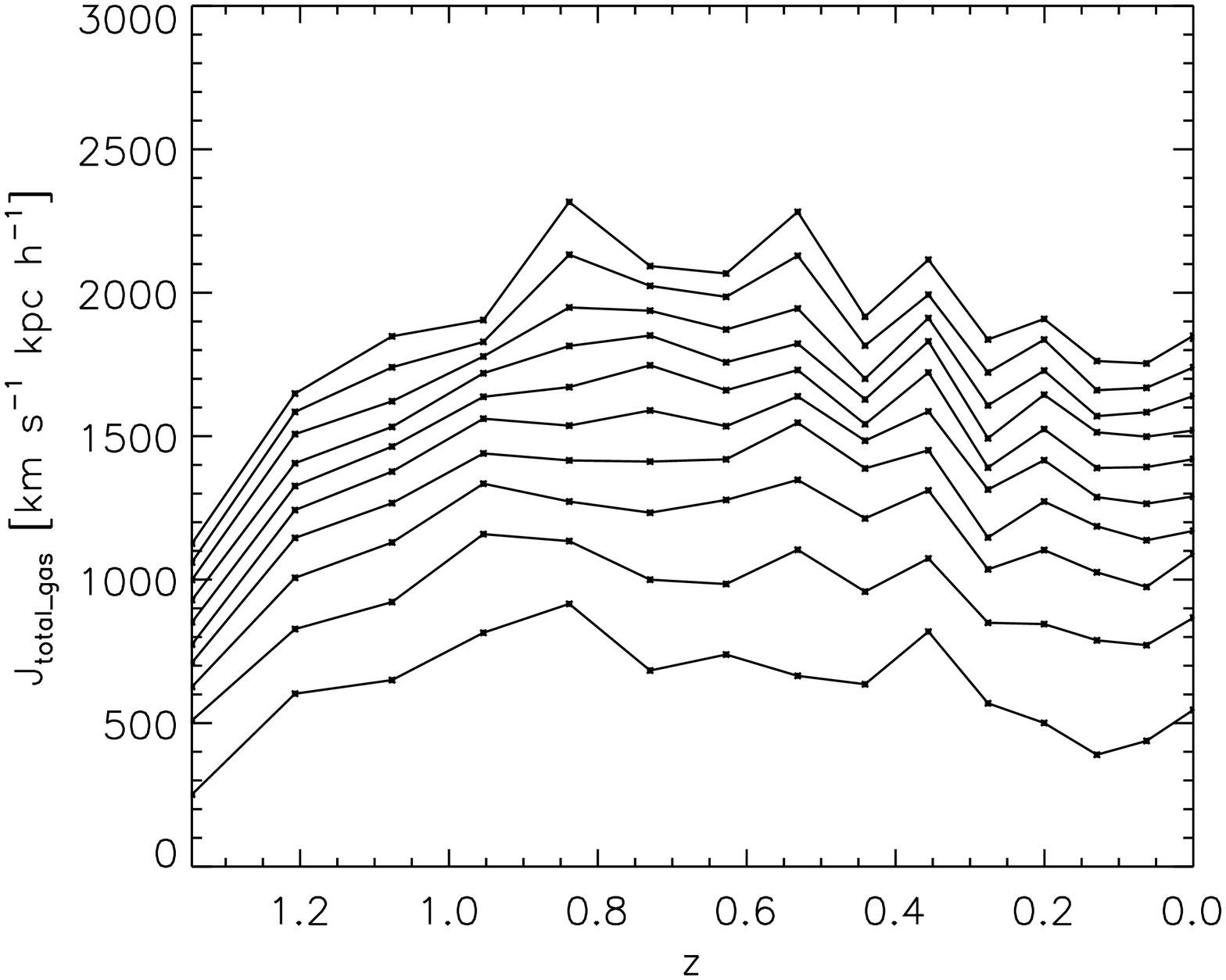}}
\hspace*{-0.2cm}\resizebox{6cm}{!}{\includegraphics{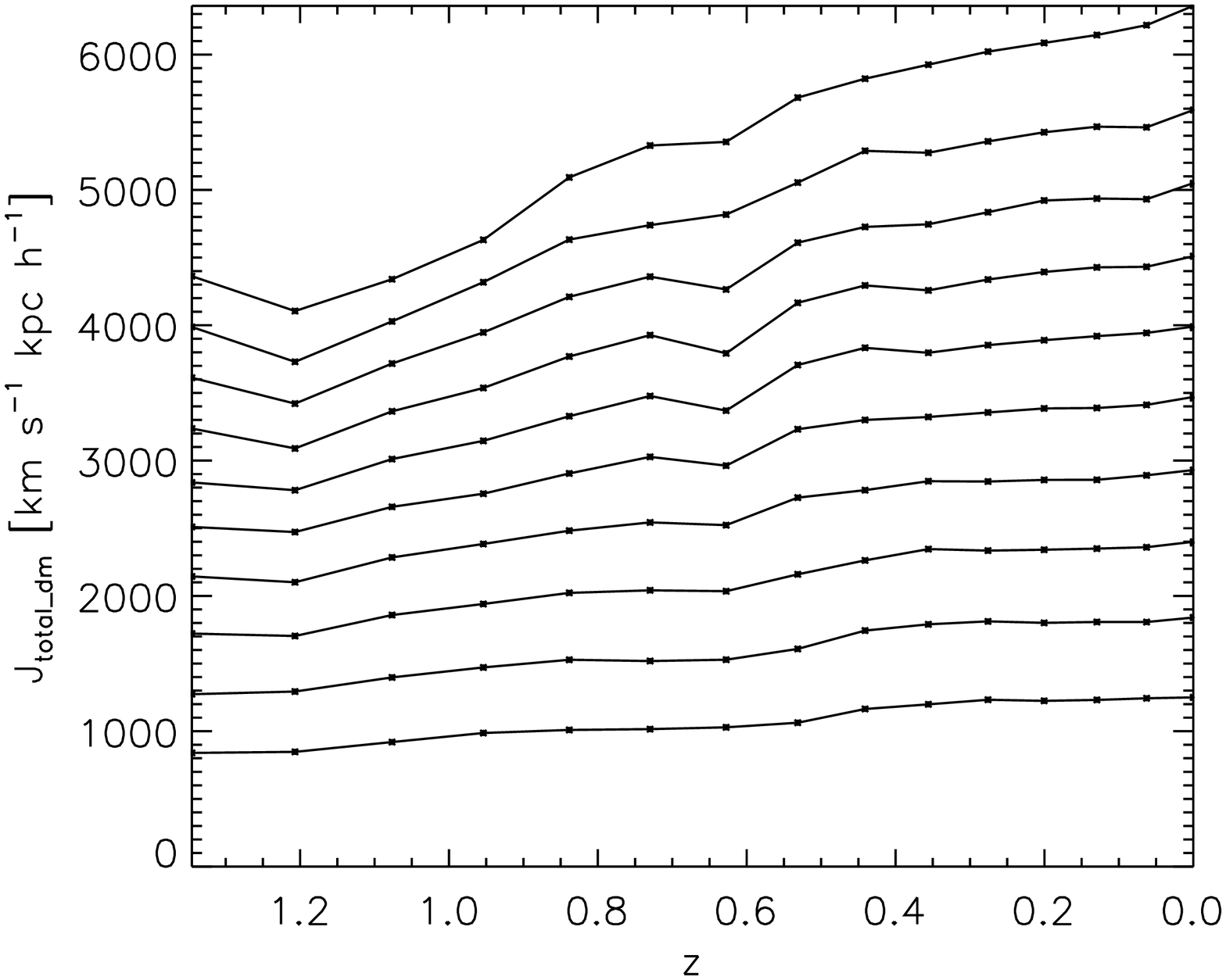}}\\
\hspace*{-0.2cm}\resizebox{6cm}{!}{\includegraphics{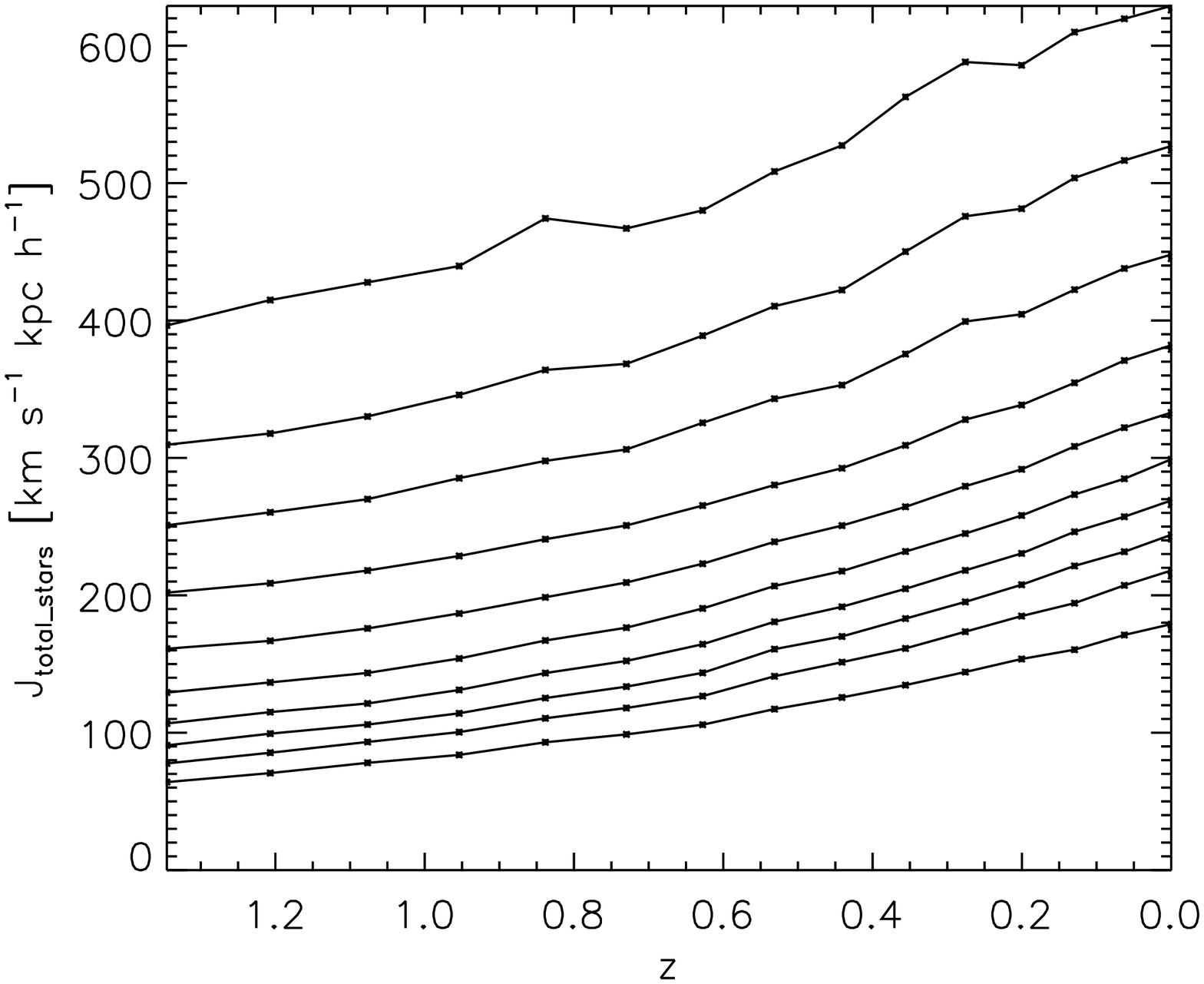}}
\hspace*{-0.2cm}\resizebox{6cm}{!}{\includegraphics{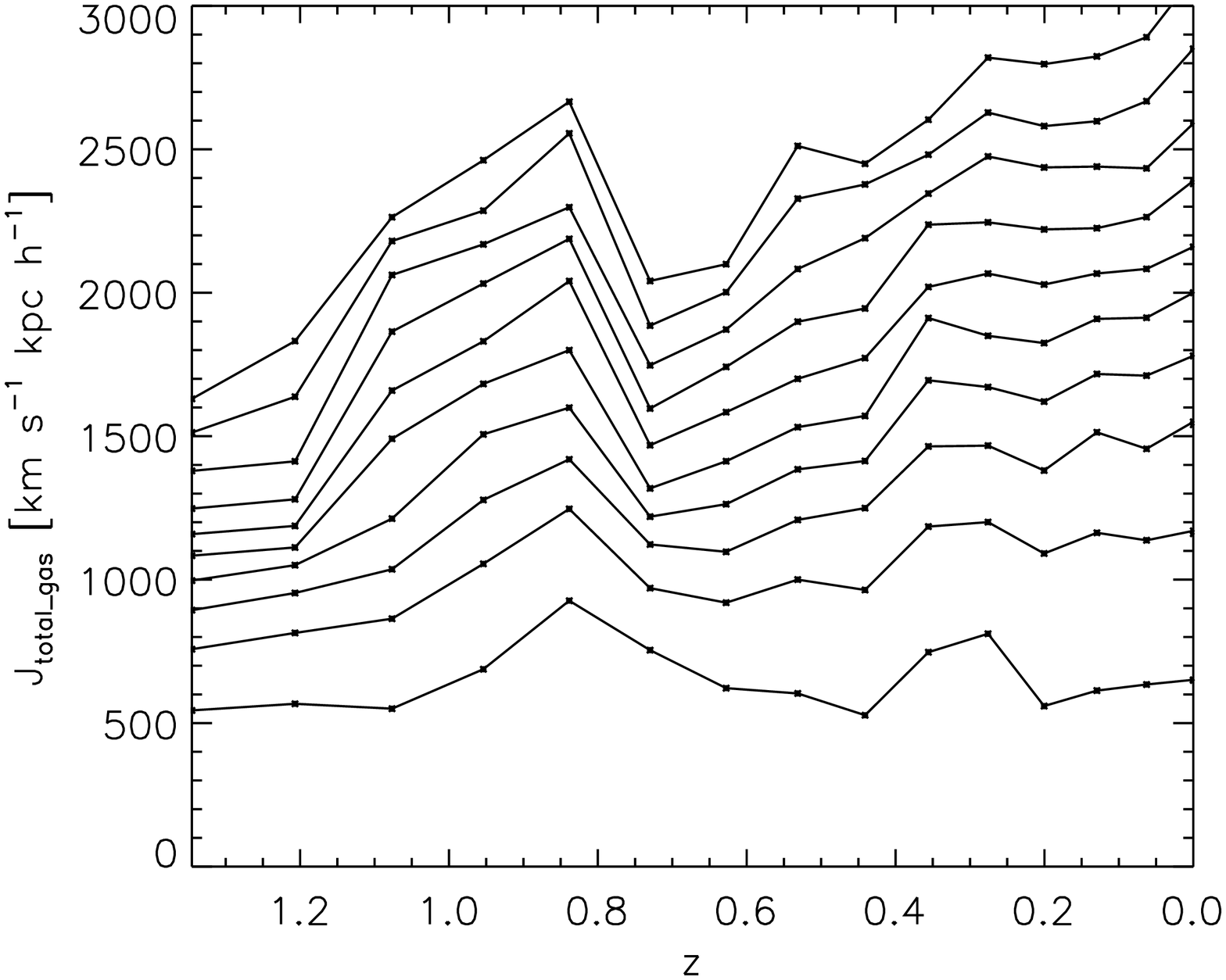}}
\hspace*{-0.2cm}\resizebox{6cm}{!}{\includegraphics{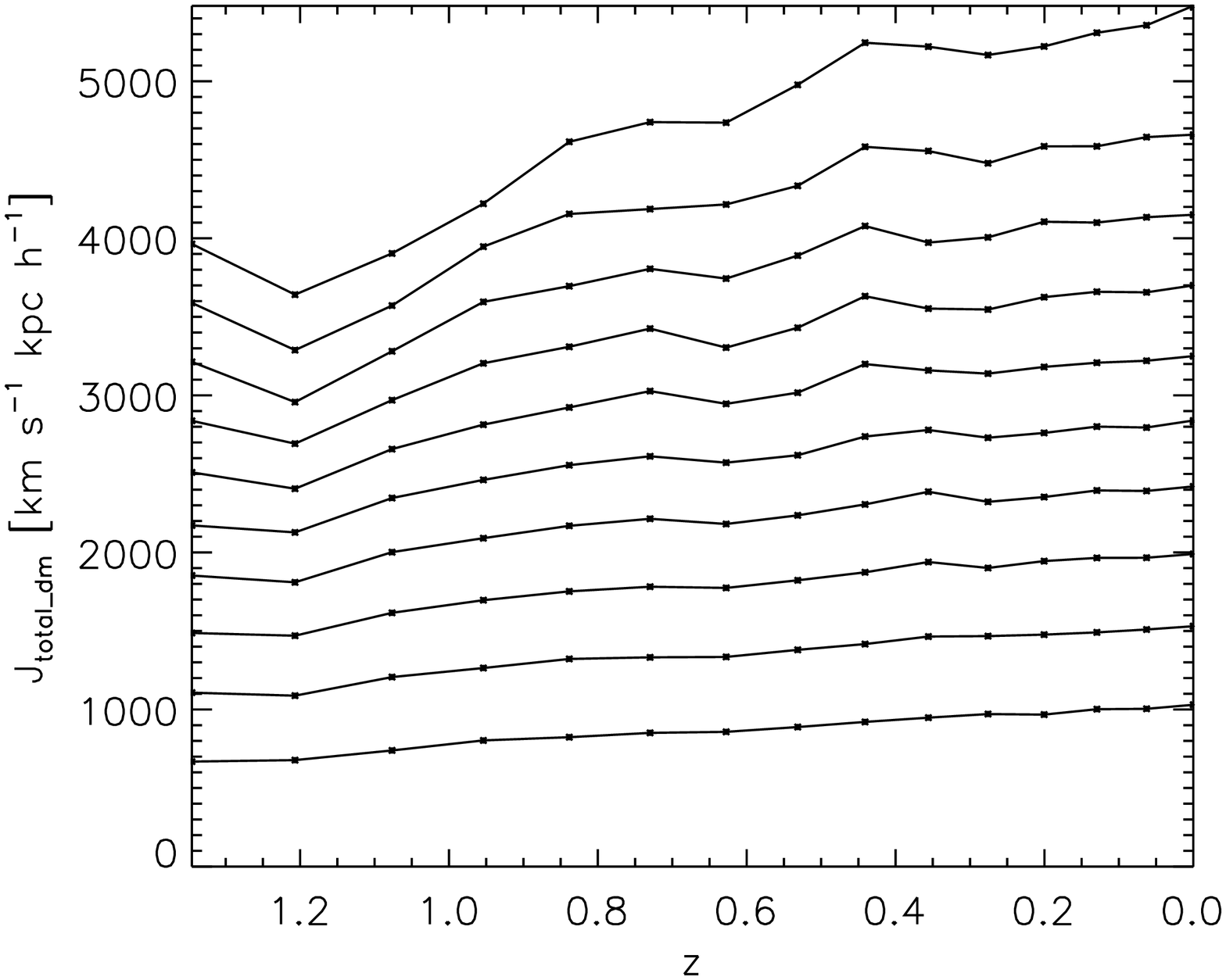}}
\caption{Time evolution of the specific angular momentum  of  stellar (left panels) and gas (middle panels) components within 1.5 the optical radius ($ \approx 12 h^{-1}$ kpc ) and
of the DM halo (right panels) within the virial radius measured at  $z\approx 1.6$, in  mass bins containing a growing fraction of the total mass a function of redshift for the NF (upper panels) and E-0.7 (lower panels) runs. The lines represent a growing fraction of mass, from $10\%$ for the lower one up to $100\%$ for the upper one.} 
\label{Jtotal}
\end{figure*}

In Fig.~\ref{Jtotal}, we show the distributions for the NF and E-0.7 cases.
The stellar mass that was within 1.5 the optical radius  at $z \approx 1.6$ shows an increase of the specific angular momentum in each
mass bin, even in the lowest one.
Both systems show the same trend although the acquisition of angular momentum is much larger for the stars in the NF halo for all mass bins, producing a stronger stellar migration\footnote{Note that the detection of stellar migration does not affect the fact that the discs are mainly formed inside-out as shown in Fig.~\ref{histories} and S08.} (Roskar et al. 2008).
The associated gas mass in the NF case at the same redshift shows also an increase in its specific angular momentum  which is, then, partially lost at low redshifts.  In E-0.7, the gas component in outer regions gains  larger fraction of angular momentum. This could be explained  by the action of the SN feedback which triggers important gas outflows (S08).
Hence, we found that baryons determining the galaxy at $z \approx 1.6$ acquired angular momentum which produced its expansion.
Because baryons dominate the central regions, the global potential well  also changes, probably acting against further DM contraction.

Finally, we also measured the specific angular momentum content of the DM halo which host the galaxy at $z \approx 1.6$.
As it can be seen from Fig.~\ref{Jtotal}, these DM particles increase their angular momentum content as a function of redshift and in all mass bins, even in the most  central one. 
The DM particles in the NF case acquire a larger fraction of angular momentum than those in the E-0.7 case.
 Note that these DM particles represent the halo at $z \approx 1.6$ without substructure so the increase of angular momentum is expected because of the angular momentum transfer from the infalling 
satellites via dynamical friction (the total angular momentum of the  DM halo is expected to be conserved  as 
shown in S08).

\section{Rotation curves}

A well-known problem of numerical simulations is the inability of CDM scenarios to produce systems with flat rotation curves comparable to that of the Milky Way because of the catastrophic concentration of baryons at the central region. Dutton et al. (2007, 2008), among others, found that, in the absence of baryons, there appears to be a reasonable agreement between theory and observation because the former predicts $V_{max} \approx   V_{200}$, where  $V_{max}$ is the maximum of the total circular velocity and $V_{200}$
is the circular velocity at the virial radius. However, when the effects of baryons are taken into account through the AC hypothesis, results do not match observations since $V_{max}$ is significantly increased by a factor of two (Navarro \& Steinmetz 2000; Dutton et al 2007). These values are too high for matching the Tully-Fisher relation.
Dutton et al (2008) proposed that a net halo expansion that reverse the contraction would be required in order to lower the $\frac{V_{max}}{V_{200}} $ ratio. A mechanism suggested in this work  is SN feedback since a net halo expansion could result from the rapid removal of the disc mass. However, while Gnedin \& Zao (2002) found this effect to be too weak to reconcile observations and theory, Read \& Gilmore (2005) claimed that if this process is repeated several times, a reduction in the halo density could be accounted for. 

\begin{figure*}
\hspace*{-0.2cm}\resizebox{6cm}{!}{\includegraphics{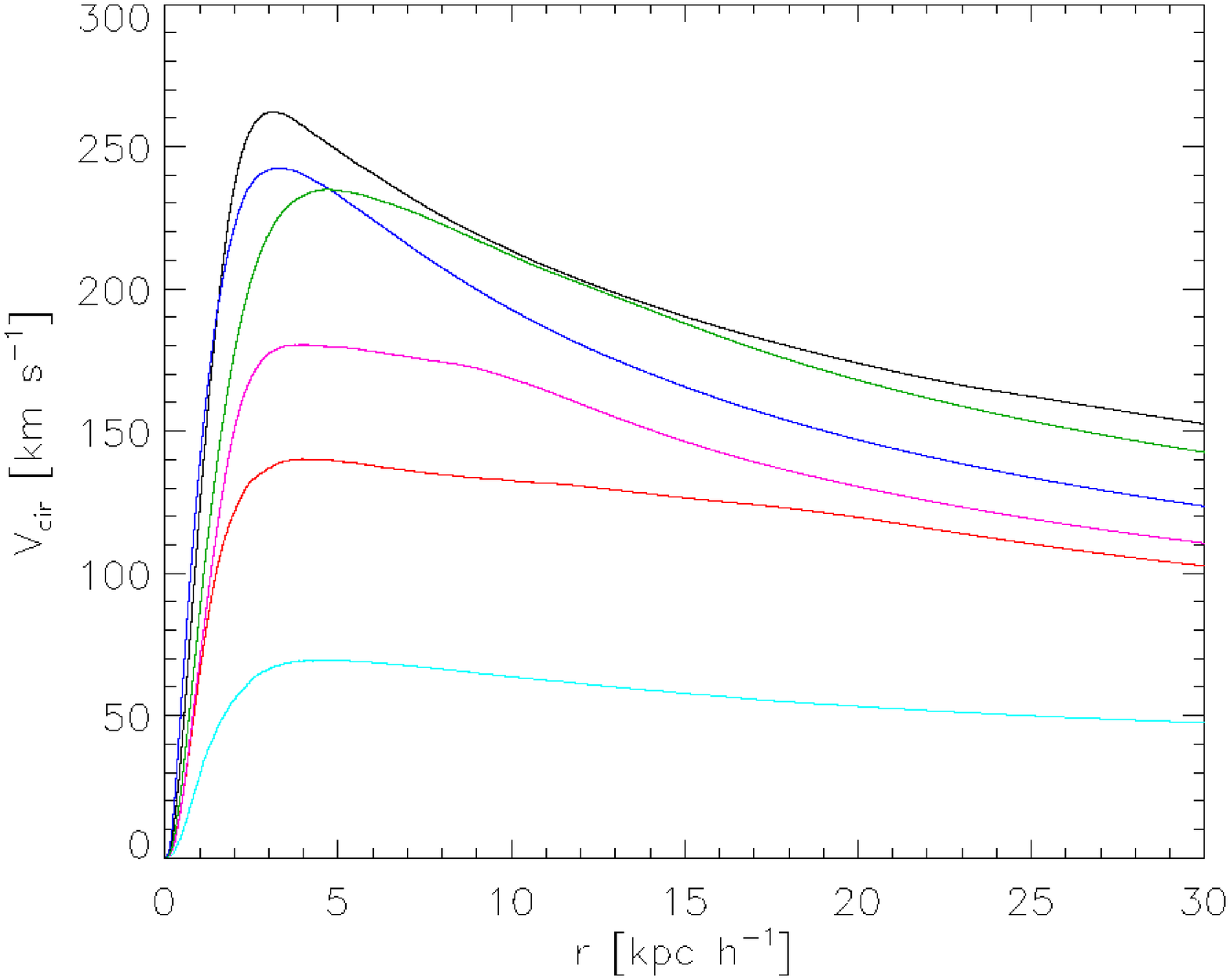}}
\hspace*{-0.2cm}\resizebox{6cm}{!}{\includegraphics{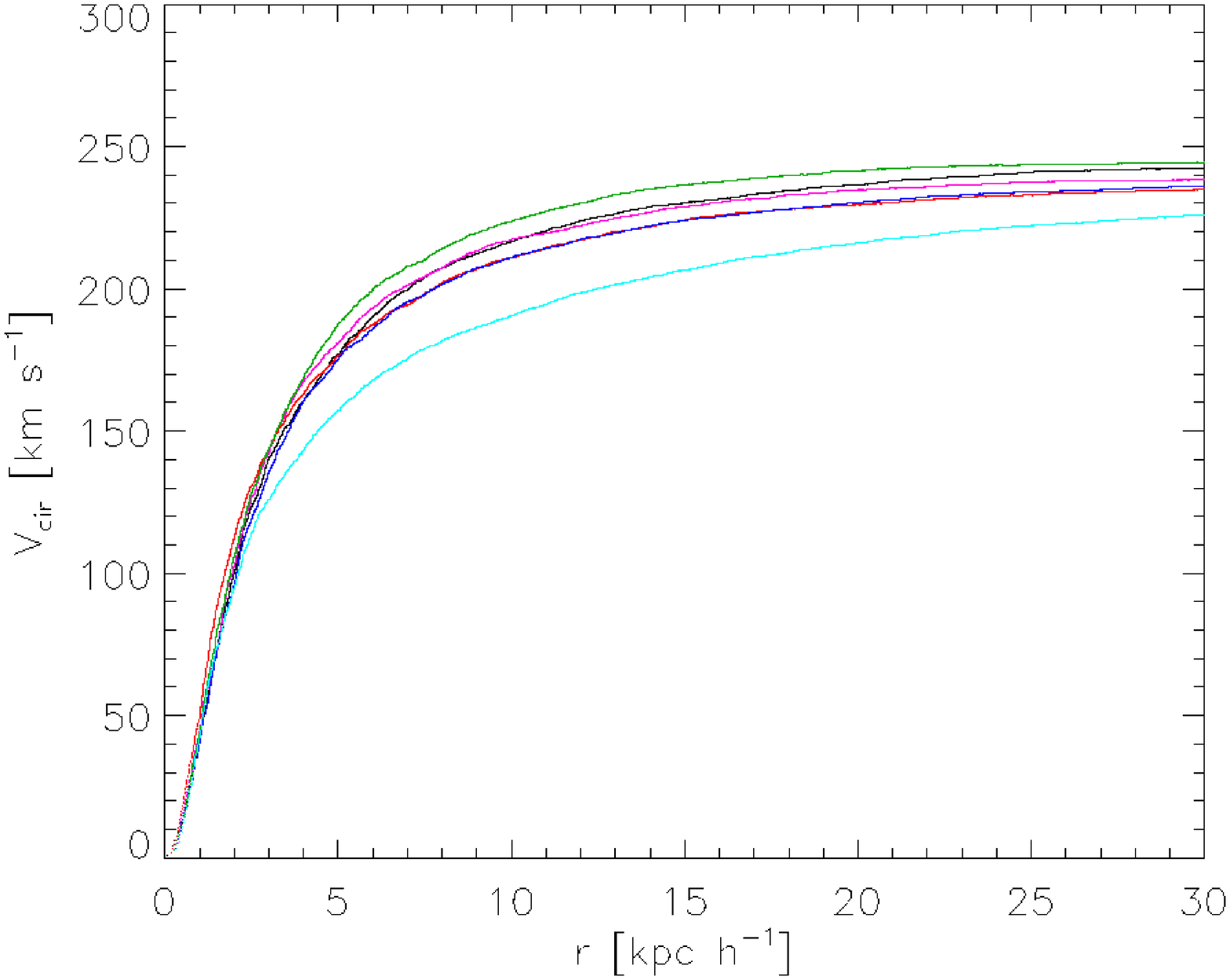}}
\hspace*{-0.2cm}\resizebox{6cm}{!}{\includegraphics{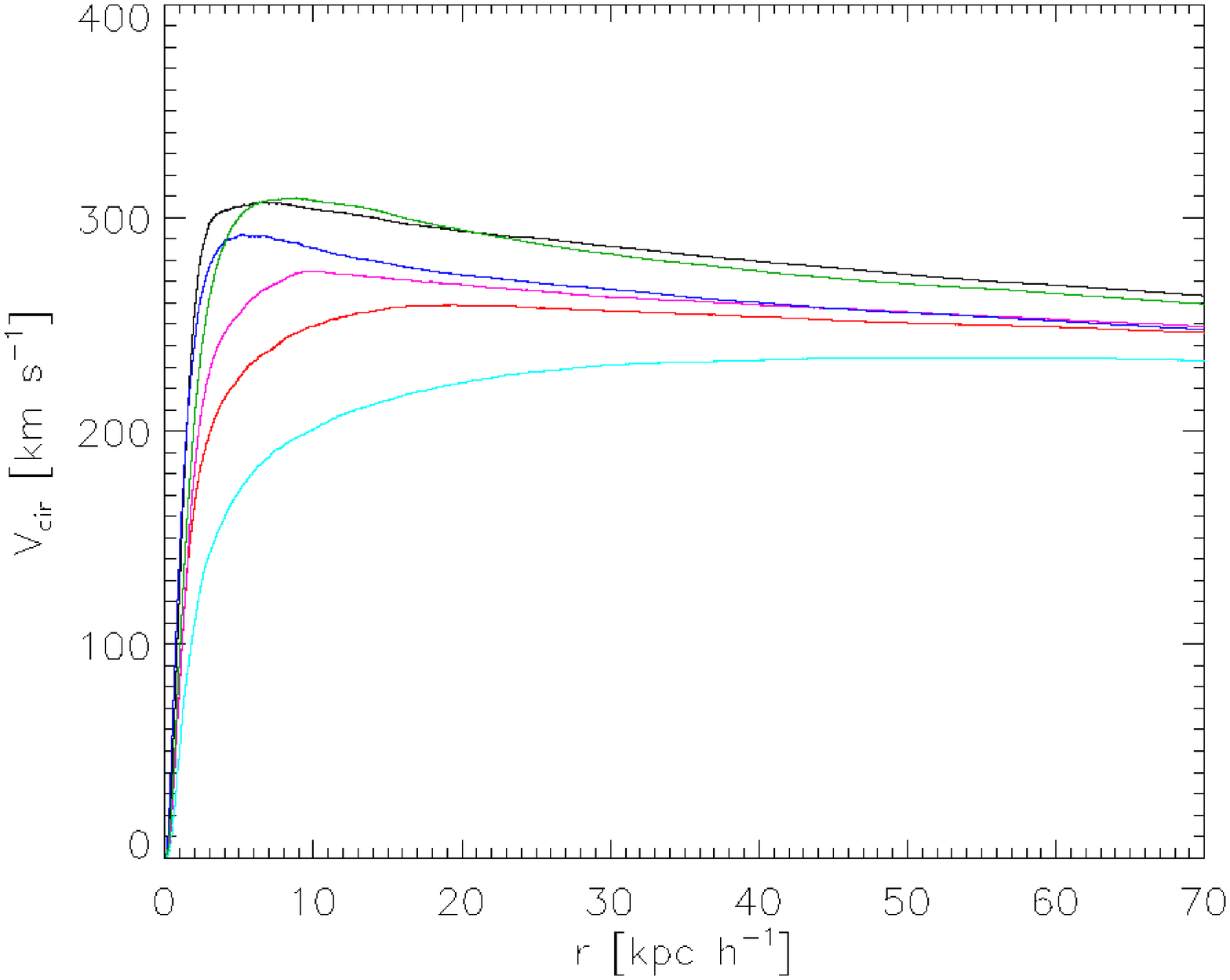}}
\caption{Baryonic (left panel), dark matter (middle panel) and total  (right panel) circular  velocities for the NF (black), E-0.7 (magenta), F-0.9 (red), C-0.01 (blue), E-0.3 (green) and E-3 (cyan) runs.} 
\label{Vcir}
\end{figure*}

From our simulations, we can estimate the circular velocities for each component and analyse how the  $\frac{V_{max}}{V_{200}} $ratio
varies for different combinations of the SF and the  SN feedback parameters. The circular velocity of baryons shown in Fig.~\ref{Vcir} (left panel) reflects the presence of a concentrated dominating spheroid  (NF, C-0.01 and E-0.3) or a dominating disc component (E-0.7 and F-0.9). The dark matter
distributions (middle panel) vary between haloes in agreement with Fig. 1,  but in  the central region the shape of the total circular velocity (right panel) is
determined mainly by the baryonic component.  As it can be seen from this plot, only when an important disc component is at place, the total velocity  distribution gets flat in the baryonic dominated region.

\begin{figure*}
\hspace*{-0.2cm}\resizebox{6cm}{!}{\includegraphics{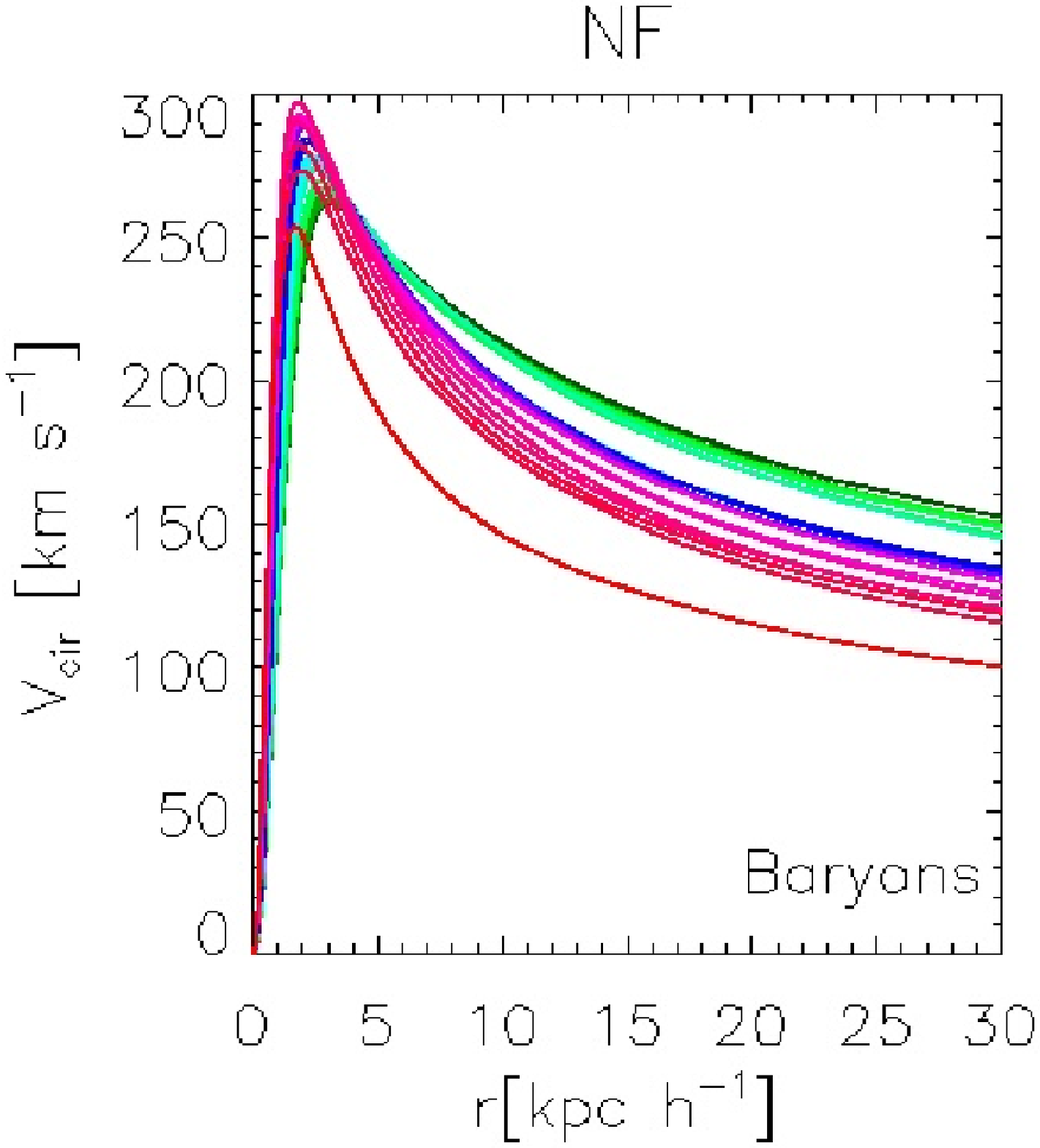}}
\hspace*{-0.2cm}\resizebox{6cm}{!}{\includegraphics{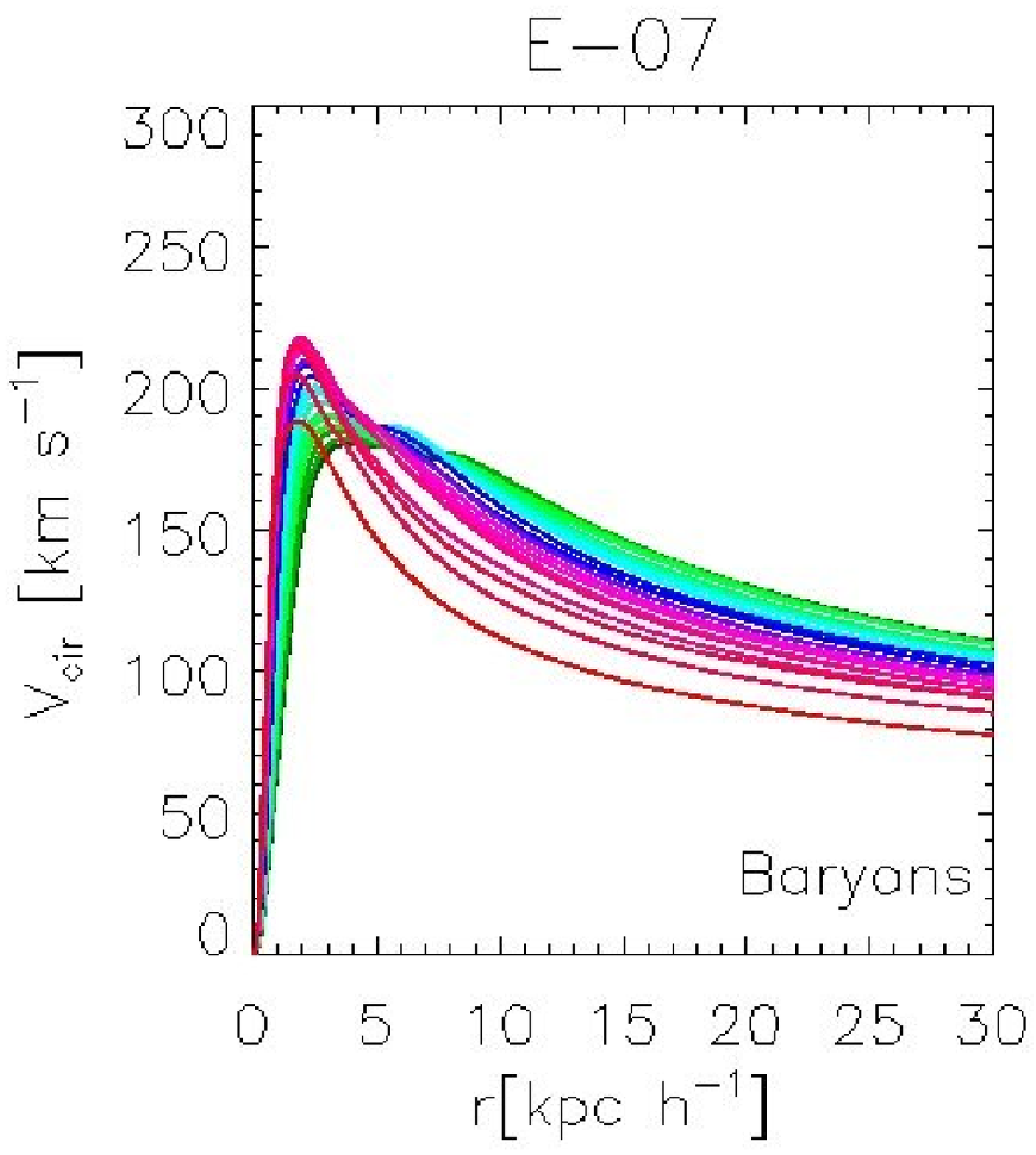}}\\
\hspace*{-0.2cm}\resizebox{6cm}{!}{\includegraphics{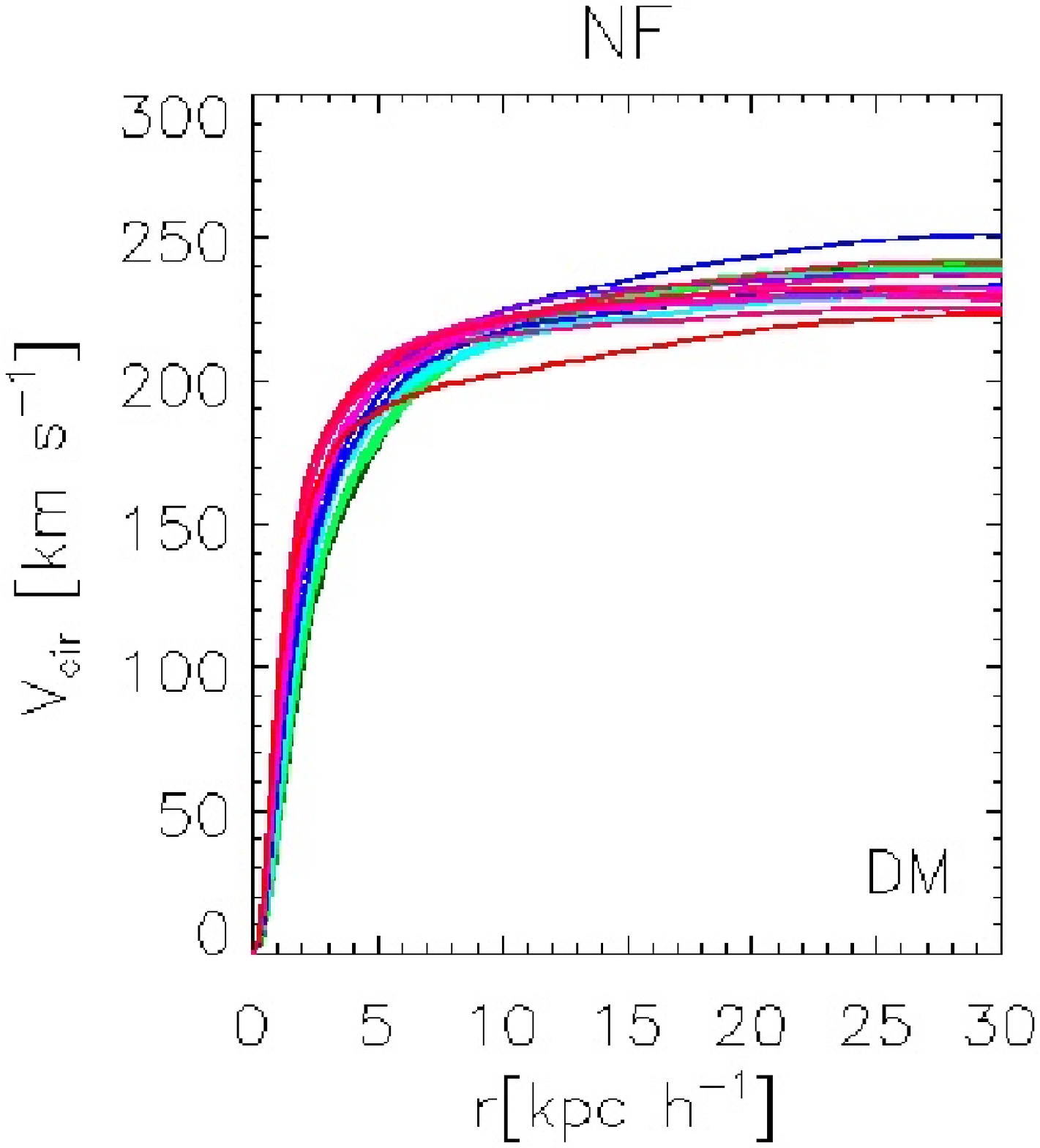}}
\hspace*{-0.2cm}\resizebox{6cm}{!}{\includegraphics{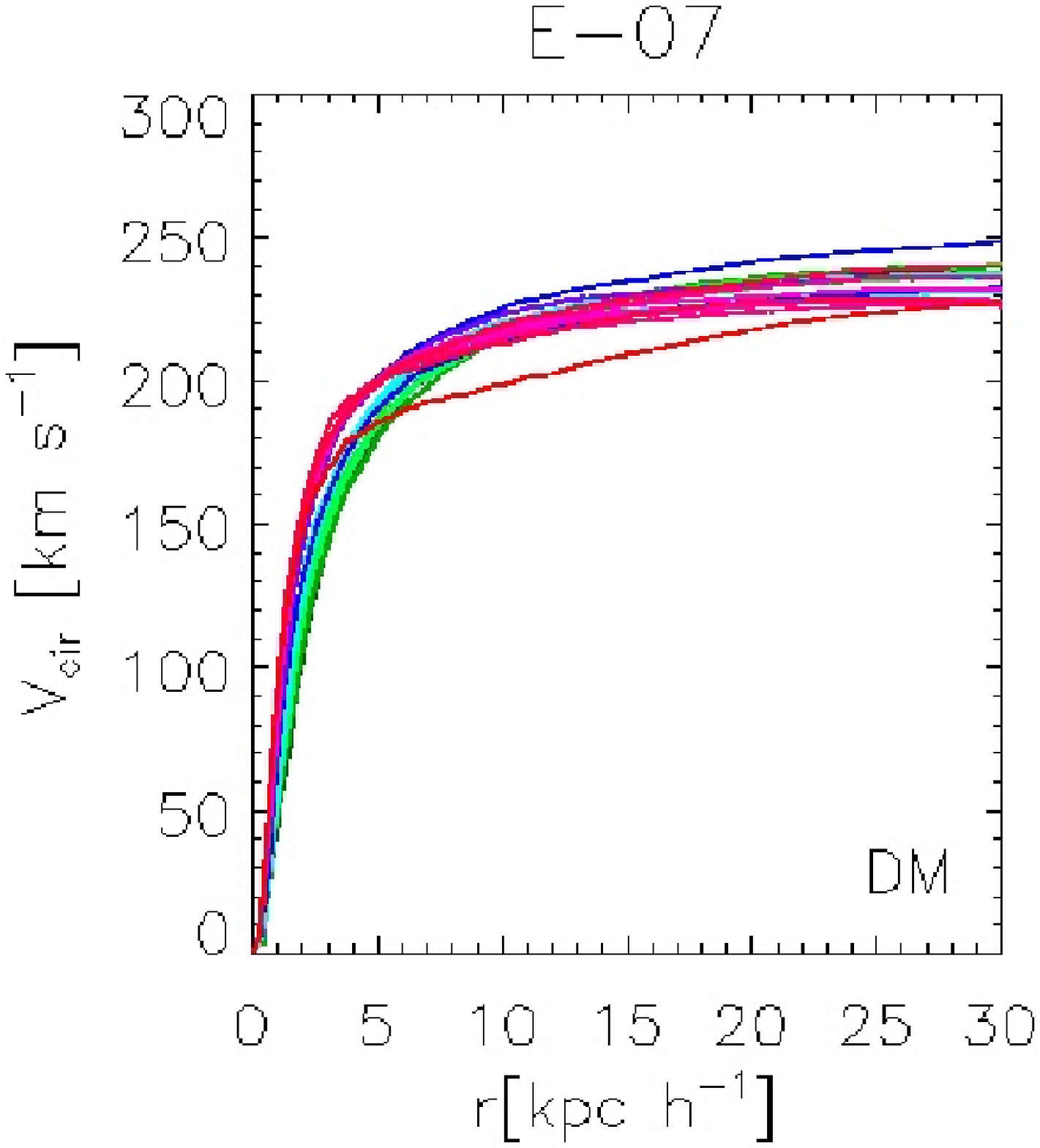}}
\caption{Baryonic (upper panels) and dark matter (lower panels) circular velocities as a function of redshift for the NF (left panels) and E-0.7 (right panels) simulations. The redshift decreases  from red to green colors starting at $z \approx 2 $ (red) and ending at $z=0$ (light green). } 
\label{Vcir_E07_red}
\end{figure*}

In Fig.~\ref{Vcir_E07_red} (upper panels), we show  the evolution of the circular velocity for the baryonic component of the NF and E-0.7 cases. In the E-0.7 case, the curves become flatter in an inside-out process (see Fig.~\ref{histories} and S08). We can also see how the baryons  in the very central region moves outwards, contributing to produce a flatter curve.
In the case of  the NF run, the baryonic circular velocity is very sharp from high redshift and it also shows an outward
displacement of baryons located in the central region  which is explained by the increase
in the angular momentum content of the the stellar component as shown in the previous section.

\begin{figure}
\hspace*{-0.2cm}\resizebox{7cm}{!}{\includegraphics{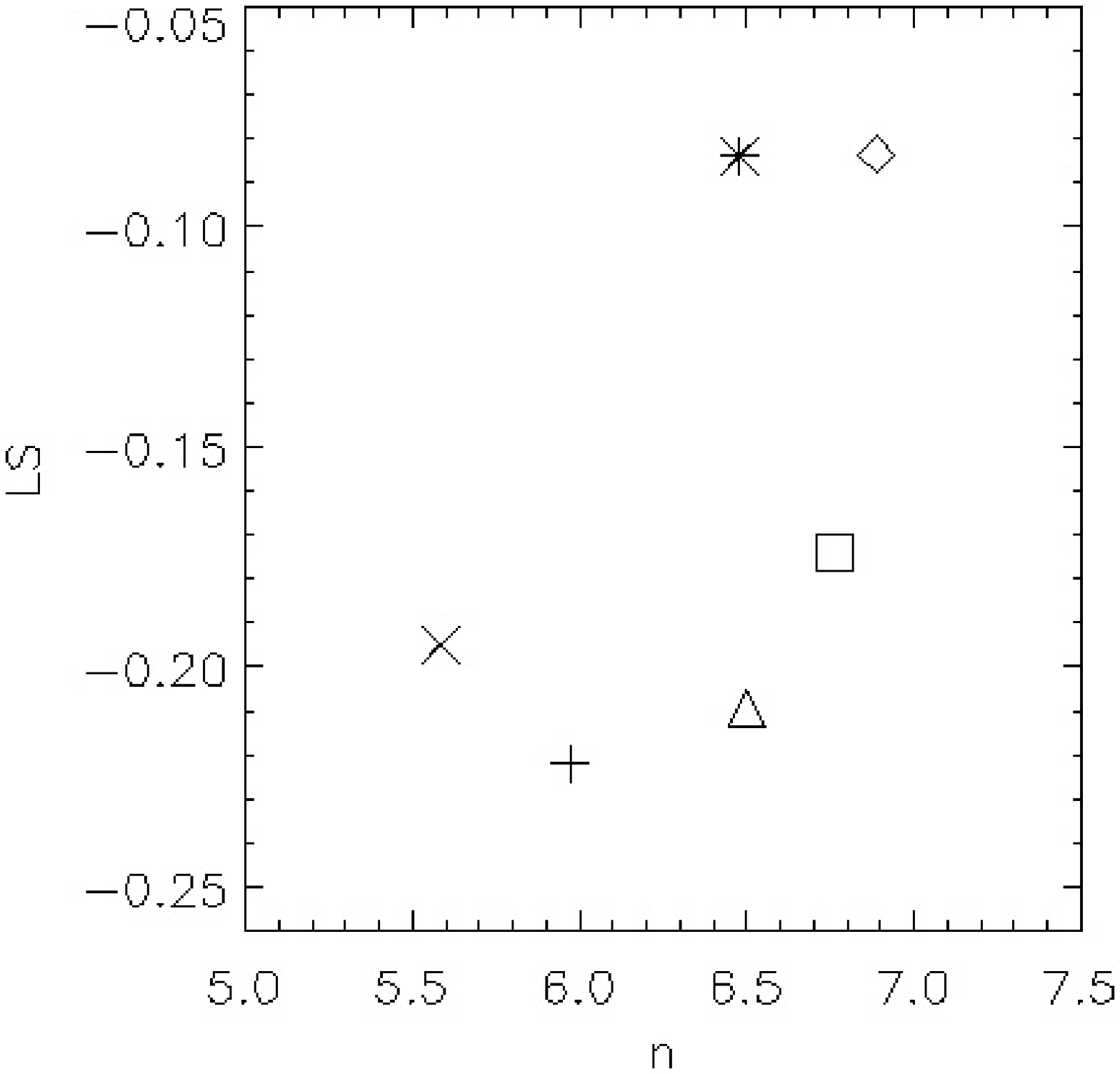}}
\hspace*{-0.2cm}\resizebox{7cm}{!}{\includegraphics{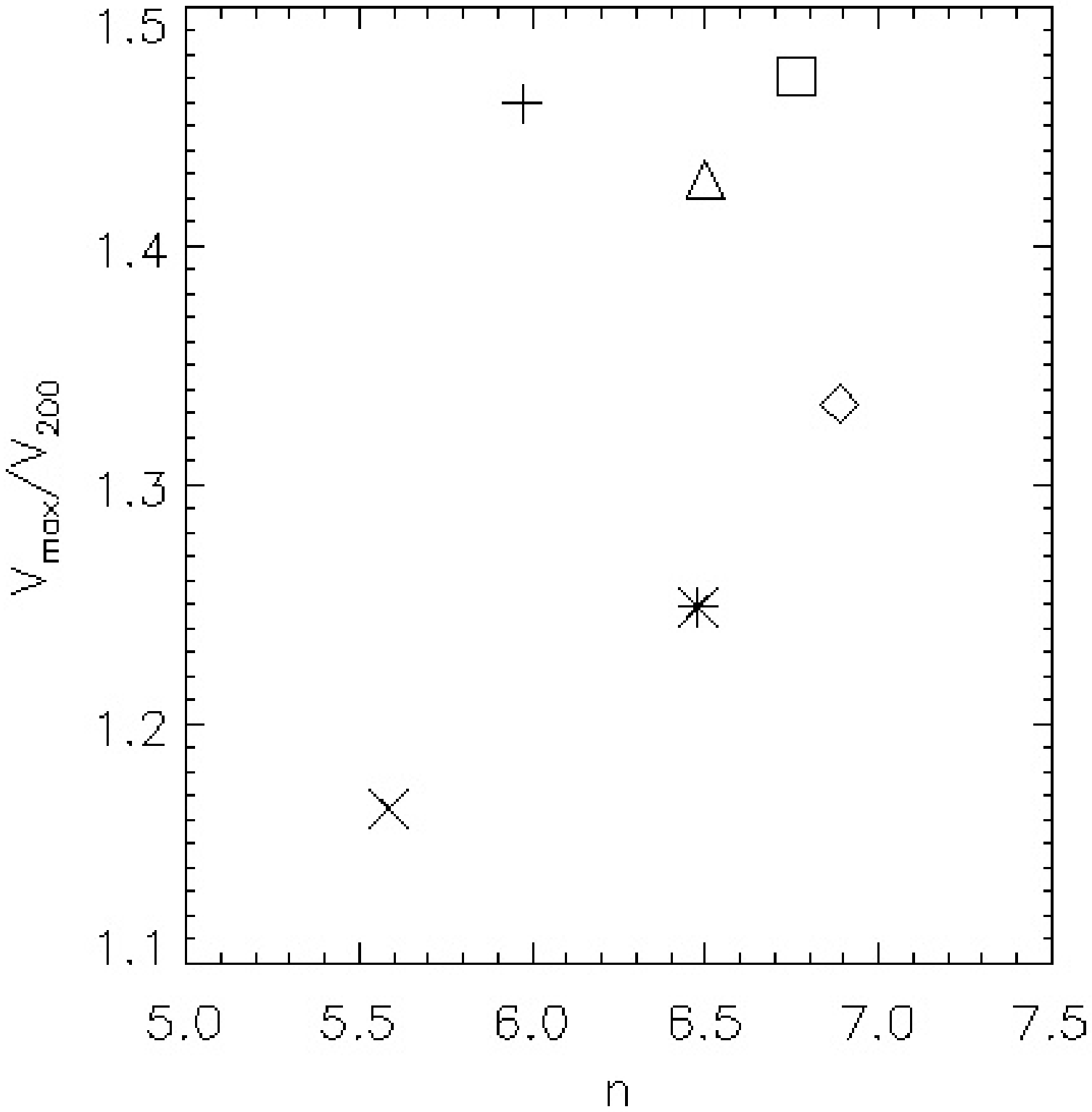}}
\hspace*{-0.2cm}\\
\caption{Baryonic logarithmic slope  (upper panel) and the ratio between the maximum total circular velocity and the virial velocity as a function  the Einasto's shape parameter ($n$) for  NF (plus), E-0.7 (diamond), F-0.9 (asterisk), C-0.01 (triangle), E-0.3 (square), E-3 (cross) runs.} 
\label{Vmax200}
\end{figure}

 We have quantified the flattening of the baryonic circular velocity curve by measuring its logarithmic slope ($LS$) between the radius at the maximum velocity curve and at the optical radius for each simulated halo. From Table \ref{tab1}, we can see that  the lower absolute values correspond to the E-0.7 and F-0.9 runs when the disc component is the dominating one.
 In Fig.~\ref{Vmax200} (upper panel), we show 
$LS$ as a function of the shape parameter $n$ of the dark matter profiles. As we can see, galaxies with lower absolute values of $LS$ tend to have  haloes with the largest $n$ parameters. The less concentrated cases are found
when either spheroid-dominating systems were able to form (NF and C-0.01) or in the E-3 run because of its low baryonic content due to the action of very violent galactic winds.

We also estimated the $\frac{V_{max}}{V_{200}} $ ratio as a function of shape parameter $n$ of the dark matter profiles as displayed in  Fig.~\ref{Vmax200} (lower panel). As it can be seen,  we found values for this ratio between $\approx 1.15$ and $1.5$. Again, those systems where a disc-like galaxy was able to form show 
the lowest ratios, indicating the existence of flat circular velocity curves in agreement with observational results.  The small disc in E-3 has the lowest ratio which is produced mainly for the fact that
the halo has weakly contracted during its assembly due to the very small fraction of baryons retained in the central region.

\subsection{Adiabatic contraction prescription}

The prediction of the effects of baryons on the DM haloes  based on AC  hypothesis (e.g. B86) is widely used.  However the main assumptions of this approximation, namely that halo particles move on circular orbits,  is not realistic. Several studies (Gnedin et al. 2004; Sellwood \& McGaugh 2005) have reported the AC hypothesis to overestimate the level of compression. Possible alternative models have been  developed to calculate the contraction of the DM halo originated by the accumulation of baryons in the central region.
However, as we claimed in Pedrosa et al. (2009), the response of the DM halo to the presence of baryons
depends strongly on the way baryons are assembled (see also Romano-D\'{\i}az et al. 2008). Actually, this has been the main discussion of the current paper.
In this section, we aim at comparing  different proposed prescriptions found in the literature to predict
the effects of baryons, and the statistical motivated prescription of Abadi et al. (2009, hereafter A09).

\begin{figure*}
\hspace*{-0.2cm}\resizebox{6cm}{!}{\includegraphics{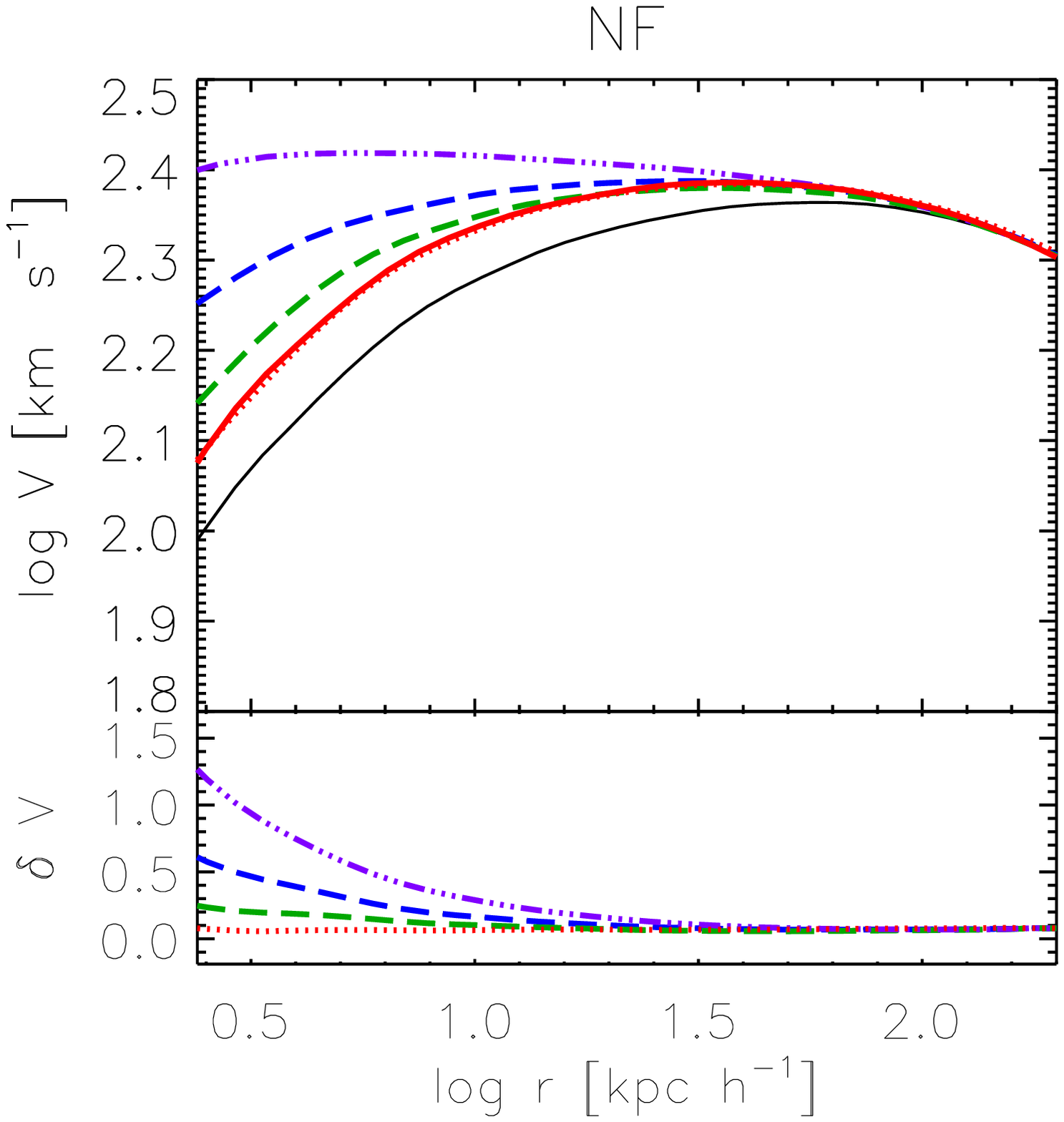}}
\hspace*{-0.2cm}\resizebox{6cm}{!}{\includegraphics{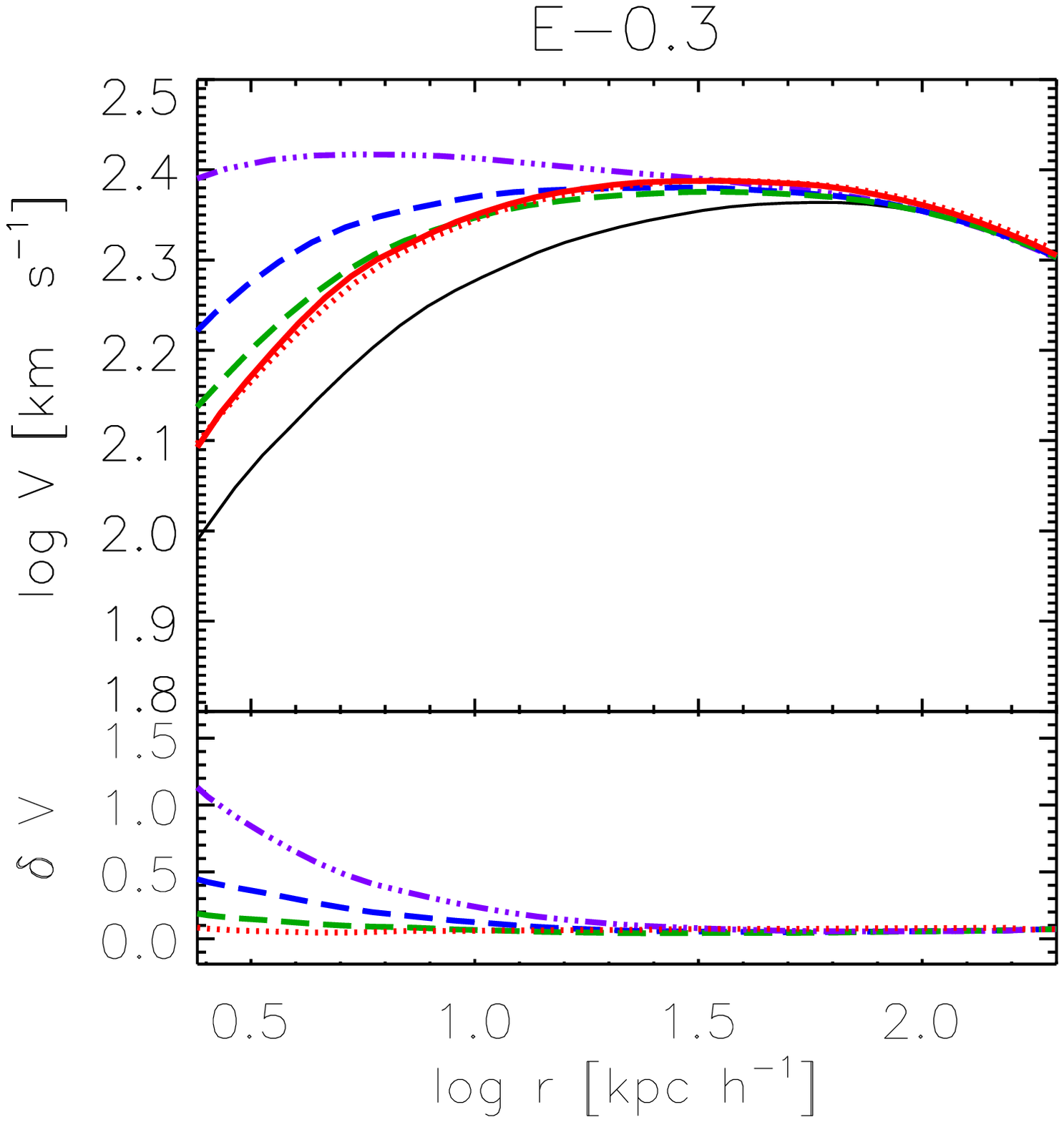}}
\hspace*{-0.2cm}\resizebox{6cm}{!}{\includegraphics{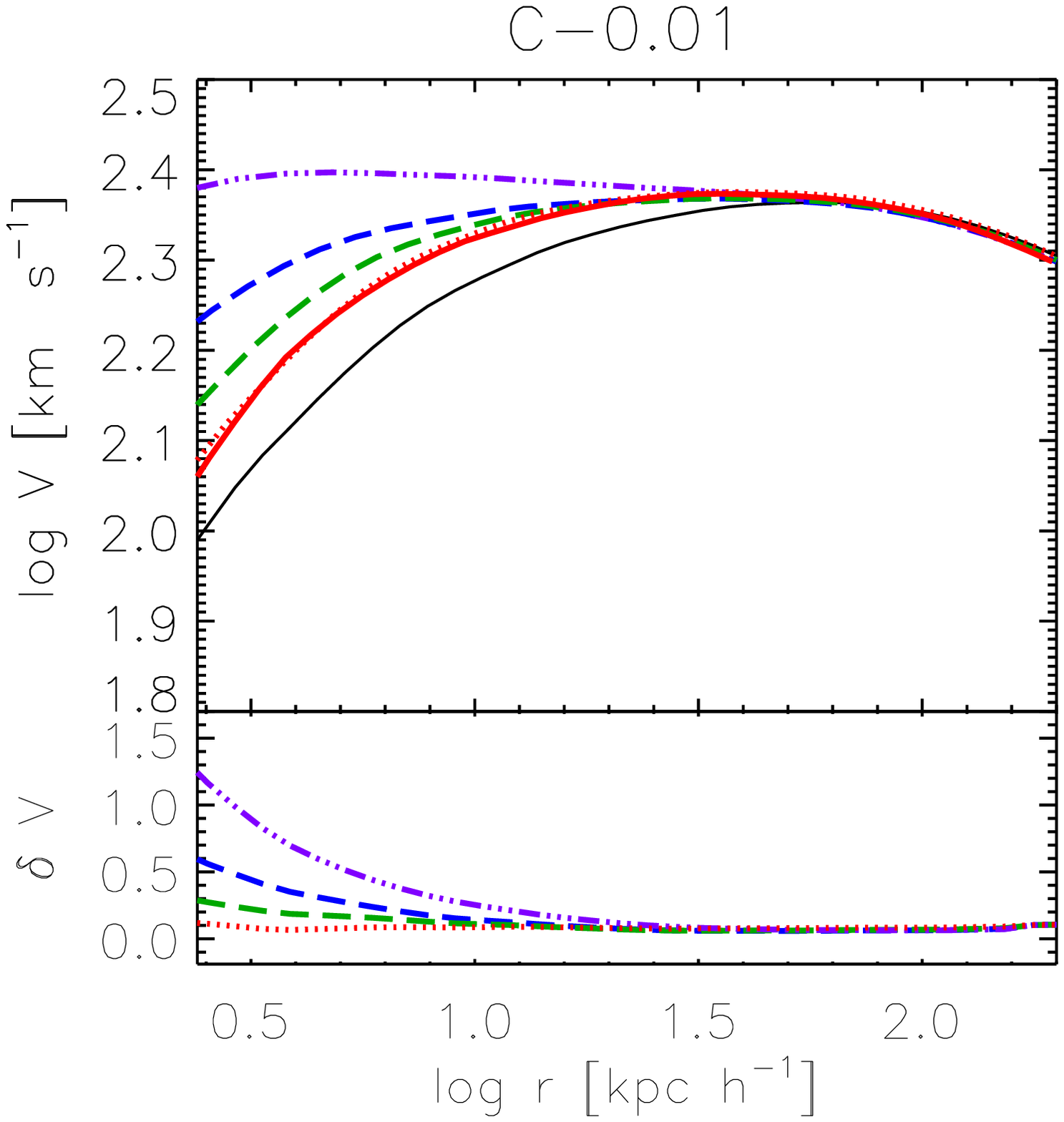}}
\hspace*{-0.2cm}\resizebox{6cm}{!}{\includegraphics{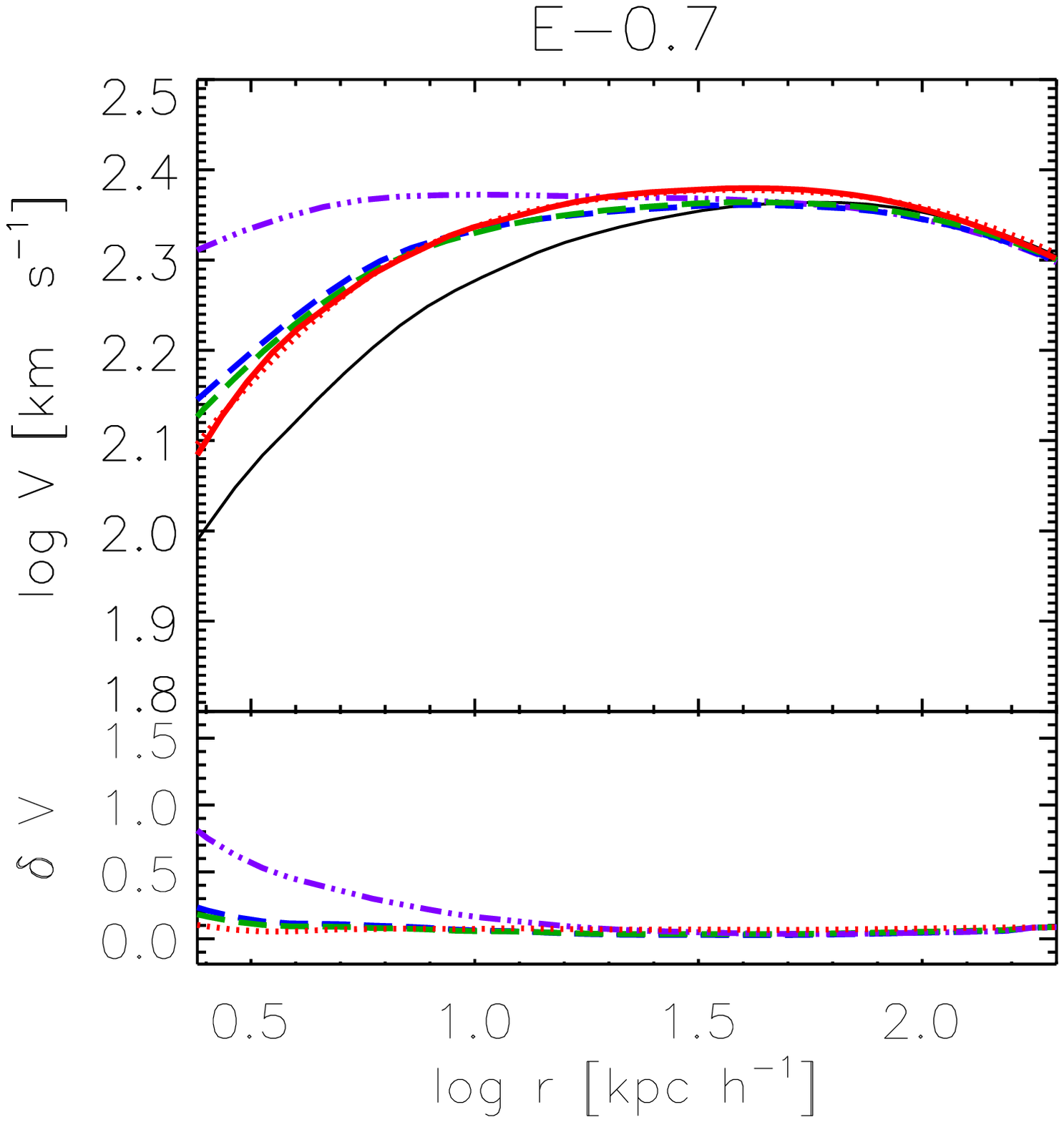}}
\hspace*{-0.2cm}\resizebox{6cm}{!}{\includegraphics{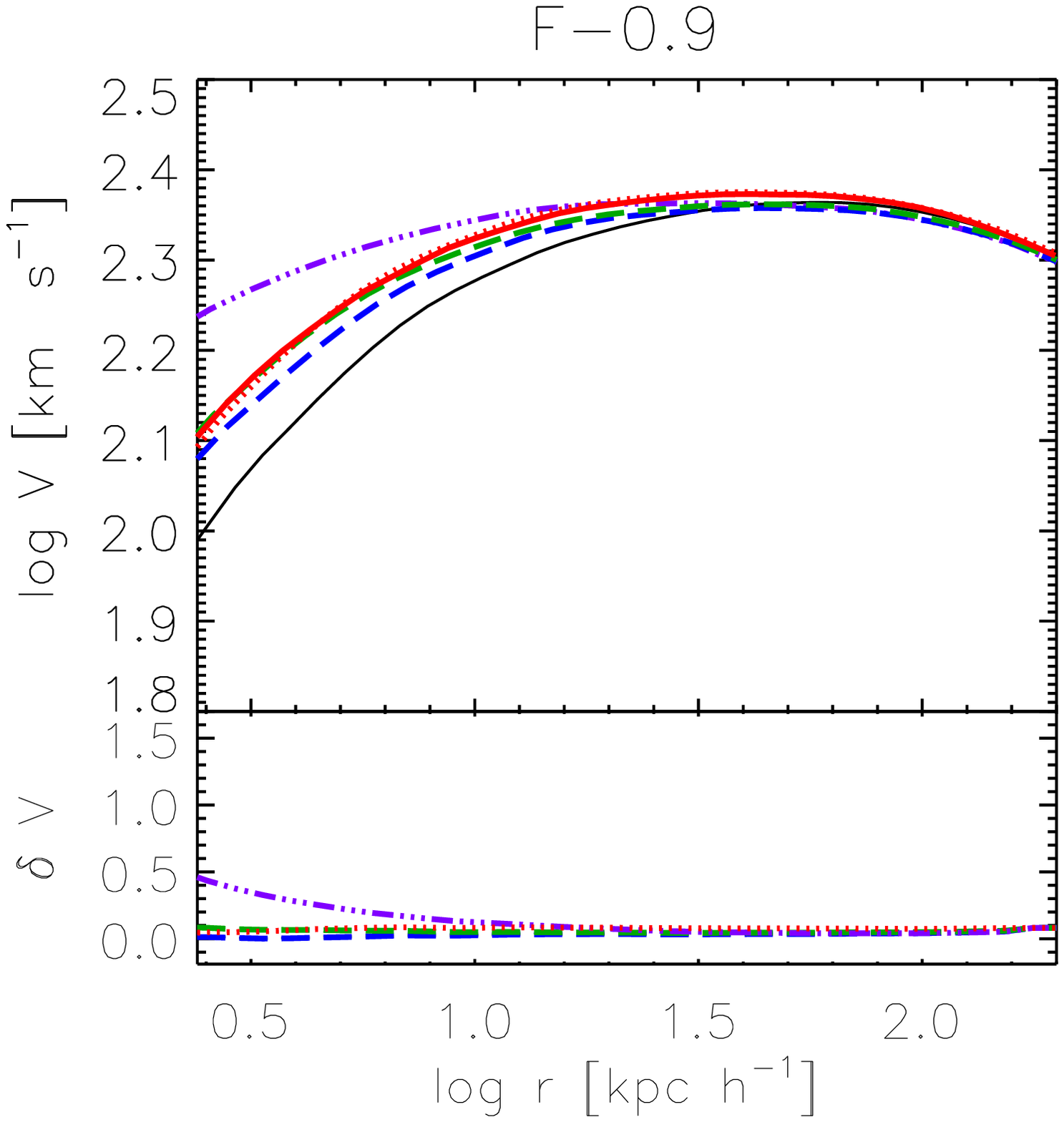}}
\hspace*{-0.2cm}\resizebox{6cm}{!}{\includegraphics{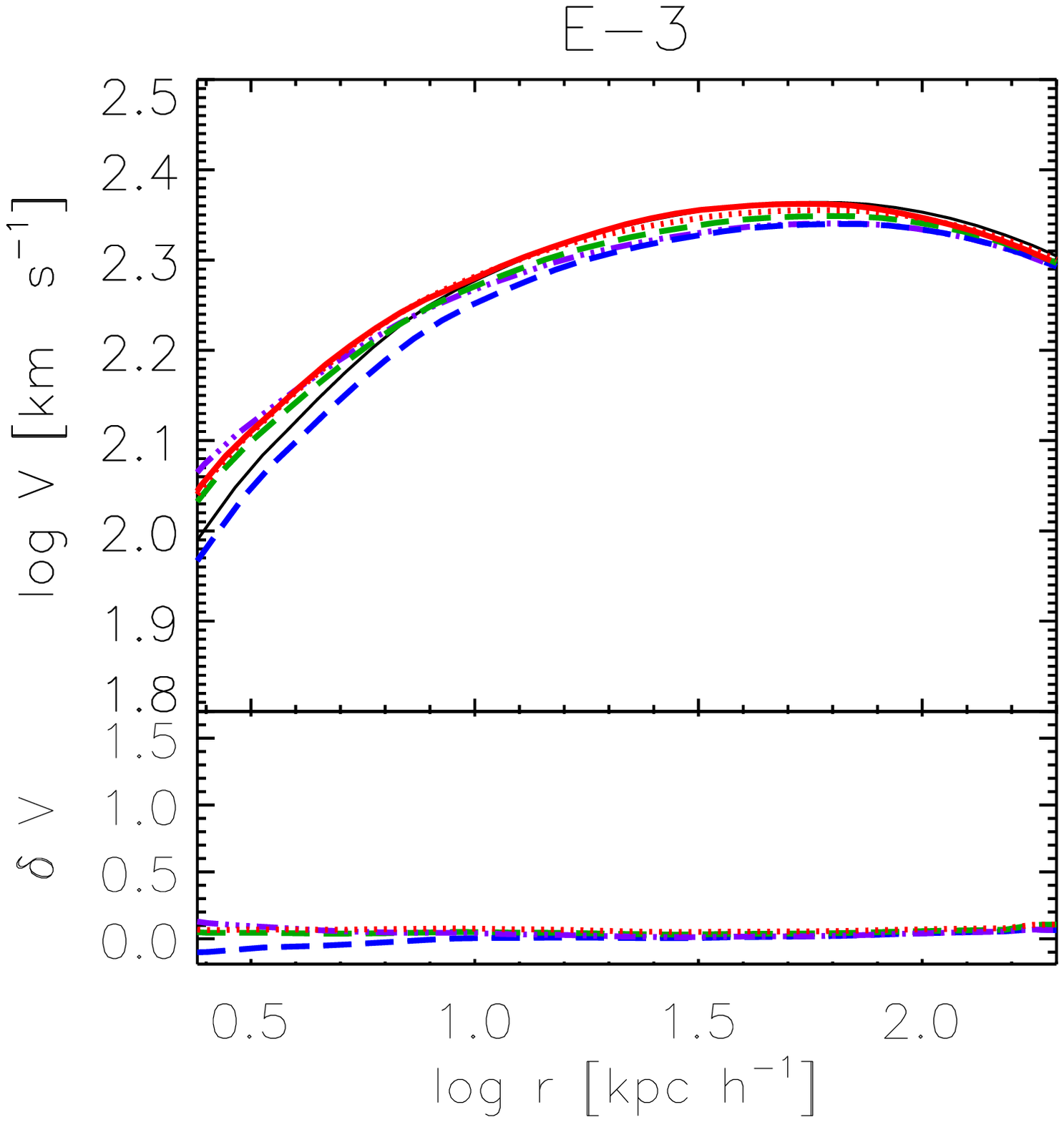}}
\hspace*{-0.2cm}
\caption{ Circular velocity obtained from our simulations with baryons (red lines), the DM-only run (black line), the B86  (dotted- dashed violet lines), the prescriptions of Gnedin et el. (2004) (blue dashed lines) and the A09 (green dashed lines), and our approximation (equation 2, red dotted lines). In the small, lower plots we show the residuals for each  velocity curve respect to that of the DM-only run.}
\label{AC2}
\end{figure*}

In Fig.~\ref{AC2}, we show the DM circular velocity obtained for each of our runs (red solid lines), the corresponding one from the DM-only run (black lines) and the velocities estimated by applying different AC models to the DM-only run taking into account the  baryonic distributions of the corresponding hydrodynamical runs. We found that, as expected, the B86 model (violet lines) largely overpredicts the level of concentration and also changes the shape of the DM distribution when compared to the DM profiles obtained from the cosmological runs.  The recipes of Gnedin et al. (blue lines) and A09 (green lines)  also overpredict the level of contraction, although the disagreement is not so large. Tissera et al. (2009) found a similar behaviour for their simulated haloes, which are roughly one order of magnitude higher in numerical resolution.

From the mass distribution of our simulations  we can estimate the best fitting for the relation between $\frac{r_{f}}{r_{i}}$ and $\frac{M_{i}}{M_{f}}$ (A09), where $r_i$ and $M_i$ correspond to the radii that contains a given number of particles and the total mass within that radii in the DM-only run, and $M_f$ is the final total (baryonic and DM) mass at $r_f$ estimated in the same way from each of our simulations including baryons. Following A09, we assume a function of the form:

\begin{equation}
\frac{r_{f}}{r_{i}} =  1 + a \times ((\frac{M_{i}}{M_{f}})^x - b)
\label{eq2}
\end{equation}

\begin{figure}
\hspace*{-0.2cm}\resizebox{7cm}{!}{\includegraphics{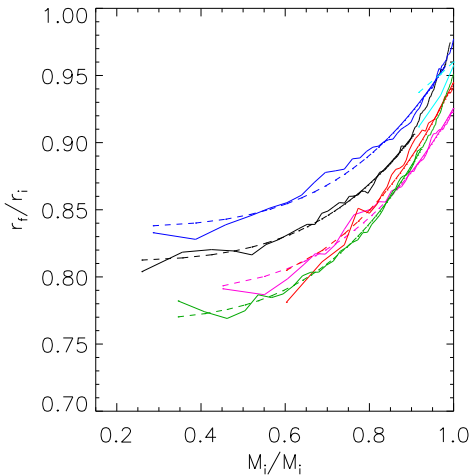}}
\hspace*{-0.2cm}
\caption{$\frac{r_{f}}{r_{i}}$ ratio vs $\frac{M_{i}}{M_{f}}$, NF (black lines), E-0.7 (magenta line), F-0.9 (red line), E-0.3 (green line), C-0.01 (blue line), E-3 (cyan line). The dashed lines correspond to the fits obtained according to Eq.2}
\label{AC1}
\end{figure}

We found that $x=4 $ is the exponent that best represent our mass relations. Keeping $x=4$ fixed, we fit the  $a$ and $b$ parameters in order to reproduce the level and shape of the contraction that we obtained in our simulations (Fig. \ref{AC1}). 
For the spheroid-dominated systems (NF, C-0.01 and E-3), we found that they are better reproduced with values of $a=0.14$ and $b=1.25$, while for haloes hosting important disc structures (E-0.7, F-0.9 and E-0.3), we get  $a=0.15$ and $b=1.40$. We have included the E-0.3 case among the ones with disc structure as it was able to develop a small and thick disc with a $D/S =0.6$ and, it presents a similar behaviour for the relation between radius and masses ratios than the other two runs with disc galaxies. The largest contraction predicted for haloes hosting a disc structure is in agreement with our previous results and discussion.

Consistently with previous findings,  Fig. ~\ref{AC2} shows that the response of the DM halo to the presence of baryons does not depend solely on the amount collected at the central region but is the result of the joint evolution of baryons and DM as they are assembled. Although it can be seen from the residuals in Fig. ~\ref{AC2} that our prescription provides a good prediction of the level of contraction, a larger sample of  galaxies is needed in order to claim its general validity.

\section{Conclusions}

We have studied the DM distribution in a set of runs of a  $\approx 10^{12} M_{\sun}$ halo extracted from a cosmological simulation, where the physics that regulates the  SF activity and SN feedback was varied allowing the formation of galaxies with different morphologies at $z=0$.
Since the underlying DM merger tree is the same in all our runs, the differences in the properties of the DM and baryons can be directly ascribed to the variations in the baryonic physics.
For the same reasons, all simulations have been run with the same numerical resolution so they are affected by resolution in a similar way. Because we are studying the DM distribution principally in the central regions, we have estimated all quantities outside three gravitational softenings.
And as a further test of the robustness of our results against numerical artifacts, we have studied the DM profiles in two haloes selected from a fully cosmological simulation with higher numerical resolution. These haloes host galaxies with different morphologies and reproduce remarkably well  our findings (appendix A).
  
 Our main results can be summarized as follows:

\begin{enumerate}

\item[1. ]We found that the Einasto's model provides the best fit for the spherically-averaged density profiles of our DM haloes. When baryons are present, the haloes become more concentrated in the central regions. However, the amount of baryons collected within the inner regions does not by itself  determine the response of the DM halo to the assembly process of the galaxy.

\item[2. ]When baryons are included, the velocity dispersion increases in the central region compared to the dissipationless case and no ``temperature inversion'' is observed, except for the  E-3 run where the fraction of baryons remaining in the halo is very small due to its strong SN feedback. The slope of the inner velocity dispersion profiles increases with increasing baryonic mass collected at the centre. We found that haloes hosting spheroidal galaxies tend to have weaker levels of velocity anisotropy than the DM-only run. While those haloes with an important disc galaxy show the highest levels of velocity anisotropy.

\item[3. ]The formation history plays an important role in the final distribution of its DM halo. We observed that those systems that are able to develop inside-out-formed discs, although they host in the central regions a lower amount of baryons than  galaxies that formed  old extended spheroids, have  more concentrated DM  profiles.
Since all our runs shared the same merger tree, the differences between them can be directly ascribed to their different baryonic evolutions  which are determined mainly by the SN feedback.

\item[4. ]We followed the evolution of the DM distribution in the central regions with redshift via the concentration parameter $\Delta_{v/2}$. We found that all haloes increase their concentration as they grow in time. As expected, the dissipationless run has the lowest concentration at all times. Also, we observed that  haloes present a flattening in this relation with respect to the critical relation indicating an expansion of the mass distribution in the central regions. This flattening differs among the different haloes being larger for those systems hosting an spheroidal galaxy. This trend can be linked with close approaches of satellites and their properties. 

\item[5. ] The analysis of  the satellites within the virial radius of the progenitor objects as a function of redshift yields that in the NF case, the satellites are clearly more massive and have been able to survive further in the halo, since stars are more gravitationally bounded. The systems that, at lower redshifts, developed a disc component (e.g E-0.7 and F-0.9) show less massive satellites at all redshifts as  a consequence of the action of SN feedback. As expected, the most diffuse satellites correspond to the E-3 case. By analysing the mass of the satellites within the virial radius, we found noticeable mergers or disintegration episodes which can be correlated with features in the specific
angular momentum content of the mass components. 

We studied the specific angular momentum content of main baryonic component and its halo from  $z \approx 1.6$, when the $\Delta_{v/2}$ starts to
clearly indicate an expansion of the central DM concentration. We selected the  NF and E-0.7 haloes as case studies.
For these cases, we found that even the inner $10\%$ of the stellar mass gains angular momentum as a function of redshift, although
the increase is more important for the NF case. The gas component shows a similar behaviour but, in this case, it is in the E-0.7 run where the
gas acquired larger amount of angular momentum  induced by the SN feedback which we know is successful at
driving galactic outflows (S08).
The DM halo (without substructure) increases its angular momentum content at all mass bins. Again, the larger profits are measured for
the NF case which has the most massive orbiting satellites. Hence, the mass components identified at $ z \approx 1.6$  in NF case  gained more angular momentum up to $z=0$ than their counterparts in E-0.7, except for the gas component which is anyway less massive than the other ones. The fact that stars migrate also contributes to change the inner potential well since they are the dominating mass component in the central region. This, on its turn,  probably acts
against further DM contraction.  Both effects could explain  the evolution of $\Delta_{v/2}$ and the fact that, when the halo hosts a disc-dominated galaxy, it is more concentrated than when it hosts a spheroid-dominated one.

\item[6. ]The baryonic rotation curves of our simulated galaxies reflect the presence of a concentrated dominating spheroid or a dominating disc component. The total circular velocity in the central regions is mainly determined by the baryonic component. When an important disc component is present the total velocity distribution gets flat in the central regions. The evolution of the baryonic circular velocity with time of the E-0.7 run shows that as the disc forms, the curve became flatter out to larger radii. We  quantified the flattening of the curve through the LS and correlated it with the shape parameter. Galaxies with lower values of LS tend to have haloes with larger $n$ parameter. 
We found values for $\frac{V_{max}}{V_{200}} $ between $\approx 1.15$ and $1.5$. The systems where a disc galaxy was able to form present the lowest ratios, in agreement with observational results.

\item[7. ]We have compared our simulated haloes with different AC prescriptions. We found, as expected, that the Blumenthal et al (1986) model overpredicts the level of concentration and also changes the shape of the DM distribution. The recipes of Gnedin et al. (2004) and A09 are an  improvement over the B86 approach. However, they overpredict the level of contraction when the haloes host an spheroid-dominated galaxy.
 From our analysis, we have obtained a prescription that provides a better representation for the contraction of our haloes,
depending on the morphology of the galaxy. However, the 'universality' of this prescription should be
tested with a larger statistical sample.

All our findings indicate that the response of the DM halo to the presence of baryons is the result of the joint evolution of baryons and DM during the assembly of the galaxy and in this sense,  the cosmological context for galaxy formation cannot be ignored.

\end{enumerate}

\section*{acknowledgement}
SP and PBT  acknowledge productive discussions with M. Abadi,  O. Valenzuela, M.E. De Rossi, T. Tecce and C. Artale.
This work was partially supported by PICT 32342(2005), PICT  Max Planck 245(2006) of Foncyt and DAAD-Mincyt collaboration(2007).
We thank the anonymous referee for her/his useful comments.

\appendix

\section{Numerical resolution analysis}

In order to analyse  the effects of numerical resolution on our results, we used a fully cosmological simulation corresponding  to a cubic box of a comoving 10 Mpc $h^{-1}$ side length and consistent with a $\Lambda$-CDM models (  $\Omega = 0.3, \Omega_{\Lambda} = 0.7, \Omega_{\rm bar} = 0.04, H_{0}=100 \, h \, \textrm{km} \, s^{-1} \textrm{Mps}^{-1}$ with  $h=0.7$
). This simulation has a mass resolution of $5.93\times 10^{6} h^{-1}\  $M$_{\odot}$ and $9.12\times 10^{5} h^{-1}\  $M$_{\odot}$ for the dark matter and gas components, respectively. The SN energy adopted for this simulation is 0.7 $\times10^{51}$ erg per  event and this  energy is pumped in equal fractions into the cold and hot phases. We have also run the pure dynamical counterpart (DM-only case) for purpose of comparison.
 
For this analysis, we selected two haloes (G1 and G2) of  $\approx 10^{12} M_{\sun}$ which host galaxies with different morphologies in order to validate the general trends obtained in the previous sections. Halo G1 has 846730 total particles (417129 of DM, 192527 of gas and 237074 of stars) within the virial radius, while Halo G2 has 333933 total particles (171789 of DM, 52177 of gas and 109967 of stars). We would like to stress the point that these galaxies do not share the same merger tree as it was the case for the ones in  the main set (Table \ref{tab1}). As a consequence and accordingly to our results, the effects of baryons on the DM distributions are expected to be different. However, the main patterns should be present and we will focus on their quantification. 

\begin{figure*}
\resizebox{5cm}{!}{\includegraphics{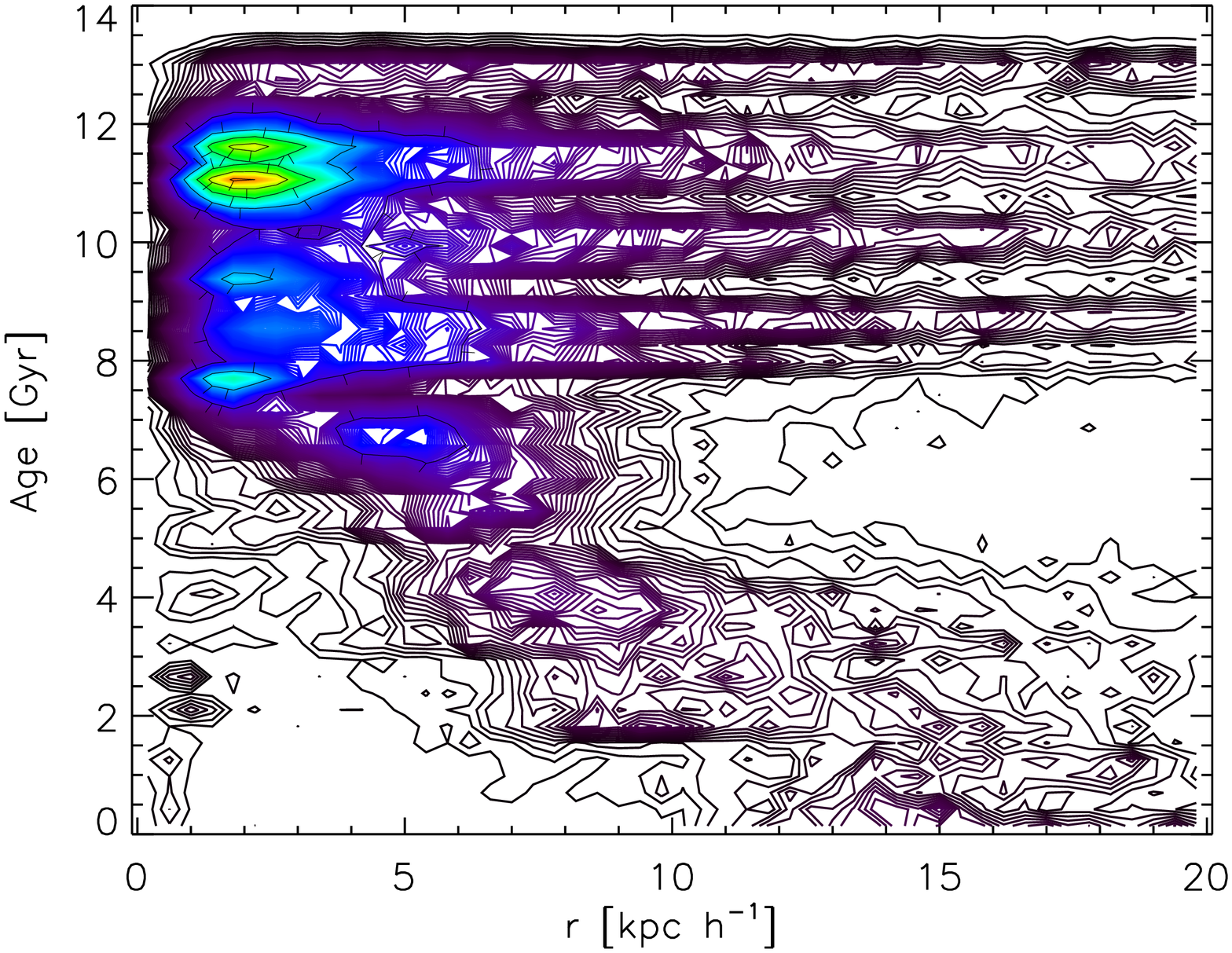}}
\resizebox{5cm}{!}{\includegraphics{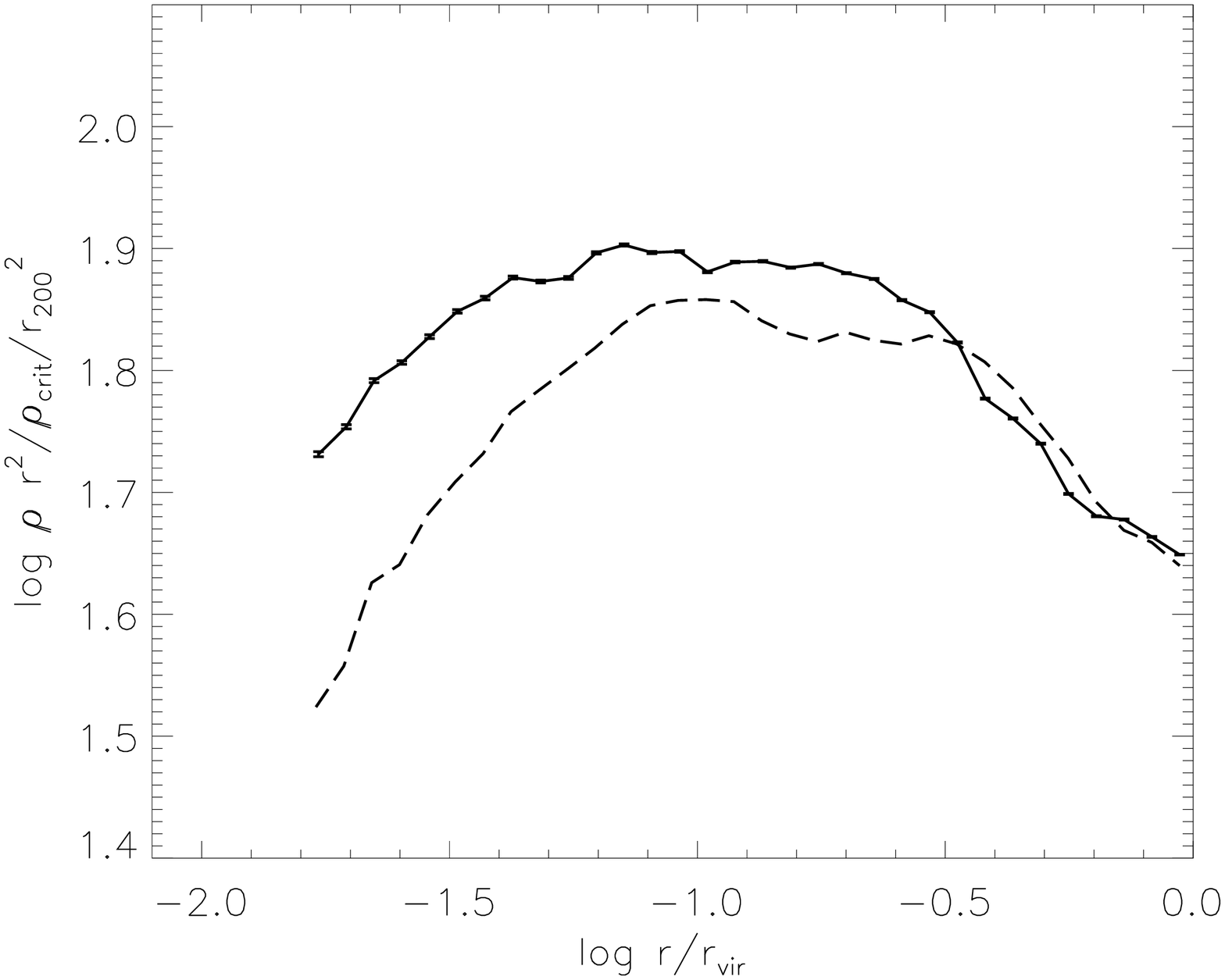}}
\resizebox{5cm}{!}{\includegraphics{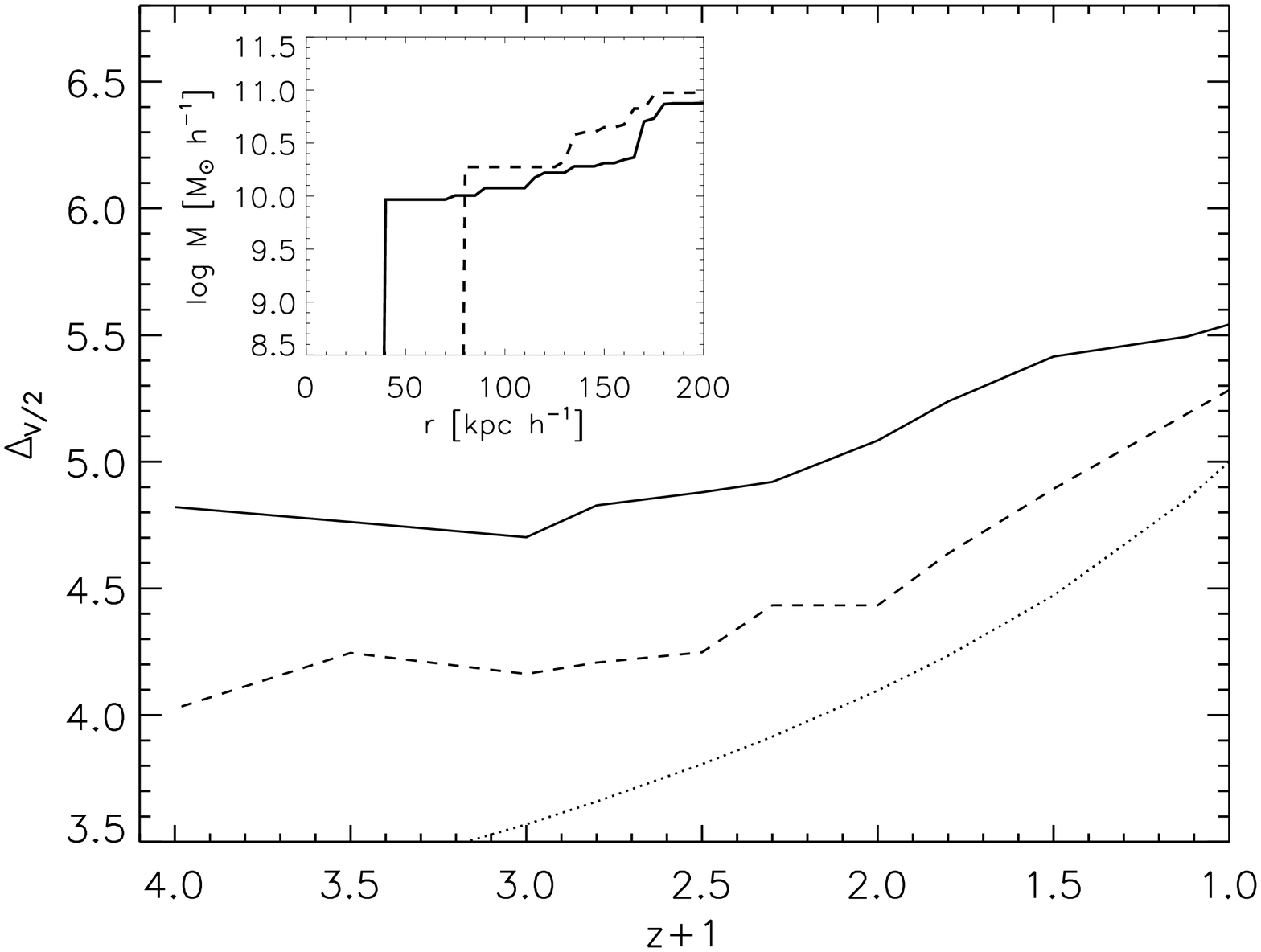}}\\
\resizebox{5cm}{!}{\includegraphics{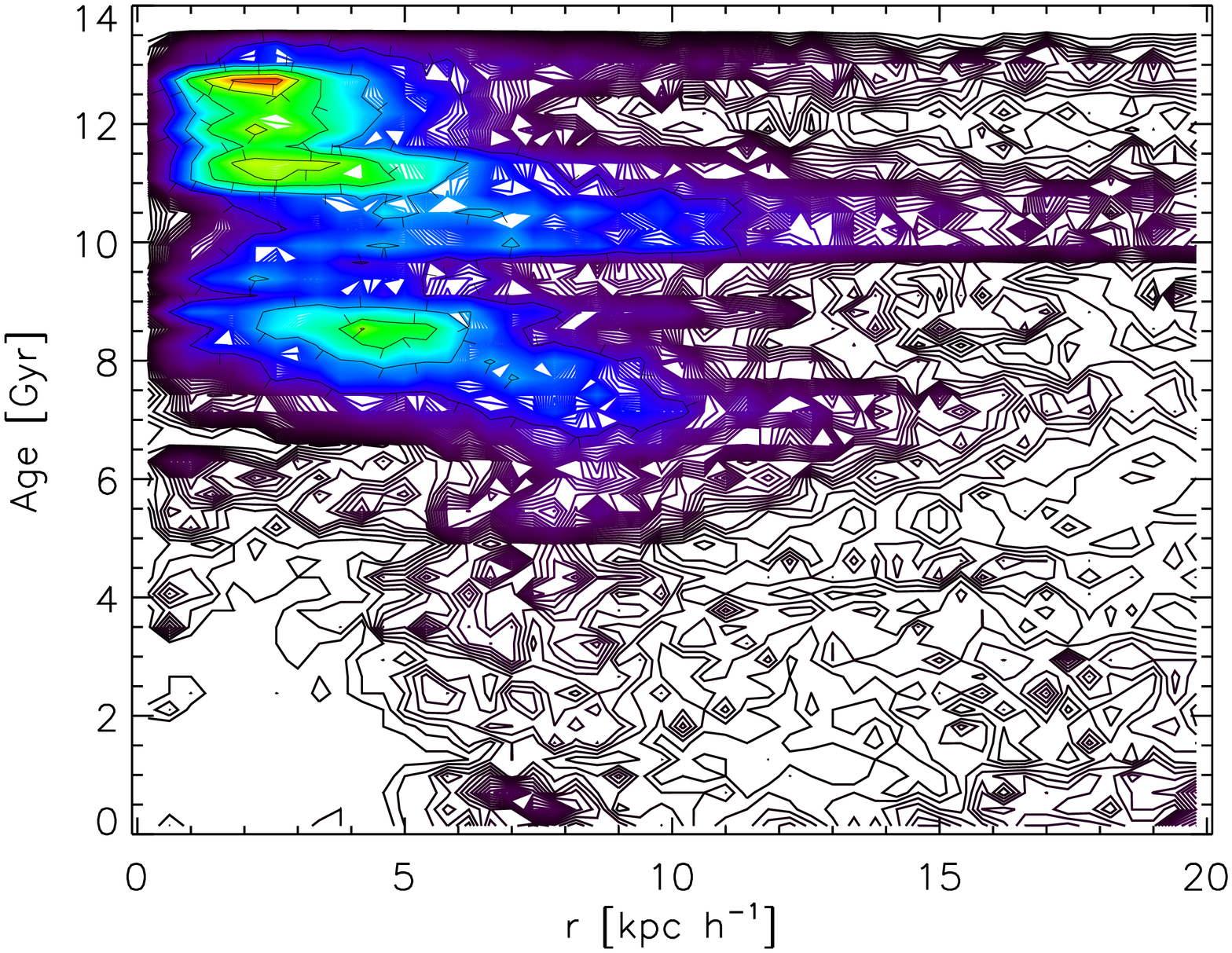}}
\resizebox{5cm}{!}{\includegraphics{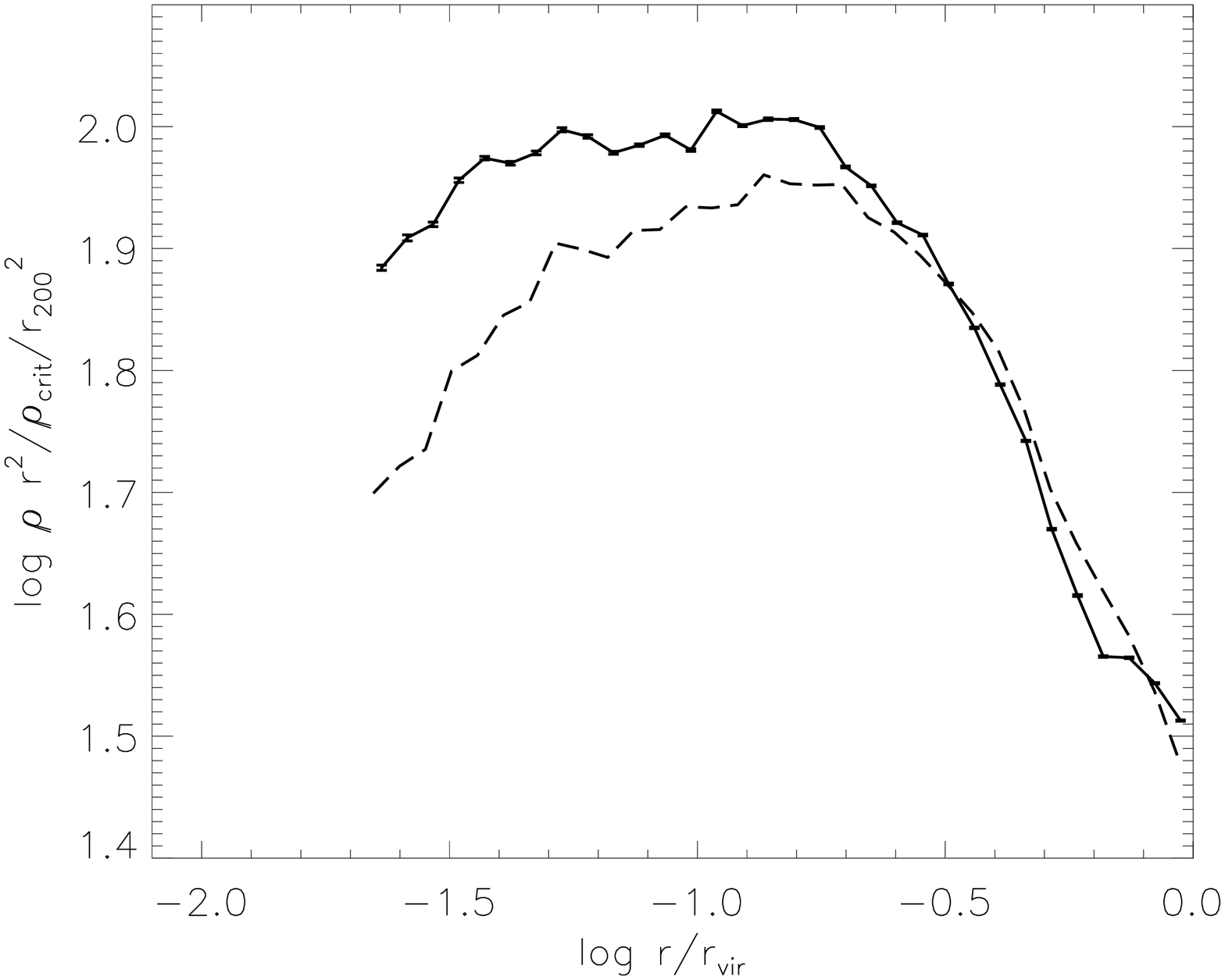}}
\resizebox{5cm}{!}{\includegraphics{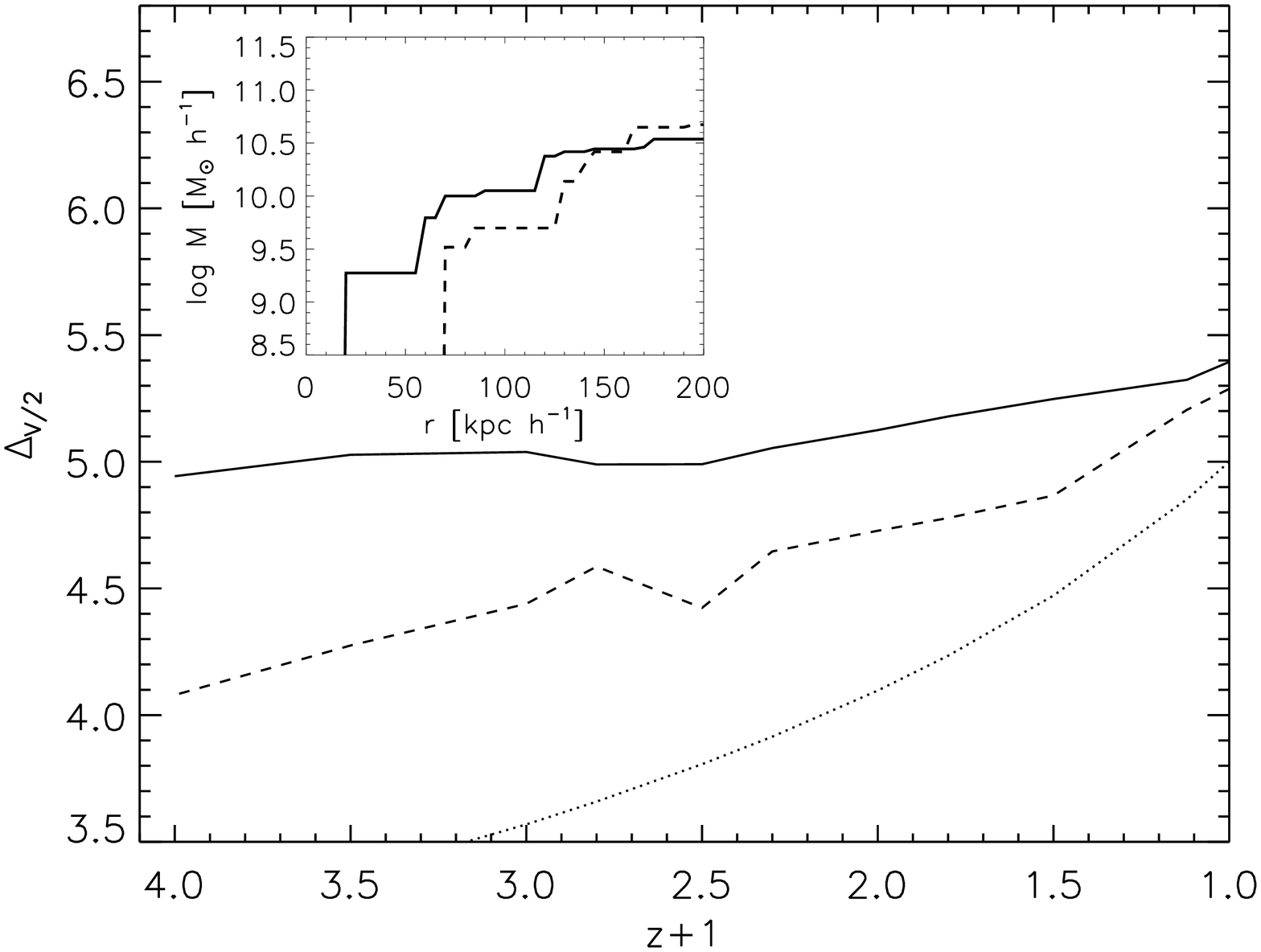}}
\hspace*{-0.2cm}
\caption{Age-radial maps for the stellar components (right panels), spherically-averaged DM profiles (central panels) and  the $\Delta_{v/2}$ parameters (left panels) as a function of redshift for the G1 (upper) and G2 (lower) haloes in a fully cosmological simulation (solid lines). The dashed lines correspond to the DM-only cases. Inset plots: cumulative mass of satellites at $z \approx 0.1$. } 
\label{profiles2}
\end{figure*}

In  Fig. \ref{profiles2} we display the age-radial distance maps of the stars (left panels),  DM profiles (middle panels) and $\Delta_{v/2}$ parameters (right panels)  for G1 (upper panels)  and G2 (lower panels) galaxies. 
From the age-radial maps we can appreciate that G1 has a more  important disc structure populated by younger stars than G2 where most of the stars are old and located in the central region.
As expected their DM profiles (solid lines) are more concentrated than their pure dynamical counterparts (dashed lines) as shown in the middle panels of Fig. \ref{profiles2}\footnote{The Einasto fitting parameters for G1 are $n=8.55$ and $r_{-2}=22.45$ and for its DM-only counterparts, $n=6.45$ and $r_{-2}=36.07$. In the case of G2, we get $n=5.36$ and $r_{-2}=14.41$ while for its DM-only we have  $n=4.29$ and $r_{-2}=22.22$.}. 
In order to assess the level of contraction and the evolution of the central density as a function of redshift, we estimated the 
$\Delta_{v/2}$ parameter for the progenitor systems. As it can be appreciated from the right panels,  both haloes are always more concentrated than the DM-only cases as expected.  However, the rate of increase in the concentration with time  in comparison to their pure dynamical counterparts is different. 
In the G1 case, $\Delta_{v/2}$ shows globally  a slightly sharper increase with redshift than in its pure dynamical counterpart. Conversely, in G2 
the relation is  significantly more flatten than its dynamical counterpart, suggesting a more important  expansion of the central mass distribution in the latter case.  In our previous discussion, we detected a correlation between the flattening in $\Delta_{v/2}$, the presence of more massive satellite systems and the amount of angular momentum transfer to the mass in the central region. We also found  a lower rate of increase in $\Delta_{v/2}$ associated with  systems dominated by an spheroidal galaxy.
Accordingly to these results,  the satellite systems in G1 and G2 should  show  clear differences: G1, which has an important disc, should have satellites less massive than its dynamical counterparts and vice versa, for G2.
To check this point, we  estimated the cumulative total mass of the  satellites within the virial radius  as a function radius. As an example, we show these distributions for $z \approx 0.1$ as insets in the right panels of Fig. \ref{profiles2} but these behaviors is common through out the evolution of the systems.
 In general, the satellites in the DM-only runs are located further away than their corresponding SPH runs at all redshifts.  In the case of G1,  we also note that the satellites in the SPH runs are less massive than it DM-only counterpart. The opposite situation is found for G2 where the satellites  are significantly more massive than its pure dynamical counterpart. 
These trends supports our main results.        

\begin{figure*}
\hspace*{-0.2cm}\resizebox{6cm}{!}{\includegraphics{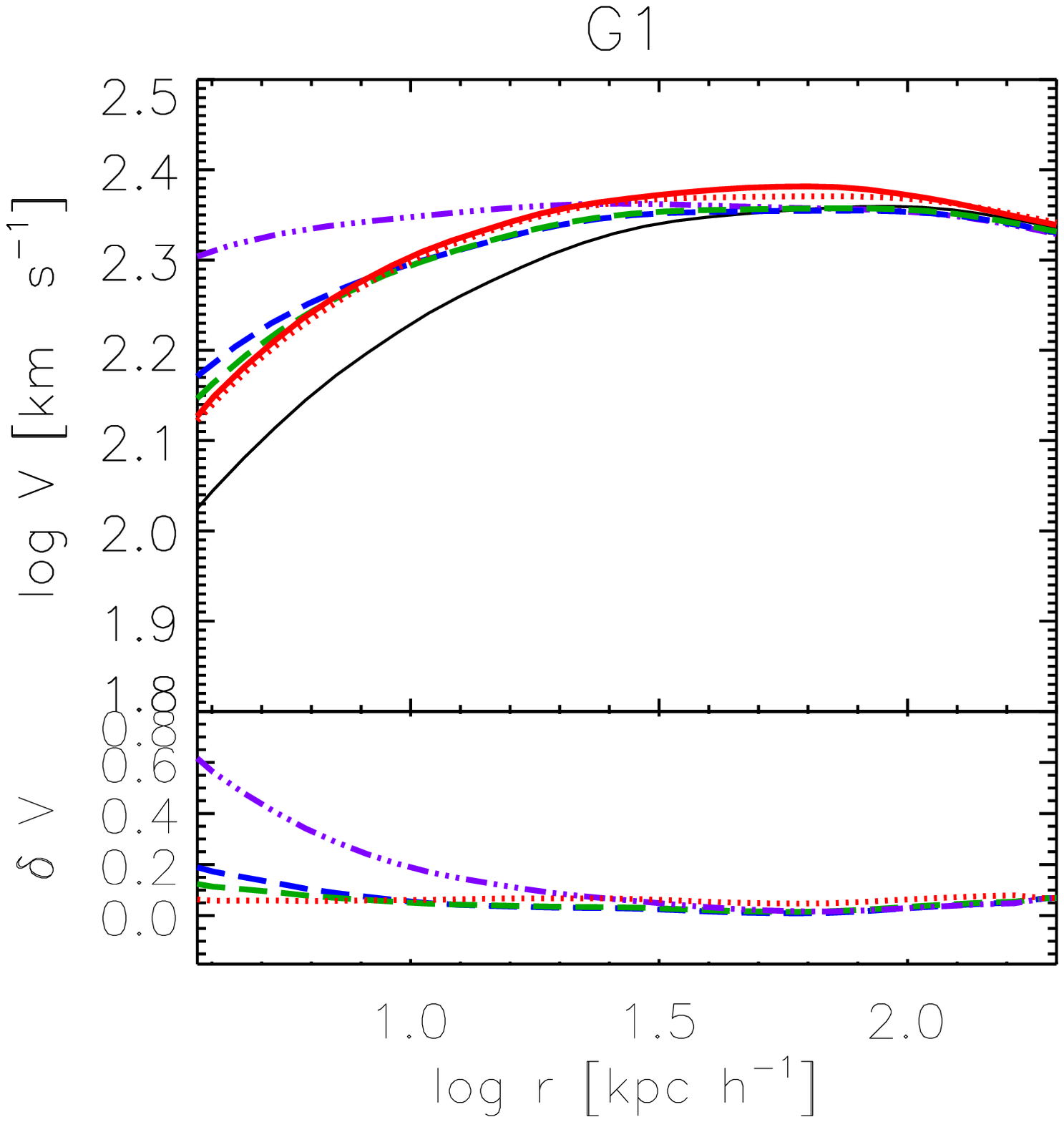}}
\hspace*{-0.2cm}\resizebox{6cm}{!}{\includegraphics{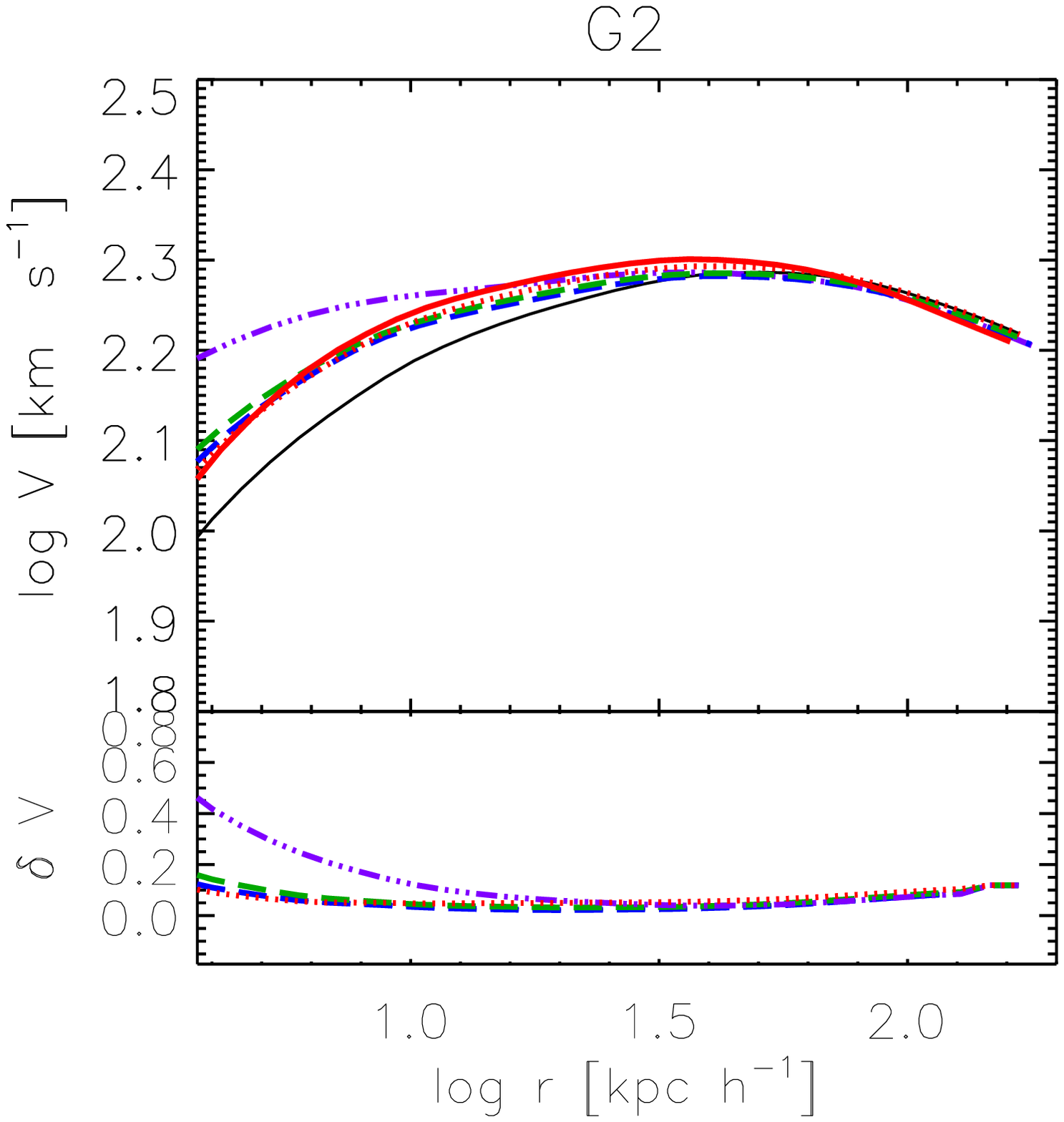}}
\hspace*{-0.2cm}
\caption{ Circular velocity obtained from haloes G1 and G2 selected from a fully cosmological simulations (red lines), their DM-only counterparts (black line), the B86  (dotted-dashed violet lines), the prescriptions of Gnedin et el. (2004) (blue dashed lines) and the A09 (green dashed lines), and our approximation (equation 2, red dotted lines). In the small, lower plots we show the residuals for each  velocity curve respect to that of the DM-only run.}
\label{AC3}
\end{figure*}

We have also tested our prescription for the adiabatic contraction in the G1 and G2 haloes as shown in Fig. ~\ref{AC3} . We found that  G1 is  better fitted by using our formula for the disc type systems. In G2, the differences between the residuals of the two fitting formulae are not so large as in the G1 case, but both formulae provide a  better prediction of the level of  contraction than all the others prescriptions.

\end{document}